\documentclass[preprint,5p]{elsarticle}

\RequirePackage[OT1]{fontenc}
\RequirePackage{amsthm,amsmath}
 
\usepackage[british,UKenglish]{babel}
\usepackage[utf8]{inputenc}
\usepackage{microtype}

\usepackage{bm}
\usepackage{bbm}
\usepackage{mathbbol}
\usepackage{amsmath}
\usepackage{mathtools}
\usepackage{amsfonts}
\usepackage{amssymb}
\usepackage{mathrsfs}
\usepackage{amsthm}
\usepackage{braket}
\usepackage{BOONDOX-calo}

\usepackage{tabularx}

\usepackage{color}
\usepackage{xspace}
\usepackage[dvipsnames,table]{xcolor}

\usepackage[colorlinks=true,linkcolor=blue,citecolor=blue,urlcolor=blue]{hyperref}

\usepackage{floatrow}
\usepackage[justification=raggedright, singlelinecheck=false]{caption}
\usepackage{subcaption}
\usepackage{pgfplots} 
\usepackage{tikz}
\usepackage{tikz-3dplot}
\usepackage{tikzscale}

\usepackage[toc]{appendix} 

\usepackage{floatrow}
\usepackage{stackengine}

\newcommand*{\eg}{e.g.\@\xspace}
\newcommand*{\ie}{i.e.\@\xspace}

\newcommand{\CS}{\textit{Code$\_$Saturne}\xspace}
\newcommand{\EM}{Euler--Maruyama\xspace}

\newcommand{\Ito}{It\^o\xspace}
\newcommand{\Stra}{Stratonovich\xspace}

\newcommand{\av}[1]{{{\displaystyle{\langle} {#1}\displaystyle{\rangle}}}}
\newcommand{\Id}{\mathbb{1}}

\newcommand{\EE}{\mathbb{E}}
\newcommand{\var}{{\mathbb{V}}\!{\textrm{ar}}}
\newcommand{\suba}{_{\scriptstyle a}}
\newcommand{\subs}{_{\scriptstyle s}}
\newcommand{\subba}{_{\scriptstyle ba}}
\newcommand{\subbs}{_{\scriptstyle bs}}
\newcommand{\supa}{^{\scriptstyle a}}
\newcommand{\sups}{^{\scriptstyle s}}
\newcommand{\supt}{^{\intercal}}
\newcommand{\Tr}{\text{Tr}}
\newcommand{\Pdf}{\mathcal{P}}
\newcommand{\dt}{\Delta t} 

\newcommand{\ErrS}{\mathcal{E}\textrm{rr}^{\textrm{str}}}
\newcommand{\ErrW}{\mathcal{E}\textrm{rr}^{\textrm{wk}}}

\newcommand{\nua}{\nu_{\!a}}
\newcommand{\nuf}{\nu}
\newcommand{\nus}{\nu_{\!s}}
\newcommand{\nuas}{\nu_{\!{a,s}}}
\newcommand{\taukol}{\tau_{\eta}}

\newcommand{\Xp}{{\bm{X}_{\! p}}}
\newcommand{\Up}{{\bm{V}_{\! p}}}
\newcommand{\Vf}{{\bm{V}_{\! f}}}
\newcommand{\Uf}{{\bm{U}_{\! f}}}
\newcommand{\Us}{{\bm{V}_{\! s}}}
\newcommand{\Zp}{{\bm{Z}_{\! p}}}
\newcommand{\Xpi}{{X_{\! p,i}}}
\newcommand{\Upi}{{V_{\! p,i}}}
\newcommand{\Ufi}{{U_{\! {f,i}}}}
\newcommand{\Vfi}{{V_{\! {f,i}}}}
\newcommand{\Usi}{{V_{\! s,i}}}

\newcommand{\Upj}{{U_{\! p,j}}}
\newcommand{\Ufj}{{U_{\! {f,j}}}}
\newcommand{\Vfj}{{V_{\! {f,j}}}}
\newcommand{\Usj}{{V_{\! s,j}}}
\newcommand{\pb}{\bm{p}}
\newcommand{\qb}{\bm{r}}
\newcommand{\bphi}{\bm{\phi}}
\newcommand{\Ac}{\mathbb{A}}
\newcommand{\Aij}{\mathbb{A}_{ij}}
\newcommand{\Bc}{\mathbb{B}}
\newcommand{\Cijkl}{\mathcal{C}_{ijkl}}
\newcommand{\Dc}{\mathcal{D}}

\newcommand{\Ic}{\mathbb{I}}

\newcommand{\Ktilde}{\mathbb{\widehat{K}}}
\newcommand{\Mc}{\mathbb{M}}
\newcommand{\Ntilde}{\mathbb{\widehat{N}}}
\newcommand{\Oc}{\mathbb{O}}
\newcommand{\Rc}{\mathbb{\Omega}}
\newcommand{\Rtilde}{\mathbb{\widehat{\Omega}}}
\newcommand{\Sc}{\mathbb{S}}
\newcommand{\Wc}{\mathbb{W}}
\newcommand{\Wca}{\mathbb{W}^{\scriptstyle a}}
\newcommand{\Wcs}{\mathbb{W}^{\scriptstyle s}}
\newcommand{\tWcs}{\widetilde{\mathbb{W}}^{\scriptstyle s}}
\newcommand{\Zc}{\mathbb{Z}}

\newcommand{\wa}{\bm{\mathcal{w}}^{\scriptstyle a}}

\newcommand{\avomega}{\braket{\bm{\omega}}}
\newcommand{\bmphiO}{\bm{\phi}_{\perp p}}
\newcommand{\phiP}{\phi_{\parallel p}}

\newcommand{\tumbE}{\text{TuR}_{\EE}}
\newcommand{\tumbV}{\text{TuR}_{\var}}
\newcommand{\spinE}{\text{SpiR}_{\EE}}
\newcommand{\spinV}{\text{SpiR}_{\var}}

\newcommand{\htumbV}{\widehat{\text{TuR}}_{\var}}

\newcommand{\Tlong}{{T_{\textrm{\tiny{Eq}}}}}

\newcommand{\Gc}{\reflectbox{\rotatebox[origin=c]{180}{$\mathbb L$}}}
\newcommand{\Wb}{\bm{W}}

\newcommand{\Hess}{\text{Hess}}

\theoremstyle{remark}

\definecolor{pyblue}{rgb}
{0.12156862745098039, 0.4666666666666667, 0.7058823529411765}
\definecolor{pyorange}{rgb}
{1.0, 0.4980392156862745, 0.054901960784313725}
\definecolor{pygreen}{rgb}
{0.17254901960784313, 0.6274509803921569, 0.17254901960784313}
\definecolor{pyred}{rgb}
{0.8392156862745098, 0.15294117647058825, 0.1568627450980392}
\definecolor{pyviolet}{rgb}
{0.5803921568627451, 0.403921568627451, 0.7411764705882353}
\definecolor{pybrown}{rgb}
{0.5490196078431373, 0.33725490196078434, 0.29411764705882354}
\definecolor{pypink}{rgb}
{0.8901960784313725, 0.4666666666666667, 0.7607843137254902}
\definecolor{pygrey}{rgb}
{0.4980392156862745, 0.4980392156862745, 0.4980392156862745}
\definecolor{pylime}{rgb}
{0.7372549019607844, 0.7411764705882353, 0.13333333333333333}
\definecolor{pyazure}{rgb}
{0.09019607843137255, 0.7450980392156863, 0.8117647058823529}
\definecolor{myblue}{rgb}{0.1, 0.5, 0.84}

\usepackage{lineno}

\newcommand{\Dt}{\Delta t}

\newcommand{\nbpbs}{\widetilde{\bm p}_{\scriptstyle bs}}

\newcommand{\hnpbsj}{\widehat{p}_{\scriptstyle bs,j}}

\newcommand{\hnbpbs}{\widehat{\bm p}_{\scriptstyle bs}}

\newcommand{\bmpa}{\bm{p}_{\scriptstyle a}}

\newcommand{\bmps}{\bm{p}_{\scriptstyle s}}

\newcommand{\Wm}{\mathbb{W}}
\newcommand{\Wms}{\mathbb{W}^{\scriptstyle s}}
\newcommand{\Wma}{\mathbb{W}^{\scriptstyle a}}

\newcommand{\qu}{\mathbb{q}}

\newcommand{\Rev}[1]{{{\leavevmode\color{black}{#1}\color{black}\xspace}}}

\journal{{\huge{{\!\!\!\!$\color{white}\blacksquare\blacksquare$}}}}

\begin{document}
%%%%%%%%%%%%%%%%%%%%%%%%%%%%%%%%%%%%%%%%%%%%%%%%%%%%%%%%%%%%%%%%%
%%%%%%%%%%%%%%%%%%%%%%%%%%%%%%%%%%%%%%%%%%%%%%%%%%%%%%%%%%%%%%%%%

\begin{frontmatter}



\title{Lagrangian stochastic model for the orientation of inertialess \Rev{spheroidal} particles in 
turbulent flows: an efficient numerical method for CFD approach}

\address[inria]{Universit\'{e} C\^{o}te d'Azur, Inria, CNRS, Sophia-Antipolis, France}

\author[inria]{Lorenzo Campana \corref{cor1}}
\ead{lorenzo.campana@inria.fr}

\author[inria]{Mireille Bossy}
\ead{mireille.bossy@inria.fr}

\author[inria]{Christophe Henry}
\ead{christophe.henry@inria.fr}

\cortext[cor1]{Corresponding author}

%----------------------------------------------------------------------------------------
%	ABSTRACT
%----------------------------------------------------------------------------------------
\begin{abstract}
In this work, we propose a model for the orientation of \Rev{inertialess spheroidal particles 
suspended in} turbulent flows. This model consists in a stochastic version of the Jeffery equation 
that can be included in a statistical Lagrangian description of particles suspended in a flow. It \Rev{is 
compatible and coherent with turbulence models that are widely used in CFD codes for the 
simulation of the flow field in practical large-scale applications}. In this context, we propose and 
analyze a numerical scheme based on a splitting scheme algorithm that decouples the orientation 
dynamics into its main contributions: stretching and rotation. We detail its implementation in an 
open-source CFD software. We analyze the weak and strong convergence of both the global 
scheme and of each sub-part. Subsequently, the splitting technique yields to a highly efficient 
hybrid algorithm coupling pure probabilistic and deterministic numerical schemes. Various 
\Rev{numerical} experiments were implemented and the results were compared with analytical 
predictions of the model to \Rev{assess the algorithm efficiency and accuracy}.
\end{abstract}


\end{frontmatter}
{{\small \tableofcontents}}
\clearpage

% \linenumbers

%******************************************************************************
% Introduction
%******************************************************************************
\section{Introduction}\label{sec:intro}

\subsection{General context}\label{sec:intro:context}

Investigating the dynamics of non-spherical particles suspended in turbulent flows is paramount to 
several industrial, biological and environmental applications. To name a few examples, one can cite 
clouds in the atmosphere (with the presence of complex-shaped ice-crystals 
\cite{pruppacher1998microphysics}), plankton in the ocean (such as diatom chains 
\cite{karp1998motion}), fibres in papermaking industries \cite{lundell2011fluid} or even bacteria 
\cite{koch2011collective}. One of the main difficulties is that non-spherical particles display both 
translational and rotational dynamics, which depend on particle properties (especially their shape 
and inertia). 
\Rev{Furthermore, in many of these applications, the flow is highly turbulent and it has a profound effect on the rotational dynamics of anisotropic particles: for instance, elongated slender fibers were shown to have their symmetry axis strongly aligned with the vorticity vector in homogeneous isotropic turbulence \cite{pumir2011orientation} while preferential particle orientations occurs in near-wall turbulence \cite{zhao2015rotation}.
}

Despite significant progress over the last decades (see the reviews by \citet{voth2017anisotropic} and \citet{du2019dynamics}), the dynamics of non-spherical particles is still faced with various challenges. \Rev{This includes the development of new experimental techniques (especially to better capture the dynamics of complex-shaped particles or to track individual particles within dense suspensions \cite{voth2017anisotropic}), as well as numerical models (particularly to handle arbitrarily shaped particles \cite{voth2017anisotropic}, fluid-structure interactions with flexible fibers \cite{du2019dynamics} or models that are compatible with large-scale simulations).}

\Rev{
In the following, we focus on the development of a stochastic model that reproduces the key 
features of the dynamics of isolated inertialess spheroids immersed in turbulent flows and that is 
compatible with large-scale simulations of dispersed two-phase flows
}.

\subsection{Modeling the dynamics of spheroids in suspension: existing approaches and 
limitations}\label{sec:intro:limits}

\Rev{Various types of Lagrangian models have been suggested in the literature to capture the dynamics of spheroidal particles suspended in turbulent flows. These models differ depending on the 'level of description' used as well as on the 'information content' \cite{minier2016statistical}. In particular, these models vary depending on how they handle: (a) the simulation of turbulent flows; (b) the representation of non-spherical particles (\ie, in terms of degrees of freedom); (c) the coupling between continuous (fluid) and discrete (particle) phases. In the following, we briefly recall the key existing models in the multiphase flow community.

Simulations of turbulent flows can be performed either by solving explicitly all the spatial and temporal scales involved (as in Direct Numerical Simulations, DNS) or by using reduced descriptions of turbulence (such as Large-Eddy Simulation, LES, or Reynolds-Averaged Navier-Stokes approaches, RANS) \cite{pope2000turbulent}. LES consists in solving explicitly only the largest scales and modeling sub-grid scales while RANS approaches describe only the average velocity and its fluctuation. Thanks to the reduced number of degrees of freedom to be solved, these reduced turbulence models are widely used in the CFD community, especially since these approaches remain tractable even when dealing with practical situations (\eg flows in complex geometries or atmospheric flows). When particles are immersed in the fluid, an additional complexity arises depending on the ratio between the particle size $d_p$ and the smallest fluid scales (here the Kolmogorov dissipative scale $\eta$) \cite{kuerten2016point}: first, when $d_p^*=d_p/\eta\gg1$, particles are treated as finite-sized objects around which the fluid flow is explicitly solved (particle-resolved DNS, or PR-DNS \cite{uhlmann2005immersed}); otherwise, particles can be treated as infinitely small points (leading to the so-called Point-Particle DNS, or PP-DNS \cite{allende2018stretching, dotto2020deformation}). 

Meanwhile, spheroidal particles are usually described as a single object in the multiphase flow community (in contrast with molecular approaches which describe a particle as an ensemble of bounded molecules, as in \cite{yamamoto2013molecular}). Yet, the number of degrees of freedom depends highly on the description chosen:
\begin{itemize}
 \item First, when dealing with flexible thin filaments (\ie elongated, thin and deformable rods), one of the most detailed descriptions is based on the slender body theory \cite{du2019dynamics}. It is derived from the beam theory for thin solids and it describes the relation between elastic and drag forces on thin filaments. 
 \item Second, a reduced description can be used drawing on bead-rod models \cite{hamalainen2011papermaking, dotto2020deformation}, where a filament is represented by a 1D chain of linked rigid bodies. Going further along this idea of reducing the degrees of freedom, elongated spheroids can be further simplified with only three beads (trumbbells) or even two beads (dumbbells) connected to each other. The interest of these reduced descriptions is that they allow to capture some of the information at relatively low computational costs. For instance, dumbbells can accurately predict the elongation of deformable particles \cite{vincenzi2007stretching} while trumbbells can also provide data on the bending of semiflexible particles \cite{ali2016semiflexible}.
 \item Third, when dealing with rigid bodies, the representation can be further simplified by describing the motion of a spheroid, whose size is fixed (as in \cite{mortensen2008dynamics, marchioli2016relative}). 
\end{itemize}
 
Simulating the dynamics of isolated spheroids in turbulent flows further requires to couple a model 
for the fluid phase and for the particle phase. To that end, various formulations can be used. For 
very large spheroids, the flow can be solved with PR-DNS and the hydrodynamic forces/torques 
acting on each spheroid are obtained by direct integration of the velocity across the particle surface 
\cite{jain2021impact}. Alternatively, point-particle approximations are widely used for small 
spheroids in combination with Jeffery's equation for the translational and rotational dynamics of rigid 
spheroids \cite{jeffery1922motion}. Jeffery's equation requires knowledge of the velocity gradient at 
the particle position, which is naturally obtained from PP-DNS in both homogeneous isotropic 
turbulence \cite{ali2016semiflexible} and wall-bounded turbulent flows 
\cite{marchioli2010orientation, marchioli2016relative}). These studies have already provided detailed 
information on the dynamics of spheroids in turbulent flows. For instance, some studies have 
highlighted the complex rotational behavior of such axisymmetric particles in both homogeneous 
isotropic turbulence (HIT, see \cite{parsa2012rotation, gustavsson2014tumbling}) and in turbulent 
channel flows \cite{zhao2015rotation}. In particular, these studies have revealed the importance of 
investigating the complex rotation of spheroids, characterized by: the spinning rate (\ie the rotation 
of the particle around its own symmetry axis) and the tumbling rate (\ie the rotation along the two 
other directions). 

Yet, the coupling of a model for the fluid phase and another one for the particle phase imposes to 
use models that are compatible in terms of the `information content'. This places an additional 
constraint on the development of a model for the dynamics of spheroids. For instance, coupling 
Jeffery's equation for small rigid spheroids to a turbulence model is not straightforward since the 
instantaneous velocity gradient at the particle position is not directly accessible (as would be the 
case if coupled to a DNS). This has led to the development of a number of Lagrangian stochastic 
models for the velocity gradient tensor (see the review by \citet{meneveau2011lagrangian} and 
papers like \cite{pumir2011orientation, chen2016large, johnson2016closure, 
johnson2018predicting}). For instance, one of the first attempts was to couple an existing model for 
the velocity gradient \cite{pumir2011orientation} to an LES in HIT to study the orientation of 
anisotropic particles \cite{chen2016large}. In these works, the particle dynamics is exactly solved for 
an observation time of the order of the Kolmogorov time scale by using the modeled information 
about the velocity gradient tensor.

In the case of hybrid approaches that couple a RANS simulation of the fluid phase to Lagrangian tracking methods for the dispersed phase, an additional constraint comes from the fact that the model should remain valid even for large time steps, possibly much larger than the Kolmogorov time $\taukol$. To the authors' knowledge, this remains an unexplored field since such stochastic Lagrangian models have only been developed in the case of spherical particles (see reviews \cite{minier2001pdf, minier2016statistical}).
}

\subsection{Objectives}\label{sec:intro:objectives}

Drawing on the limitations introduced above, this paper is focused on \Rev{developing a new model that reproduces the key features of the orientation of isolated inertialess rigid spheroids immersed in turbulent flows}. The present methodology relies on an existing Lagrangian stochastic model for spherical particles (see \eg \cite{minier2001pdf}). It then proposes an extension to describe the orientation dynamics of spheroids as well as the rotation dynamics (specifically the tumbling and spinning rates). In addition, a numerical scheme is suggested to implement this model in an existing CFD code (namely \CS). For that purpose, an efficient and accurate numerical method based on splitting schemes is suggested to obtain an algorithm that matches the requirements of CFD software for industrial applications (in terms of efficiency, compatibility, tractability and convergence).

In that context, the aim of this paper is four-fold:
\begin{enumerate}
	\item To extend existing stochastic Lagrangian models for spherical particles, in order to handle \Rev{spheroidal} particles of arbitrary axisymmetric shape;
	\item To develop a new formulation that allows to obtain the rotation statistics (including tumbling and spinning rates) even in the case of stochastic differential equations (which do not allow to compute time derivatives of a process);
	\item To evaluate the accuracy of the numerical scheme retained as well as its convergence in simple ideal cases;
	\item To assess the robustness of the numerical scheme in a realistic applicative setting.
\end{enumerate}

\subsection{Layout of the paper}\label{sec:intro:layout}
 
For that purpose, the paper is organized as follows. The refined model for the dynamics of isolated 
inertialess spheroids is presented in Section~\ref{sec:SDE_model}. In particular, we start by briefly 
recalling the existing stochastic Lagrangian models for the translational dynamics of inertialess 
spherical particles in Section~\ref{sec:SDE_model:translation}. Then, the stochastic Lagrangian 
model for the orientation of spheroids is detailed in Section~\ref{sec:SDE_model:orientation}. The 
numerical method proposed to solve such SDEs is described in Section~\ref{sec:num}. More 
specifically, we give first an overview of the splitting algorithm (see Section~\ref{sec:num:splitting}). 
\Rev{Next, we introduce the numerical schemes retained to each sub-equation of the splitting 
approach, with particular attention to the two Brownian differential equations to be integrated}, 
followed by a presentation of both weak and strong convergence test results in the ideal case of HIT 
(see Section~\ref{sec:num:valid}). Finally, the splitting scheme is tested in the case of a 
homogeneous shear flow (see Section~\ref{sec:assess}), where we assess the impact of shear both 
on the long-time equilibrium for the particle orientation PDF (see Section~\ref{sec:assess:orient}) 
and on the tumbling \& spinning rates (see Section~\ref{sec:assess:tumb-spin}).

%******************************************************************************
% Lagrangian stochastic model for the orientation
%******************************************************************************

\section{Lagrangian stochastic model for the orientation}
 \label{sec:SDE_model}

\begin{figure*}[ht!]
	\centering
	\resizebox{\textwidth}{!}{\includegraphics{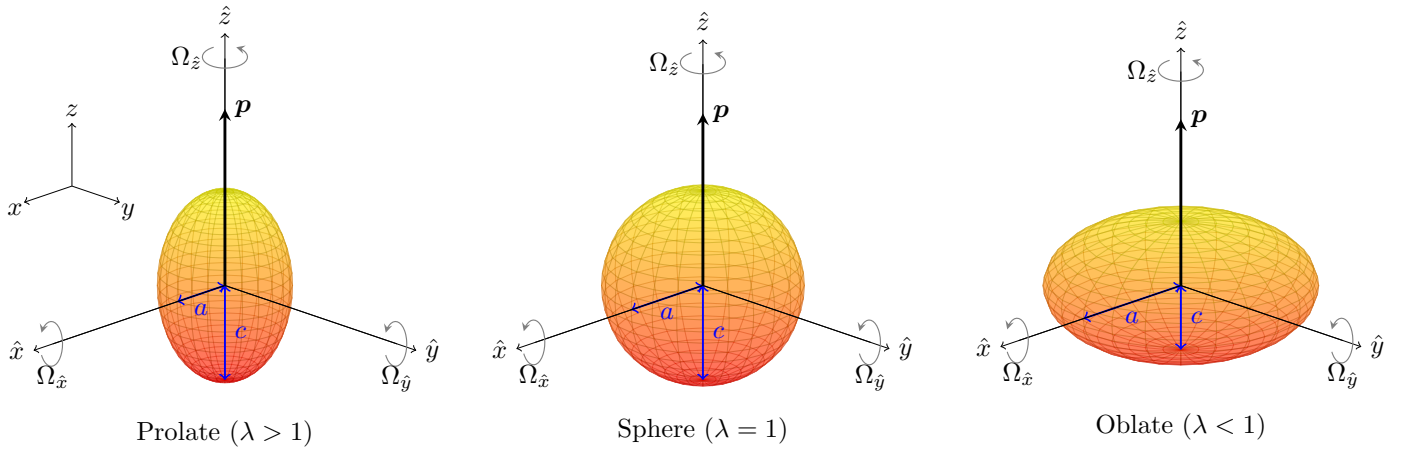}} 
	\caption{\label{fig:spheroids}
		Illustration of ellipsoids with different aspect ratio $\lambda = c/a$, with $a$ equals to the 
		equatorial radius of the spheroid and the semi-axis $c$ is the distance from center to pole 
		along the symmetry axis, always aligned with the $\hat{z}$-axis of the particle fixed coordinate 
		system, comprising the orientation vector $\pb$. The rotations of axisymmetric particles are 
		decomposed into a component ($\Omega_{\hat{z}}$) along the symmetry axis, called spinning, 
		and components perpendicular to the symmetry axis ($\Omega_{\hat{x}}$, 
		$\Omega_{\hat{y}}$), called tumbling.}
\end{figure*}

\Rev{We consider here the case of an isolated spheroid, which can be characterized by two 
parameters (see also Fig.~\ref{fig:spheroids}): its semi-major length $c$ (i.e. along its symmetry 
axis) and its semi-minor length $a$. Defining the spheroid aspect ratio $\lambda = c/a$ allows to 
distinguish between prolate spheroids ($\lambda>1$), spheres ($\lambda=1$) or oblate spheroids 
($\lambda<1$). Alternatively, one can rely on the shape parameter, which is defined as} $\Lambda = 
\frac{\lambda^2-1}{\lambda^2+1}$. The interest of this shape parameter $\Lambda$ is that it is 
bound between $-1$ (infinite disk) and $+1$ (infinite rods), with $\Lambda = 0$ corresponding to 
spheres.

The first question to address when developing a model is to identify what are the important variables that we want to capture. Here, we are interested in tracking the dynamics of spheroids in turbulent flows. This means that some level of information is required for the translational and rotational motion of such particles. In the following, for the sake of clarity and completeness, we first briefly review existing models for the translational dynamics of spheroids (see Section~\ref{sec:SDE_model:translation}) before detailing a stochastic model for their orientation (see Section~\ref{sec:SDE_model:orientation}).

 \subsection{Existing Lagrangian approaches for the translation dynamics}
  \label{sec:SDE_model:translation}
  
In the framework of Lagrangian models, a system of equations is written to describe the time-evolution of variables attached to each particle: this constitutes the state vector $\Zp$ associated to a particle.

\Rev{Here, we focus on the case of rigid spheroids suspended in a fluid flow solved with a 
turbulence model. Spheroids are assumed to be much smaller than the smallest active scale of the 
fluid velocity (\ie the Kolmogorov length scale $\eta = (\nuf^{3}/\varepsilon)^{\frac{1}{4}}$, with 
$\varepsilon$ the turbulent dissipation rate and $\nuf$ the fluid kinematic viscosity). This allows to 
use the point-particle approximation and, since the particle does not deform, to track the motion of 
its center of mass only \cite{kuerten2016point}. This means that the state vector simplifies here to 
$\Zp=(\Xp, \Up, \Us)$, with $\Xp$ the center-of-mass position, $\Up$ its velocity and 
$\Us=\Uf(t,\Xp(t))$ the fluid velocity seen which corresponds to the fluid velocity sampled at the 
particle position at a given time. Furthermore, we consider the case of spheroids that are sufficiently 
dilute to neglect their interaction and their feedback on the flow. Thus, spheroids behave as tracers, 
\ie particles that follow the fluid streamlines such that their velocity is equal to the fluid velocity. 
Hence, the state vector of inertialess spheroids can be further reduced to $\Zp=(\Xp, \Vf)$, where 
$\Vf$ is the instantaneous fluid velocity sampled at the particle position. 

The corresponding system of stochastic differential equations (SDEs) for the time-evolution of the variables attached to each spheroid is (more details in \cite{minier2001pdf, minier2016statistical}):
\begin{subequations}
 \label{eq:SLM_tracer}
 \begin{align}
  d\Xpi =& \, \Ufi \, dt \label{eq:SLM_tracer_xp} \\
  d\Vfi =& \, - \frac{1}{\rho_{\! f}} \frac{\partial \braket{P_{\!f}}}{\partial x_i} dt + G^*_{ij} \left(\Vfj - \braket{\Ufj}\right) dt  \nonumber \\
  & + B_{s,ij} \, dW_j(t),
 \end{align}
\end{subequations}
It is worth noting that this stochastic Lagrangian model is introduced because of the consistency in 
terms of ``information content'' between the fluid simulation and the Lagrangian tracking. In fact, we 
rely here on a RANS approach to compute reduced information on the flow field. RANS models are 
based on a Reynolds decomposition of the fluid velocity in terms of its average and fluctuating part, 
\ie $\Uf = \braket{\Uf}+\bm{u}'_{\! f}$. This implies that RANS simulations provide information on the 
mean flow field $\braket{\Uf}$, the mean pressure field $\braket{P_{\!f}}$, the Reynolds-stresses 
$\braket{u'_{\!f,i} u'_{\!f,j}}$ and the turbulent dissipation $\epsilon$. Hence, the instantaneous fluid 
velocity at any point in space $\Uf(\Xp(t),t)$ is not directly available but has to be modeled. As 
detailed elsewhere \cite{minier2001pdf}, the Langevin model for the fluid velocity of tracers $\Uf$ 
samples an instantaneous fluid velocity for each particle using information from the flow field 
(namely the mean pressure $\braket{P_{\!f}}$, the mean velocity $\braket{\Uf}$ and the turbulent 
dissipation $\epsilon_f$). The precise expressions used to close the matrices $G^{*}_{ij}$ and 
$B_{s,ij}$ have been detailed elsewhere \cite{minier2001pdf,minier2016statistical}. Details on the 
numerical implementation of such a stochastic Lagrangian approach are available in 
\cite{peirano2006mean} as well as in the open-source software \CS. At this stage, it is worth noting 
that the algorithm implemented in \CS for the translational dynamics is written for the more general 
case of inertial spherical particles (the case of tracers with vanishing inertia is naturally recovered by 
setting the particle relaxation time to $\tau_p =0$, as briefly recalled in in~\ref{app:A}).}

\subsection{Extended Lagrangian stochastic model for the orientation} 
  \label{sec:SDE_model:orientation}

\Rev{Since we are interested in capturing the key features for the dynamics of spheroids, the state vector should contain information not only on the translational velocity but also on the rotational velocity. This raises the question of which variable associated to the particle rotational velocity should be introduced in the model.} 

To address this question, we start by briefly recalling existing models for the orientation of spheroids 
in the context of fully resolved turbulent flows (\ie based on DNS). Then, we derive a stochastic 
model for spheroid orientation based on the stochastic modeling of the velocity gradient tensor 
$\nabla \Uf$. The associated rotational stochastic dynamics and tumbling, spinning statistics are 
further analyzed.

\subsubsection*{Jeffery's equation for the orientation}
 \label{sec:SDE_model:orientation:jeffery}

% Many numerical and theoretical studies are based on Jeffery's equation \cite{jeffery1922motion} to predict the time-evolution of a spheroid orientation along its trajectory in a turbulent flow. Indeed, Jeffery's equation has been used in DNS to capture the dynamics of a spheroid orientation vector $\pb(t)$ in time (see \eg \cite{shin2005rotational, pumir2011orientation, zhang2001ellipsoidal}).

% More precisely, for inertialess spheroidal particles, their orientation $\pb$ corresponds to the unit vector denoting the direction of the principal axis $c$. When expressed in terms of the particle shape parameter $\Lambda$ (see Fig.~\ref{fig:spheroids}), Jeffery's equation for their orientation dynamics gives \cite{jeffery1922motion}

Jeffery's equation expresses the time evolution of the spheroid orientation vector $\pb(t)$ \cite{jeffery1922motion}
\begin{align}\label{eq:jeffery}
 \begin{aligned}
  & \frac{d\pb}{dt} = \Bc \, \pb - \left( \pb\supt \Bc \, \pb \right) \, \pb , 
  \quad \mbox{with } \Bc = \Oc + \Lambda \, \Sc, \\
  &\qquad \Oc(t) = \tfrac{1}{2}(\nabla \Uf - \nabla\Uf\supt)(\Xp(t),t), \\
  &\qquad \Sc(t) = \tfrac{1}{2}(\nabla\Uf+\nabla\Uf\supt)(\Xp(t),t). 
\end{aligned}
\end{align}
$\Sc$ and $\Oc$ denote the rate-of-strain tensor and the rate-of-rotation tensor, which are the symmetric and antisymmetric parts of the velocity gradient tensor along a generic Lagrangian trajectory $\Xp(t)$. 

Equation~\eqref{eq:jeffery} is a non-linear vector equation, and it could be seemingly complex to solve. However, the non-linearity is only a geometric constraint to preserve the unitary norm of $\pb$. Due to the antisymmetric nature of the tensor $\Oc$, the contribution $(\pb\supt\Bc\,\pb)$ restricts naturally to $\Lambda(\pb\supt\Sc\pb)\pb$. This non-linear stretching along $\pb$ is thus continuously subtracted from the contribution $\Bc \, \pb$ to prevent any elongation of $\pb$. \citet{bretherton1962motion} observed that one might equivalently model the orientation of the particle with any vector $\qb$, which obeys the same linear terms without compensating for any elongation:
\begin{equation}
 	\label{eq:linear_jeffery}
	 \frac{d\qb}{d t} = \left(\Oc +\Lambda \Sc\right) \qb, 
\end{equation}
formally solved by the time-ordered exponential form. The rotation rate $\Oc$ rotates $\qb$, and 
the strain $\Sc$ aligns and stretches $\qb$ towards its strongest eigendirection. Owing to the 
common linear terms in Eq.~\eqref{eq:jeffery} and~\eqref{eq:linear_jeffery}, $\qb$ has the same 
angular dynamics as $\pb$, but stretched and compressed by $\Sc$. The orientation $\pb$ is then 
recovered at any instant by normalizing $\qb$ to its unit length: 
\begin{equation}
	 \label{eq:norm_q}
	 \pb(t) = \frac{\qb(t)}{\|\qb(t)\|}.
\end{equation}
In the context of DNS, the preferential orientation of tracer spheroidal particles has been investigated in terms of the alignment with the eigensystem of strain and rotation rate in HIT \cite{guala2005evolution, parsa2012rotation, pumir2011orientation,gustavsson2014tumbling, ni2014alignment, chevillard2013orientation} and in turbulent channel flows \cite{andersson2015anisotropic, challabotla2015shape, zhao2016spheroids}. In these studies, the translational and rotational dynamics of spheroids are tracked simultaneously (sometimes including inertial effects as in \cite{shin2005rotational, zhang2001ellipsoidal}).

\subsubsection*{Stochastic model for the orientation}
 \label{sec:SDE_model:orientation:stochastic}

Drawing on the fact that a spheroid orientation is driven by the fluid velocity gradients encountered 
by the particle along its Lagrangian trajectory, we propose here a stochastic model that allows to 
represent some of the main statistics on spheroid orientation using limited information on the 
velocity gradient tensor (here at large observation time scales). For that purpose, we proceed in 
three main steps. First, assuming some basic structure on the velocity gradient tensor, we introduce 
a stochastic model for this tensor seen along the trajectory of tracers (step A below). Second, we 
derive the corresponding stochastic Jeffery equation (step B). Third, deducing the stochastic 
underlying dynamics for the angular velocity, we analyze the corresponding tumbling and spinning 
rates (step C).

\paragraph{A) Stochastic model for the velocity gradient tensor}
A small spheroid rotates in response to the velocity gradients along its Lagrangian trajectory, which is defined as:
\begin{equation}
	\Aij(t) =\frac{\partial \Ufi}{\partial x_j} (\bm{X}(t),t).
\end{equation} 
$\Aij$ fluctuates rapidly in turbulent flows and is dominated by small-scale motion (typically around or below the Kolmogorov scale $\eta$). This means that, in the context of CFD approaches based on turbulence models (such as RANS), additional models are required to reproduce the key features of the velocity gradients at small scales (which are not explicitly solved and thus not directly accessible).

%This modelling challenge has led to a number of Lagrangian stochastic models for the velocity gradient tensor in the literature (see the review \cite{meneveau2011lagrangian}). For instance, one of the first attempts was to couple an existing model for the velocity gradient \cite{pumir2011orientation} to an LES in HIT to study the orientation of anisotropic particles \cite{chen2016large}. Similarly, another model for the velocity gradient \cite{johnson2016closure} was used to study the deformation of droplets in a turbulent channel flow \cite{johnson2018predicting}. In these works, the particle dynamics is exactly solved for an observation time of the order of the Kolmogorov time scale $\taukol$ thanks to the modelled information about the velocity gradient tensor.

Unlike previous Lagrangian models for the velocity gradient tensor \cite{meneveau2011lagrangian, pumir2011orientation, chen2016large, johnson2016closure}, our objective here is to propose a stochastic model for the orientation $\bm{p}$, which aims at reproducing spheroids' orientation on a time-scale possibly much larger than the Kolmogorov one. To do so, we extend the stochastic orientation model approach used in \cite{campana2022stochastic} for a rod-like particle ($\Lambda=1$, and $\Bc=\Ac$) to spheroids, considering that the shape parameter $\Lambda$ only acts on the symmetric part. This follows from the observation that, in the case of spheres ($\Lambda=0$), the symmetric part of the velocity gradient does not play any role in the particle orientation. Hence, from the analysis developed in \cite{campana2022stochastic}, a stochastic version of Eq.~\eqref{eq:linear_jeffery} can then be written as
\begin{equation}\label{eq:sde_jeffery_linear_strat} 
	\begin{aligned}
	dr_i(t) =& 
	\left(\braket{\Oc_{ij}} +\Lambda \braket{\Sc_{ij}}\right) \, r_j \, dt \\
	&+\big(\left(\Dc_{ijkl} \, \partial\Wc_{kl}\right)\supa 
	+\Lambda \left(\Dc_{ijkl} \, \partial\Wc_{kl}\right)\sups \big) r_j. 
	\end{aligned}
\end{equation}
\Rev{The introduction of a stochastic model in Equation~\eqref{eq:sde_jeffery_linear_strat}  is justified in \cite{campana2022stochastic} by considering that the maximum of the integral times associated with the fluctuations of the velocity gradient $\tau_I$ is sufficiently small  compared to $\taukol$ so that the temporal dynamics can be reduced into a Brownian dynamics (via the central limit theorem).}
Equation~\eqref{eq:sde_jeffery_linear_strat} is expressed in the \Stra sense, and 
$\Wc$ is a $3\times3$ Wiener matrix whose entries are independent, one-dimensional Wiener processes. Note that we make use of the symbol $\partial$ in front of the Wiener processes $\Wc_{ik}$ for the stochastic \Stra integral sense, while we use the symbol $d$ for the stochastic \Ito integrals. 

The symmetric and antisymmetric parts of the fluctuation contribution $ \Dc_{ijkl} \partial\Wc_{kl}$ are denoted using the superscripts '$s$' and '$a$' respectively. 
% SUggestion pour remplacer le paragraphe:
\Rev{
To obtain a closed set of equations, we follow  \cite{campana2022stochastic} which relates the tensor $\Dc_{ijkl}$  to the autocorrelation tensor $\Cijkl$ of $\Ac$, such that $\Dc_{ijmn}\Dc_{klmn}=2\Cijkl$. In its generic form, this tensor is composed of 81 components in 3D requiring  information not yet  available in turbulence models.  
%and would require tedious developments. 
Here, a reduced description of this tensor is obtained 
%
%
%by expressing each component as a  product  of the maximum of the integral times $\tau_I$ with the instantaneous variance velocity gradient tensor (more details can be found in \cite{campana2022stochastic}). 
%
%The general form of the effective correlation tensor $\Cijkl$ 
%is strongly simplified
 by discarding the cumulative in time effects of the velocity gradient fluctuations 
$\partial_j u'_{\! f,i}$ and, considering instead its  equal-time covariance matrix (see 
\cite{campana2022stochastic} for more details): 
\begin{equation}\label{Aeq:C_tens}
	\Cijkl = \tau_{\textrm{I}} \braket{\partial_j u'_{\! f,i}(0) \partial_l u'_{\! f,k}(0)}.
\end{equation}
%The latter expression is weighted by the Lagrangian integral time scale of the velocity gradient 
%fluctuations $\tau_{\textrm{I}}=\max(\tau_{\textrm{I}}^{ijkl})$ 

This choice is motivated by its compatibility with RANS approaches, which provides information on $\taukol$ and $\braket{\nabla \Uf}$. Finally, fluctuations are assumed to be isotropic, consistently with high-Reynolds-number turbulence models where the behaviour of the velocity gradient can be approximated by an assumption of local isotropy~\cite{nelkin1994universality}. This approximation of isotropy is further justified by the assessment of the model in simple homogeneous cases (like HIT or shear flows).
}

The assumed isotropic form for the tensor $\Cijkl$ does not require a-priori calibration of the fluctuation tensor $\Dc_{ijkl}$ using DNS data. Instead, the coefficients entering this tensor are derived analytically, with imposed additional constraints (incompressibility, homogeneity and dissipation of kinetic energy). This gives the following effective square root of correlation tensor (see~\ref{A:D_tensor} for more details):
\Rev{
\begin{equation}\label{eq:corr_tensor}
\begin{aligned}
\sqrt{\tfrac{\taukol}{\alpha}} \ \Dc_{ijkl} 
&= \tfrac{1}{2}\left(\tfrac{1}{\sqrt{5}} 
		+\tfrac{1}{\sqrt{3}}\right)\delta_{ik}\delta_{jl}
		-\tfrac{1}{3 \sqrt{5}}\delta_{ij}\delta_{kl} \\
&\qquad \qquad +\tfrac{1}{2}\left(\tfrac{1}{\sqrt{5}}-\tfrac{1}{\sqrt{3}}\right)\delta_{il}\delta_{jk}.	
	\end{aligned}
\end{equation}
As shown in \ref{A:D_tensor}, $\alpha$ is a flow calibration parameter. 
In the ideal flow cases considered in the following, $\alpha$ will be set to 1. }

\paragraph{B) Stochastic Jeffery equation}
The stochastic differential equation (SDE) for the instantaneous elongation is written from Eq. \eqref{eq:sde_jeffery_linear_strat} in its \Stra form (see \ref{A:ito_stra})
\begin{align}
 \label{eq:model_elongation_strato}
 &d\qb(t) = \av{\Bc}(t)\qb(t)dt +  \partial \widetilde{\Zc}(t)\qb(t)  \\
 &\qquad\text{with }\av{\Bc} = \av{\Oc} + \Lambda \av{\Sc}, \, \text{and } \, \widetilde{\Zc} = \nua  \Wca + \Lambda \nus \tWcs.\nonumber
\end{align}
In Eq.~\eqref{eq:model_elongation_strato}, we have introduced the antisymmetric part of the Brownian 
matrix $\Wca =\tfrac{1}{2}(\Wc-\Wc\supt)$, the symmetric part $\Wcs= \tfrac{1}{2}(\Wc+\Wc\supt)$, 
and its modification $\tWcs= \Wcs - \frac{1}{3} \Tr(\Wc)\Id$, where $\Id$ is the identity matrix. 
Meanwhile, $\nus$ corresponds to the dimensional coefficient of the Brownian stretching $\tWcs$ and 
$\nua$ stands for the coefficient of the Brownian rotation $\Wca$. Their values are given by
\begin{align}
 \label{eq:def_nua_nus}
\left(\nus, \nua\right) = \sqrt{\tfrac{\alpha}{\taukol}}\left(\tfrac{1}{\sqrt{5}}, \tfrac{1}{\sqrt{3}}\right) =  \sqrt{\alpha}\left(\frac{\varepsilon}{\nuf}\right)^{\frac{1}{4}}\left(\tfrac{1}{\sqrt{5}}, \tfrac{1}{\sqrt{3}}\right), 
\end{align}
where $\varepsilon$ is the turbulent energy dissipation rate and $\nuf$ is the kinematic viscosity of the fluid (two quantities that are available in RANS simulations). %In CFD simulations, $\varepsilon$ is generally modelled by some closure relation and it can display very strong variations in wall-bounded flows. Yet, this quantity is accessible in turbulence models and is actually used to couple this model to simulations based on RANS approaches, as will be seen in the implementation of the method in a CFD code (see Section \ref{sec:num:summary}).

To derive the stochastic version of Jeffery's equation~\eqref{eq:jeffery}, we apply the \Ito's Lemma 
on the renormalization function $\qb\mapsto \qb/\|\qb\|$ to the elongation stochastic 
SDE~\eqref{eq:model_elongation_strato}. We obtain the following Lagrangian stochastic model for 
the orientation (in its \Ito form, see details in \ref{A:ito_lemma}):
\begin{align}\label{eq:model_jeffery_ito0}
\begin{aligned}
 d\pb(t) = & \av{\Bc}\,\pb\, dt 
 - \Big(\pb^\mathsf{\!T} \av{\Bc}\,\pb\Big)\,\pb\, dt 
 + d \Zc\pb \\
 & - \Big(\pb\supt d \mathbb{Z} \pb\Big)\pb - \tfrac{1}{2}(\Lambda^2\nus^2 + \nua^2) \pb\, dt, 
\end{aligned}
\end{align}
with $\Zc = \nua  \Wca + \Lambda\nus  \Wcs$, or equivalently:
\begin{equation} \label{eq:model_jeffery_ito}
 d\pb(t) = \pb \times \Big( \big(\av{\Bc}\,\pb\, dt + d \Zc\pb \big) \times \pb \Big) - \tfrac{1}{2}(\Lambda^2\nus^2 + \nua^2) \pb\, dt.
\end{equation}
Its \Stra form is 
\begin{equation*} 
 d\pb(t) = \av{\Bc}\,\pb\, dt 
 - \Big(\pb^\mathsf{\!T} \av{\Bc}\,\pb\Big)\,\pb\, dt 
 + \partial \Zc\pb
 - \Big(\pb\supt \partial \mathbb{Z}\,\pb\Big)\pb, 
\end{equation*}
or equivalently:
\begin{equation}
 \label{eq:model_jeffery_strato}
 d\pb(t) = \pb \times \Big( \big(\av{\Bc}\,\pb\, dt + \partial \Zc\pb
 \big) \times \pb \Big). 
\end{equation}

\paragraph{C) Angular dynamics within the stochastic model}
Studies on anisotropic particles generally provide information not only on the evolution of the orientation $\pb$ but also on the angular velocity \cite{voth2017anisotropic}. In particular, the spinning and tumbling rates are frequently measured \cite{voth2017anisotropic}. From the stochastic model for spheroid orientation described above, we can also derive information on such quantities.

\begin{figure}[ht!]
	\centering
	\resizebox{0.7\textwidth}{!}{\includegraphics{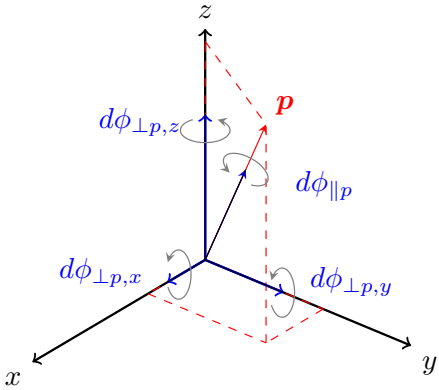}}  
	\caption{\label{fig:sketch_phi} 
		Sketch of angular increment in the Lagrangian stochastic model in Eq.~\eqref{eq:phi}. The total 
		angular increment $d\bm{\phi}$ is decomposed in spinning $d\phiP$ and tumbling $d\bmphiO$, 
		the latter has been projected along the Cartesian axis.}
\end{figure}

The rotational dynamics of spheroids in turbulent flows is related to the temporal evolution of a unit 
vector $\pb(t)$, which can be also expressed in terms of the total angular velocity of the particle 
$\bm{\Omega} = \tfrac{\bm{\omega}}{2} +\Lambda \pb \times \Sc \pb$, where $\bm{\omega}$ is the 
vorticity vector. The Jeffery's equation~\eqref{eq:jeffery} can be rewritten as $d\pb/dt = 
\bm{\Omega} \times \pb$. In the Lagrangian stochastic model, due to the presence of stochastic 
integrals, the particle angular velocity $\bm{\Omega}$ is not properly defined. Hence, associated 
statistics (like the tumbling and spinning rates) must be suitably redefined in the context of such 
stochastic models. More precisely, we identify the angular displacement vector $\bphi$, associated 
to the particle angular velocity $\bm{\Omega} \, dt =\partial \bphi$ (in \Stra formulation) as the 
quantity that properly characterizes the tumbling and spinning statistics. We identify the equation 
for $\bphi$ using
\begin{equation}
 \label{eq:sde_jeffery_cross_strat}
 d\pb = \partial \bphi \times \pb, 
\end{equation}
or its \Ito form,
\begin{equation}
 \label{eq:sde_jeffery_cross_ito}
 d\pb = d\bphi \times \pb -\frac{1}{2}(\nua^2 + \Lambda \nus^2) \pb \, dt,
\end{equation}
where the additional linear and diagonal terms stretch $\pb$ without contributing to the rotation dynamics. Then, we deduce the angular increment $d\bphi$ associated to the spheroid angular velocity
\begin{align} \label{eq:phi}
 \begin{aligned}
  d\bphi = & \tfrac{1}{2}\avomega dt 
  +\Lambda (\pb \times \braket{\Sc} \pb) dt \\
  & +\tfrac{1}{2}\nua  d\wa
  +\nus \Lambda \, \pb \times d\Wcs \pb.
 \end{aligned}
\end{align}
Here, the mean antisymmetric part of the velocity gradient tensor has been rewritten as $\braket{\Oc} \, \pb=\tfrac{1}{2}\avomega \times \pb$, where $\avomega$ is the mean vorticity vector. Analogously, the fluctuating part becomes $\Wca \, \pb = \tfrac{1}{2}\wa \times \pb$, where we have identified the increment of the vorticity vector fluctuation in terms of the antisymmetric matrix $\Wca$:
\begin{equation}\label{eq:wa_vorticity_fluctuation}
 \wa = (\Wc_{32}-\Wc_{23}, \Wc_{13}-\Wc_{31}, \Wc_{21}-\Wc_{12})^\intercal. 
\end{equation}
The angular velocity ``$\tfrac{d\bm{\phi}}{dt}$'' is often qualified with the tumbling and spinning rate \cite{byron2015shape} through its orthogonal and parallel projection on $\bm p$. By analogy, in the stochastic case, we can still define these two quantities from the projection of the angular increment $d\bm{\phi}$ on $\bm p$ (see Fig.~\ref{fig:sketch_phi}):
\begin{align}
 \begin{aligned}
    d\bmphiO(t) =& \,
	\bm{p} \times \left(d\bm{\phi} \times \bm{p}\right)=\bm{p} \times d\pb, \\
	d\phiP(t) =& \, \bm{p} \cdot d\bm{\phi}, 
 \end{aligned}
 \label{eq:phi_orthogonal_orig}
\end{align}
leading to the followings stochastic angular models: 
\begin{align}
&\begin{aligned}	
d\bmphiO(t) =  & \tfrac{1}{2}(\Id -\bm{p}\bm{p}^\intercal) 
\avomega \, dt 
+\Lambda \bm{p} \times \braket{\Sc} \bm{p} \, dt  \\
& +\tfrac{1}{2}{\nua}(\Id -\bm{p}\bm{p}^\intercal) d\wa
+\nus \Lambda \bm{p} \times d\Wcs \bm{p},	
\end{aligned} 
\label{eq:phi_orthogonal_devel}\\	
&\begin{aligned}
d\phiP(t) = \tfrac{1}{2}  \bm{p} \cdot \avomega dt
		+\tfrac{1}{2}{\nua} \bm{p} \cdot d\wa. 
\end{aligned}
\label{eq:phi_parallel_devel}
\end{align}

% Figure: Sketch Phi for the model

\subsection{Consistency in the formulation proposed.}

\Rev{To wrap up the description of the stochastic Lagrangian model, it is important to note the consistency in terms of the ``information content'' between the model for the flow field (here based on a RANS approach), the model for the translational motion of tracers (according to Eq.~\eqref{eq:SLM_tracer}) and the model for the rotational dynamics of inertialess spheroids (see Eq.~\eqref{eq:model_jeffery_ito}). In fact, both models for the translational and rotational motion of inertialess spheroids resort to Langevin models that reproduce the key effects related to turbulent fluctuations (which were filtered by the RANS approach). 

Another important remark is related to the time discretization (or time step) that will be used in the 
numerical part (described in the next section). A particular attention has been paid so that the 
orientation model does not imply additional constraints on the time step that could be used. In fact, 
the model has been developed so that large time steps can be used (possibly much larger than the 
Kolmogorov timescale $\taukol$). 

Finally, the increments of the Wiener processes appearing in Eq.~\eqref{eq:SLM_tracer} and in Eq.~\eqref{eq:sde_jeffery_cross_ito} are independent from each other.}

%******************************************************************************
% Numerical scheme for the orientation dynamics
%******************************************************************************

\section{Numerical scheme for the orientation dynamics} 
 \label{sec:num}
We propose here an algorithm for the stochastic orientation model introduced in Section~\ref{sec:SDE_model}, together with its implementation in a CFD code as well as detailed evaluations of its convergence.
 
\subsection{A splitting approach} 
 \label{sec:num:splitting}
 
The orientation SDE \eqref{eq:model_jeffery_ito0} combines a pure renormalized stretching effect with a pure rotation effect. In order to simplify its implementation and coupling in a CFD code, we propose here a numerical scheme on the vector $\pb$, without prior transformation and/or introduction of auxiliary equations and variables as it is done in some DNS (see \eg \cite{mortensen2008dynamics,siewert2014orientation}). 

Particular attention must be paid to the ability of the chosen scheme to conserve the orientation vector norm as close as possible to one, so that the angular dynamics are accurately rendered. 
Another requirement is the robustness of the numerical method to potentially large time steps. In CFD applications, the time step used for the Lagrangian dispersed-phase variables is often imposed by the integration scheme used for the computation of the fluid phase. This can lead to time steps that are orders of magnitude greater than the Kolmogorov scale. Moreover, spatial inhomogeneities may be induced, especially when dealing with wall-bounded flows \cite{hamlington_krasnov_boeck_schumacher_2012}. This can lead to possibly strong spatial variations of the $\nuas$ coefficients, without the possibility of adjusting the time step accordingly. So we need an integration scheme robust enough to handle large time steps. 

To fulfil these two criteria, we turn to a splitting method in order to separate the different effects in 
the dynamics and also to select an efficient scheme for each component. In the framework of SDEs, 
splitting schemes are little studied. But as for ODEs, splitting schemes display a very good behavior 
when it comes to preserving certain quantities kept by the equation, independently of the time step 
(see \eg \cite{buckwar2022splitting} and references therein).
The decomposition strategy of the whole equation to be integrated into a set of coupled sub-equations depends largely on the problem considered. Clear theoretical criteria to establish this splitting are in a general framework still largely to be made. From a practical point of view, the decomposition of the main equation into several coupled sub-parts must privilege the identification of (semi)-exactly integrable sub-equations, or integrable by approximation with a reliable scheme.

Here, we decompose the deterministic and stochastic effects of the dynamics on the one hand, and 
the stretching effect and pure rotation on the other hand, favoring the accurate resolution of the 
rotation sub-equations. The main \Ito equation \eqref{eq:model_jeffery_ito0} is decomposed as: 
\begin{align*}
\begin{aligned}
d\pb(t) =&\underbrace{\Lambda \av{\Sc}\,\pb dt-(\pb\supt\av{\Sc}\pb)\pb dt }_{\text{Continuously  
renormalized mean stretching }} \underbrace{+ \av{\Oc} \pb dt}_{\text{ ~ Mean rotation } } \\
& \underbrace{-\Lambda^2\tfrac{1}{2}{\nus^2} \pb dt + 
   \pb \times \big(\nus\Lambda d \Wcs\pb \times \pb \big) 
}_{\text{Continuously renormalized Brownian stretching }} \\
& \underbrace{-\tfrac{1}{2}{\nua^2} \pb dt + \nua {d\Wca}\pb}_{\text{Brownian rotation  }}. 
\end{aligned}
\end{align*} 
Introducing a time step $\dt$ and discrete times $\{t_k= k \dt, k\geq 0\}$, we replace equation \eqref{eq:model_jeffery_ito0} by the following split-step-forward Lie--Trotter composition: 
\begin{equation}\label{eq:splitting_general}
\left\{
\begin{aligned}
& \text{ Given $\pb(t_k)$, compute $\pb(t_{k+1})$}\\
&\qquad \text{by successively solving on $(t_k, t_{k+1}]$ }\\
&  \frac{d\pb\subs(t)}{dt} =  
\Lambda \left(\braket{\Sc} \pb\subs -\pb\subs \pb\subs\supt \braket{\Sc} \pb\subs\right),  \\ 
& \qquad\qquad \text{with }  \pb\subs(t_k) = \pb(t_k); \\
&  \frac{d\pb\suba(t)}{dt} = \braket{\Oc} \pb\suba(t), \\
& \qquad\qquad \text{with }  \pb\suba(t_k) = \pb\subs(t_{k+1}); \\
& d\pb\subbs(t) =  -\tfrac{1}{2}{\nus^2} \Lambda^2 \pb\subbs dt + \nus\Lambda 
   \pb\subbs \times \big( d \Wcs\pb\subbs  \times \pb\subbs \big), \\
& \qquad\qquad \text{with } \pb\subbs(t_k) =  \pb\suba(t_{k+1}); \\
&  d\pb\subba(t) =  -\tfrac{1}{2}{\nua^2} \pb\subba dt +\nua d\Wca \pb\subba,  \\
& \qquad\qquad \text{with }  \pb\subba(t_k) = \pb\subbs(t_{k+1}) \\
& \text{ and  set $\pb(t_{k+1}) = \pb\subba(t_{k+1})$.} 
\end{aligned}
\right. 
\end{equation}

We detail now the chosen schemes for each subpart. 
Considering on the one hand the state of the art for the resolution of the deterministic Jeffery 
equation, we focus our study on the two stochastic subequations. 

With $\Dt$ and $\{t_k= k \dt, k\geq 0\}$, we introduce the Brownian increments $\{\Delta \Wc_{t_{k+1},ij} =  \Wc_{t_{k+1},ij} -  \Wc_{t_{k},ij}, i,j,k\}$, i.i.d. with Gaussian law $\mathcal{N}(0, \Dt)$.
%----------------------------------------------------------------------------
% Brownian Stretching
%----------------------------------------------------------------------------
\subsection{Brownian Stretching}\label{04subsec:BS}
The Brownian stretching contribution is driven by the symmetric part of the velocity gradient fluctuations: 
\begin{equation}\label{eq:BS_nonlinear}
d\pb\subbs(t) = -\tfrac{1}{2}{\nus^2} \Lambda^2 \pb\subbs dt 
+ \nus\Lambda 
   \pb\subbs \times \big(d \Wcs\pb\subbs \times \pb\subbs \big).
\end{equation}
The role of the non-linear term $\pb\subbs \times (d \Wcs\pb\subbs \times \pb\subbs)=\pb\subbs \pb\subbs\supt \, d\Wcs \, \pb\subbs$ is to continuously preserve at each step the initial norm $\|\pb\subbs(t_k)\|$; on the other hand, the presence of non-linearities in this stochastic term introduces further difficulty related \textit{a priori} to the non-globally Lipschitz diffusion coefficient as investigated by \cite{bossy2021ananumsto} and references therein. Dealing with stochastic Wiener integrals, we are not allowed to apply implicit or symplectic Euler schemes strategy for ODEs \cite{haier2006geometric}. Instead, to integrate \eqref{eq:BS_nonlinear}, we take advantage of the 
linear drift term and propose to use a semi-implicit Euler-Maruyama scheme, combined with a 
step-by-step renormalization by a standard Projection Method on the sphere. The semi-implicit 
scheme is expected to contribute to the stability but does not preserve the norm. We call 
$\hnbpbs(t_k)$ the approximation of $\pb\subbs(t_{k})$.
\begin{align}\label{eq:scheme_BS_nonlinear.vec}
\left\{
\begin{aligned}
&\text{Given  $\hnbpbs(t_k)$, such that $\|\hnbpbs(t_k)\|=1$,}\\
&\text{ given the Gaussian trials for $\{\Delta\Wcs_{t_{k+1},i,j},i,j\}$,}\\
&\text{Prediction step : }\\
& (1+\tfrac{1}{2}{\nus^2} \Lambda^2 \dt) \nbpbs(t_{k+1}) = \hnbpbs(t_k) \\
& \qquad +  \nus\Lambda  \hnbpbs(t_k) \times \big(\Delta\Wcs_{t_{k+1}} \hnbpbs(t_k) \times    \hnbpbs(t_k)\big).   \\
&\text{Projection on the unit sphere:  } \\
& \text{set }\hnbpbs(t_{k+1}) = 
\frac{\widetilde{\bm{p}}_{\scriptstyle bs}(t_{k+1})}{\|\widetilde{\bm{p}}_{\scriptstyle bs}(t_{k+1})\|}.
\end{aligned}
\right. 
\end{align}
With the help of the vectorial product property, we verify on the right-hand side of the prediction step that
\begin{align*}
\begin{aligned}
& (1+\tfrac{1}{2}{\nus^2} \Lambda^2 \dt)^2 \|\nbpbs(t_{k+1})\|^2 \geq 1, 
\end{aligned}
\end{align*}
(see \ref{A:ana_bs}). The projection step is thus always well defined. We also show in \ref{A:ana_bs} that the theoretical rate of convergence is of order $\tfrac{1}{2}$, as in the case of generic linear SDEs. 

\subsection{Brownian Rotation }\label{04subsec:BR}
The Brownian rotation contribution is driven by the antisymmetric part of the velocity gradient fluctuations:
\begin{equation}\label{eq:BR_part} 
d\pb\subba(t) = -\tfrac{1}{2}{\nua^2}  \pb\subba \ dt +\nua d\Wma \  \pb\subba. 
\end{equation} 
Although linear, this SDE preserves the unitary norm. Except in dimension 2 (where the cosines of $\Wma$ is solution), this geometric constraint acts as a singularity, revealed when a change from Cartesian to spherical coordinates is applied. $\pb\subba(t)$ is called Brownian motion on the unit sphere and \eqref{eq:BR_part} is among its various representations \cite{van1985brownian}. Written in the \Stra form, it brings up the stochastic rotational kinematics form 
\begin{equation}\label{eq:BR_cross_strat}
	d\pb\subba(t) = \tfrac{1}{2}{\nua} \; \partial \wa \times\pb\subba, 
\end{equation}
with $\wa$ as in \eqref{eq:wa_vorticity_fluctuation}. We develop here a new algorithm, inspired from rotational kinematics solvers (see further discussion in \ref{A:ana_q}). Obtaining 
$\pb\subba(t+\dt)$ from  $\pb\subba(t)$ can be done by means of a rotation matrix $\Rc$. We 
identify such $\Rc$, considering its unit quaternion representation (reviewed in \cite{kuipers2020}). 
A quaternion is defined  as the couple of a real number and a vector $\qu = (q_0, \bm{q})$,  
with  norm  $\|\qu\|= \sqrt{q_0^2 +q_1^2 +q_2^2 + q_3^2}$. A unit quaternion is defined as a 
quaternion of unitary norm $\|\qu\|=1$. The rotation matrix corresponding to a unit quaternion is
\begin{equation}\label{eq:matrix_rotation_q}
	\Rc(\qu) = \left(q_0^2-\|\bm{q}\|\right) \Id 
	+2{\bm{q}}{\bm{q}}\supt +2 q_0\  [\bm{q}]_{\times}
\end{equation} 
where $[\;]_{\times}$ denotes the antisymmetric $3\times3$ matrix such that 
\begin{equation}\label{eq:cross_notation}
	[\bm{q}]_{\times}=
	\begin{pmatrix*}[r]
		0 	& -q_3 &  q_2 \\
		q_3 &  0   & -q_1 \\
		-q_2 &  q_1 &  0 
	\end{pmatrix*}.
\end{equation}
Moreover, given any angular velocity vector $\bm w(t)$, the associated time 
derivative of the unit quaternion is \cite{kuipers2020}
\begin{equation}\label{eq:ode_quaternion}
\frac{d\qu}{dt}(t) = \tfrac{1}{2} \bm{F}(\bm{w})\qu, \quad\text{with }\bm{F}(\bm{w}) = 
\begin{pmatrix*}[r]
0     & -\bm{w}\supt  \\
\bm{w}\supt & [\bm{w}]_{\times} 
\end{pmatrix*}.
\end{equation}
By a classical regularisation argument of the Brownian trajectories involved, we can easily deduce, from the stochastic orientation equation \eqref{eq:BR_cross_strat}, how the Brownian angular velocity $\partial \wa$ produces the following equivalent stochastic quaternion dynamics 
\begin{equation}\label{eq:stochastic_quaternion}
{d\qu}(t) = \tfrac{1}{4} \nua \bm{F}(\partial \wa)\qu, 
\end{equation}
where the matrix $\bm{F}(\partial \wa)$ structures as in \eqref{eq:ode_quaternion}. 
Equation~\eqref{eq:stochastic_quaternion} can be rewritten 
{\small
\begin{equation}\label{eq:sde_strat_quaternion}
d\qu(t) = \tfrac{1}{4}\nua \; \bm{\mathcal{Q}}(\qu) \partial\wa,\quad 
\bm{\mathcal{Q}}(\qu)=
\begin{pmatrix*}[r]
 -q_1 & -q_2 & -q_3 \\
  q_0 & -q_3 &  q_2 \\
  q_3 &  q_0 & -q_1 \\
 -q_2 &  q_1 &  q_0 
\end{pmatrix*}
\end{equation}
}
with \Ito form 
\begin{equation}\label{eq:sde_quaternion}
	d\qu(t)
	= -\tfrac{3}{16}\nua^2 \qu dt
	+\tfrac{1}{4} \nua \bm{\mathcal{Q}}(\qu) d\wa.
\end{equation} 
A semi-implicit one-step \EM scheme is used for solving \eqref{eq:sde_quaternion}: given the increments of Brownian motion $\Delta\wa_{t_{k+1}} = \wa_{t_{k+1}} - \wa_{t_k}$, we are interested in the rotation increment $\Delta\qu_{t_{k+1}} = \qu(t_{k+1})-\qu(t_{k})$. The initial unit quaternion is a parameter that can be fixed to $\qu(0) = (1,\bm 0)$, associated to $\Rc((1,\bm 0))= \Id$. These choices lead to the numerical unit quaternion increment: at each new step $(t_k, t_{k+1})$, $\Delta\qu_{t_{k+1}}$ is approximated by $\Delta \widehat \qu_{t_{k+1}}$ defined by 
\begin{equation}\label{eq:scheme_quaternions}
\begin{aligned}
& \Delta \widehat \qu_{t_{k+1}} = \Delta \widetilde \qu_{t_{k+1}}/\| \Delta \widetilde \qu_{t_{k+1}}\|, \quad \text{ with} \\ 
&\Delta \widetilde \qu_{t_{k+1}} 
= (1+\tfrac{3}{16} \nua^2 \Delta t)^{-1}\left(1 \, , \, \tfrac{1}{4} \nua \Delta \wa_{t_{k+1}} \right).
\end{aligned}
\end{equation}
The SDE \eqref{eq:sde_quaternion} has a linear structure. This implies that theoretical standard strong convergence results are already available (see \eg \cite{pages2018numerical}), and $\EE[ \sup_k \|\qu_{t_k} -\widetilde \qu_{t_k}\|^2]$ is at least in $\mathcal{O}(\Dt)$. As noted in \ref{A:ana_q}
\begin{equation}\label{eq:norm_quaternion}
\EE \|\Delta \widetilde \qu_{k+1}\|^2 
=  1 - (\tfrac{3}{16} \nua^2)^2 \Delta t^2 + \mathcal{O}(\Delta t^3)
\end{equation}
which turns into a $\tfrac{1}{2}$ rate of convergence  for $\sqrt{\EE[ \|\qu_{t_k} -\widehat\qu_{t_k}\|^2]}$. 

Finally, with the increment $\Delta \widehat \qu_{t_{k+1}}$, the one-step approximation of $\pb\subba(t_{k+1})$ starting from $\widehat\pb\subba(t_{k})$ is 
\begin{align}
\widehat\pb\subba(t_{k+1})= \Rc(\Delta \widehat \qu_{t_{k+1}})\widehat\pb\subba(t_{k}).   
\end{align}

\subsection{Renormalized mean stretching}\label{subsec2}
The deterministic stretching contribution is driven by the symmetric part of the mean velocity gradient tensor:  
\begin{equation}\label{eq:psi_1}
\frac{d \bmps}{dt}(t) = \Lambda \big(\av{\Sc} \bmps
- \bmps \bmps\supt\av{\Sc} \bmps\big). 
\end{equation}
Freezing $\av{\Sc}$ on the time interval $[t, t+ \dt)$, the exact solution of Eq.~\eqref{eq:psi_1} is given by 
\begin{align}\label{eq:split_sym_det_exact}
\begin{aligned}
\bmps(t+\dt) = \frac{\qb\subs}{\|\qb\subs\|}(t+\dt),\\
\text{with} \quad	\qb\subs(t+\dt) =  \exp(\Lambda \av{\Sc} \dt) \  \pb(t).
\end{aligned}
\end{align}
A suitable numerical solution for \eqref{eq:split_sym_det_exact} is achieved by classical exponential methods for ordinary differential equations (ODEs), for example through a diagonalization procedure. The numerical experiments shown in Section \ref{sec:assess:analyt} use a fourth-order Runge--Kutta (RK4) method. 

\subsection{Mean rotation}\label{subsec3}
The deterministic rotation contribution is associated to the antisymmetric part of the mean velocity 
gradient tensor in \eqref{eq:model_jeffery_ito}:
\begin{align}\label{eq:psi_2}
\frac{d\bmpa}{dt}(t)= \av{\Oc} \bmpa.
\end{align}
Freezing ${\av{\Oc}}$ on $[t, t+ \dt)$, the exact solution of \eqref{eq:psi_2} is given by 
\begin{equation}
\bmpa(t+\Dt)= \exp(\dt {\av{\Oc}}) \ \bmpa(t).
\end{equation}
Since $\Oc$ is antisymmetric, the exponential map is a pure 
rotation matrix, identified by Rodriguez formula \cite{marsden1998introduction}: introducing again $\avomega$, the mean vorticity vector such that $\avomega\times \pb = \av{\Oc} \pb$, we have
\begin{equation}\label{eq:euler_rodriguez}
\begin{aligned}
& \exp(\dt {\av{\Oc}} ) \\
& = \Id +\frac{\sin(\dt \|\avomega\|)}{\|\avomega\|} \av{\Oc}
+ \frac{\big(1-\cos(\dt \|\avomega\|)\big)}{\|\avomega\|^2} {\av{\Oc}}^2. 
\end{aligned}
\end{equation}
  
\subsection{Summary of the retained numerical scheme and available implementation}\label{sec:num:summary}
  
Having detailed each sub-schemes, we summarize the whole algorithm: 
\begin{equation}\label{eq:scheme_general}
\left\{
\begin{aligned}
& \text{Given $\widehat{\pb}(t_k)$, and the matrix of increments $\Delta\mathbb{W}_{t_{k+1}}$, }\\
& \text{compute $\widehat{\pb}(t_{k+1})$ by }\\
& \bmps(t_{k+1}) = \frac{\qb\subs}{\|\qb\subs\|}(t_{k+1}), \quad	\qb\subs(t_{k+1}) =  e^{\Lambda 
\av{\Sc} \dt} \widehat{\pb}(t_k), \\ 
&\bmpa(t_{k+1})= \exp(\dt {\av{\Oc}}) \bmps(t_{k+1}),  \\
& (1+\tfrac{1}{2}\nus^2 \Lambda^2 \dt) \nbpbs(t_{k+1}) = \bmpa(t_{k+1}) \\
& \qquad +  \Lambda\nus \,  \bmpa(t_{k+1})\times \left(\Delta\Wcs_{t_{k+1}} \bmpa(t_{k+1})\times    \bmpa(t_{k+1})\right).   \\
&\text{Projection on the sphere:  } \hnbpbs(t_{k+1}) = 
\frac{\widetilde{\bm{p}}_{\scriptstyle bs}(t_{k+1})}{\|\widetilde{\bm{p}}_{\scriptstyle bs}(t_{k+1})\|}.\\
& \text{set }\widehat{\pb}(t_{k+1}) =  \mathcal{R}(\Delta \widehat{\qu}_{k+1}) \hnbpbs(t_{k+1}) \\
& \qquad \text{with }  \Delta \widehat{\qu}_{k+1} \text{ defined in \eqref{eq:scheme_quaternions}}.
\end{aligned}
\right. 
\end{equation}
Having integrated $\widehat{\pb}(t_{k+1})$ from $\widehat{\pb}(t_{k})$, we deduce approximations for the angular models~\eqref{eq:phi_orthogonal_orig} and \eqref{eq:phi_orthogonal_devel} in order to cumulate the tumbling and spinning motion respectively with:
\begin{align}
\widehat{\bm{\phi}}_{\perp p}(t_{k+1}) &= \widehat{\bm{\phi}}_{\perp p}(t_k)	+ \widehat{\bm{p}}(t_k) \times \widehat{\bm{p}}(t_{k+1}), 	\label{eq:scheme_phiO} \\
\widehat{\phi}_{\parallel p}(t_{k+1}) &= \widehat{\phi}_{\parallel p}(t_k)
	+ \tfrac{1}{2} \widehat{\bm{p}}(t_{k}) \cdot \braket{\bm{\omega}} \Delta t 
	+ \tfrac{1}{2}{\nua}\widehat{\bm{p}}(t_{k}) \cdot \Delta \wa_{t_{k+1}}
	\label{eq:scheme_phiP}
\end{align}
which results to be numerically an easier task compared to the direct numerical integration of the SDE~\eqref{eq:phi}.
\medskip

This algorithm has been implemented in \CS, an open-source CFD software. It can be downloaded on \href{https://www.code-saturne.org/cms/web/}{\CS website} (any release higher than version 6.0, see also \cite{CS}). \Rev{As mentioned in Section~\ref{sec:SDE_model}, this model is based on a previous algorithm for the tracking of spherical particles (details on its implementation are available in \cite{peirano2006mean}). The previous algorithm allows to compute the translational dynamics of tracers, which is obtained by solving Eq.~\eqref{eq:SLM_tracer} (in practice, tracer particles are declared as particles with zero inertia since the model is written in a generic form for inertial particles which naturally gives the limit case of tracers with $\tau_p=0$, see~\ref{app:A}). In the scope of this study, the algorithm has been extended to solve the rotational dynamics of inertialess spheroids, given by Eq.~\eqref{eq:scheme_general}.} This requires knowledge of several quantities including: the mean velocity gradient at the particle position (both its symmetric $\braket{\Sc}$ and antisymmetric parts $\braket{\Oc}$) and the turbulent dissipation rate $\varepsilon$ sampled near the particle position. These values are easily accessed from the RANS simulation of the turbulent flow. Hence, the coupling of this model within a CFD software is straightforward, although it requires careful assessment of the convergence of the algorithm, as detailed in the next few paragraphs. 

\subsection{Convergence validation in an ideal \Rev{turbulent} case}\label{sec:num:valid}

The evaluation of the stochastic model and its numerical scheme are necessary in simple idealized cases: here, we consider first a HIT flow and second a homogeneous shear turbulent flow (HST, see Section~\ref{sec:assess}). These steps allow us to evaluate our proposed numerical scheme, by varying the dissipation and shear, without entering into the discussion of a choice of turbulence model for the resolution of the mean flow. In the HIT or HST cases, the mean gradient is either zero or according to a given shear matrix, which makes it much simpler to assess the model performances. Moreover, the statistics that can be analysed in this framework depend on the temporal increment, but remain spatially homogeneous and therefore do not require to integrate the transport equation, reducing significantly the difficulty to compute some statistics on ($\pb,\mathbf{\phi}$) conditionally to a given position. 

To simplify the presentation of the numerical experiments in this section and the next one, the parametrization of the ideal flows will be reduced to two parameters: the Kolmogorov scale $\taukol$, and the shear rate $\sigma$. \Rev{Indeed,  the  range of values for the calibration parameter $\alpha$  is $(0,1)$ and, according to model hypothesis, preferably  smaller than 1.  In the ideal flow cases considered in the following, we arbitrarily set it to 1. Since $\alpha$ multiplies $\taukol$ only, this choice maximizes the impact of the value of $\taukol$  on the accuracy of the numerical integration studied here. } The discussion of the results will not explicitly mention $\varepsilon$, even though in practice, it is this variable that contributes to the coupling of the orientation $\pb$ to the flow. 

\subsubsection*{Case studied: HIT}\label{sec:num:valid:HIT}
 
The HIT case emphasizes the stochastic components of the model and allows to investigate the 
accuracy of the splitting approach~\eqref{eq:splitting_general} against analytical solutions. In this 
case, the mean contribution of the velocity gradient tensor is zero and 
Eq.~\eqref{eq:model_jeffery_ito} becomes
\begin{equation}\label{eq:p_hit_devel}
\begin{aligned}
d\pb(t) &=  \pb \times \Big( \big( (\nua d\Wca + \nus \Lambda d\Wcs) \pb \big) \times \pb\Big)\\
&\quad -\tfrac{1}{2}(\Lambda^2   \nus^2  + \nua^2) \pb\, dt. 
\end{aligned}
\end{equation}

To make a clear distinction between the ensemble average applied to the fluid fields and the expectation operator associated to the abstract probability under which Brownian integrals are introduced in our model, we denote with the symbol $\EE$ (rather than $\av{\cdot}$) the probabilistic means and $\var$ the corresponding variance. 
With HIT flows, some statistics of $\pb$ up to the third order moments can be derived analytically. 
They are used to analyze the convergence of the stochastic algorithms. 
These solutions are computed in~\ref{A:moments_analytical}, giving:
\begin{align}\label{eq:pi_moments}
\begin{aligned}
	\EE[p_i](t) =& \EE[p_i](0) \ e^{-\tfrac{\kappa}{2} t} 
	%\label{eq:pi_moment} 
	\\
	\EE[p_ip_j](t) =& \EE[p_ip_j](0)\ e^{-\tfrac{3}{2} \kappa t} +\tfrac{1}{3} \big(1 - e^{-\frac{3}{2} \kappa t}\big) \delta_{ij}
	%\label{eq:pipj_moment}
	\\
	%\EE[p_i^2](t) =& \EE[p_i^2](0) \ e^{-\tfrac{3}{2} \kappa t}  
	%+\tfrac{1}{3} \big(1 - e^{-\frac{3}{2} c t}\big) 
	%\label{eq:pi2_moment}
	%\\
	\EE[p_i^3] (t) =& \big(\EE[p_i^3](0) - \tfrac{3}{5}\EE[p_i](0) \big) \ e^{- 3 \kappa t} 
	+\tfrac{3}{5}\EE[p_i](t) \, ,  
	\\
	%\label{eq:pi3_moment}
& \text{with  ~} \kappa= \Lambda^2 \nus^2+\nua^2. 
\end{aligned}
\end{align}
Those analytic expressions \eqref{eq:pi_moments} are also valid for Brownian rotation $\pb\subba$ and Brownian stretching $\pb\subbs$ separately, with respectively $\kappa_a =\nua^2$ and $\kappa_s=\Lambda^2 \nus^2$. When $t$ goes to infinity, those statistics converge to zero, except $\EE[p_i^2](t)$ that converges to $\tfrac{1}{3}$, according to the uniform distribution on the set $\{\pb \in \mathbb{R}^3:\|\pb\|=1\}$.

\subsubsection*{Strong and weak convergences}\label{sec:num:valid:strong-weak}

The evaluation of time integration schemes for SDEs relies on two convergence norms that we briefly recall here. Given a numerical approximation $\{ \bar{\bm{p}}(t_k), k=0,\ldots, K\}$, based here on a given constant time step $\dt$, such that $T=K\dt$, and Brownian increments $\Delta \Wc_{t_{k+1}}$, we distinguish trajectorial, called also {\bf strong convergence error}, as the $L^2(\Omega)$-expectation of the supremum in time error. The numerical scheme is said to 
be \textit{strong} $\beta$-order accurate if 
\begin{equation}\label{eq:def_strong_error}
	\ErrS(\pb) = 
	\left(\EE\underset{0 \le t \le T}{\sup} \left\|\pb(t) -\widehat{\pb}(t)\right\|^2\right)^{\frac{1}{2}} 
	\le C^{\textrm{s}} \dt^{\beta}
\end{equation} 
for some constant $C^{\textrm{s}}$ which may depend on $T$ but is independent from $\Delta t$. 
For instance, the \EM scheme for SDEs with Lipschitz coefficients features a strong convergence of order $\beta = \tfrac{1}{2}$. Except in rare cases, exact trajectories $t \mapsto p(t)$ needed to compare with the scheme simulations are not available. The evaluation of the scheme is done by constructing \Rev{a set of numerical reference} trajectories with the smallest time step and sometimes an alternative scheme. Here we construct the set of reference trajectories with the same scheme but a refined time step.

\textbf{Weak convergence} deals with the convergence of the distributions. Such convergence is analysed in terms of expectation $\EE[f(\bm{p}(T))]$ for a suitable test function $f$. Thus, the numerical scheme is said to be weak $\beta$-order accurate if
\begin{equation}\label{eq:def_weak_error}
	\ErrW(f) = \left|\EE[f(\pb(T))] - \EE[f(\widehat{\pb}(T)]\right| 
	\le C^{\textrm{wk}} \dt^{\beta}
\end{equation} 
for some constant $C^{\textrm{wk}}$ which may depend on $T$ and $f$, but is independent from $\Delta t$. The \EM scheme for SDEs with Lipschitz coefficients features a weak convergence of order $\beta = 1$ for a set of test functions whose regularity depends on properties of the diffusion coefficient. 

We have tested different orders of advancement of the splitting components without observing any impact on the result. We observe that the covariance matrix of elements of $\Wms$ and $\Wma$ is always zero which could help to explain some commutativity of the scheme components, but without yet a clear proof of this insensitivity under the change of the order of the operations. We run the numerical tests according to the splitting order presented in~\eqref{eq:scheme_general}. 

% Figure: Strong Error Brownian Full IC(ra)=1,0,0
\begin{figure*}[ht!]
	\centering	
	\subfloat[\label{subfig:strong_BF_p1}]{
		\includegraphics[width=0.32\textwidth]{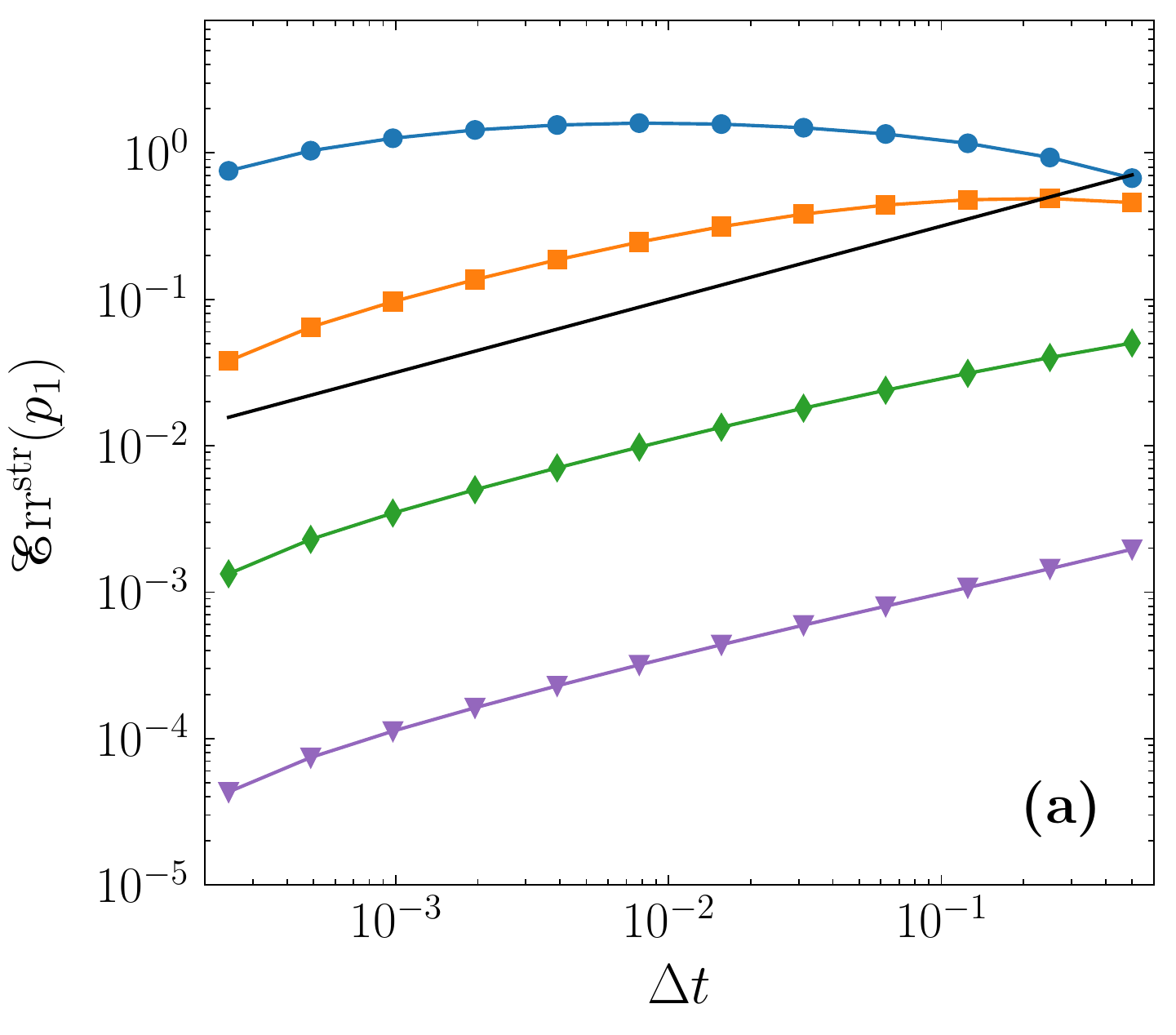}}\hfil
	\subfloat[\label{subfig:strong_BF_phiO1}]{
		\includegraphics[width=0.32\textwidth]{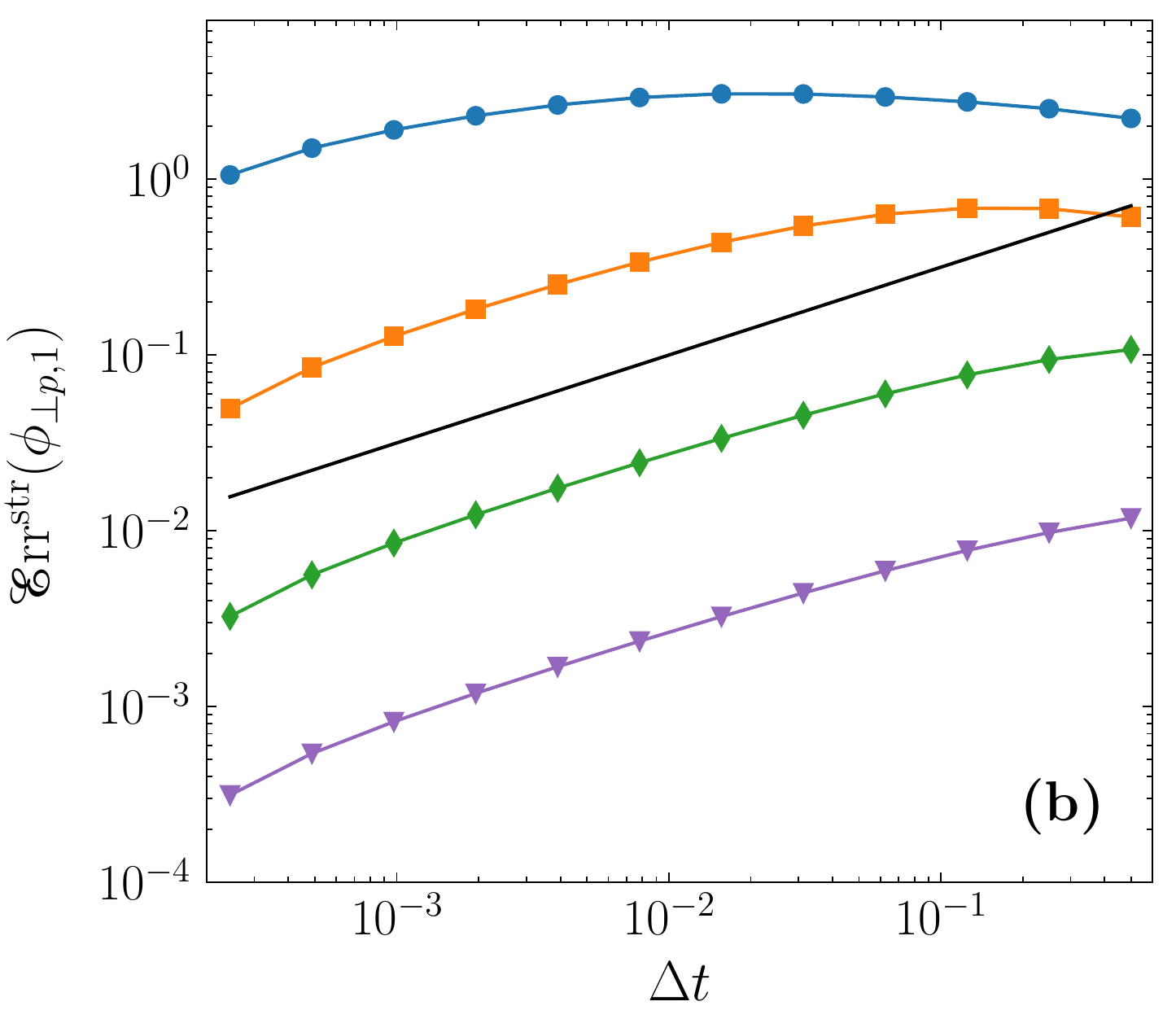}}\hfil
	\subfloat[\label{subfig:strong_BF_phiP}]{
	\includegraphics[width=0.32\textwidth]{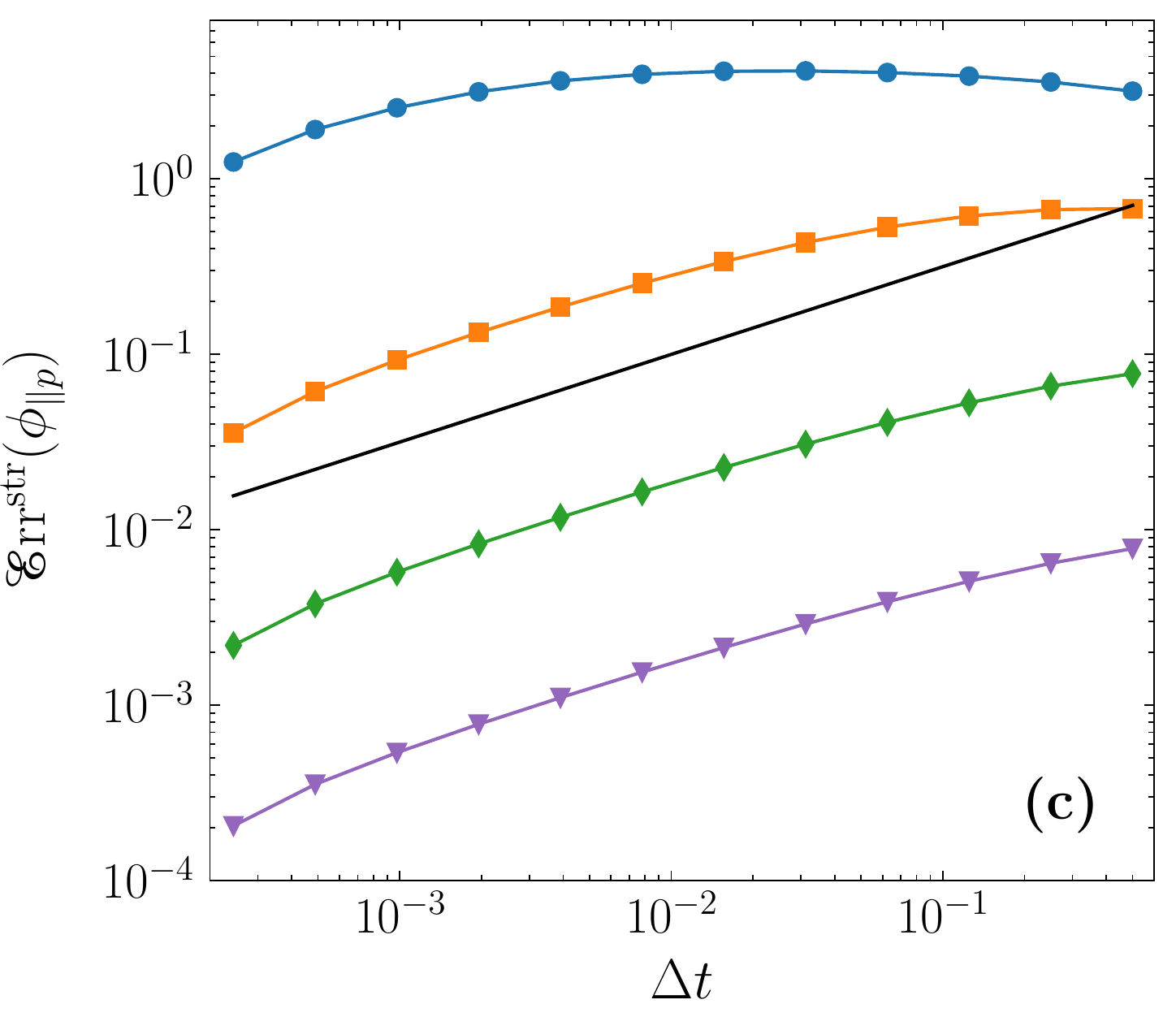}}
	\par\smallskip
	\centering
	\includegraphics[width=0.4\textwidth]{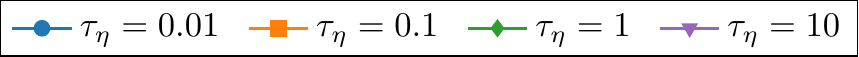}\vspace*{-1mm}
	\caption{\label{fig:strong_BF} 
		Splitting scheme in HIT: Strong error ($\ErrS$) of the splitting algorithm against the time step  
		$\Delta t$ for different  $\taukol$. 
		In \protect\subref{subfig:strong_BF_p1} the first  component of $\pb$, 
		\protect\subref{subfig:strong_BF_phiO1} the first component $\bmphiO$ of tumbling and 
		\protect\subref{subfig:strong_BF_phiP} the spinning $\phiP$. Black line indicates the slope 
		$\tfrac{1}{2}$; and the initial condition of particle orientation is $\pb(0)=(1,0,0)$. Simulation 
		performed with a shape parameter $\Lambda=1$, $N_p=5\times10^8$ particles, $\alpha=1$ and $T=0.5$. 
		The smallest $\Dt$ is $2^{-12}$ and reference trajectories  are computed with $\Dt=2^{-13}$.}
\end{figure*}

Figure~\ref{fig:strong_BF} shows the strong error for the three quantities ($p_1$, $\phi_{\perp p,1}$ and $\phi_{\parallel p}$) against $\Delta t$, for different values of $\taukol$ in Fig.~\ref{subfig:strong_BF_p1}, \ref{subfig:strong_BF_phiO1} and \ref{subfig:strong_BF_phiP} respectively. The effective order of strong convergence is $\tfrac{1}{2}$, confirming that the composition of operators in this case obeys to the convergence of the two sub-parts of the splitting algorithm. \Rev{These plots also show the impact of $\taukol$ on the strong error, which decreases as $\taukol$ increases. The rationale behind this trend is the following: when $\taukol$ increases (equivalently when $\alpha$ decreases), the two coefficients ($\nus, \nua$) appearing in Eq.~\eqref{eq:p_hit_devel} decrease. As a result, the error obtained at a given time decreases (here $T=0.5$). In addition, the strong error is dominated by the contribution from the stretching sub-part (see left panel in Fig.~\ref{fig:strong_weak_BR_BS} in~\ref{A:ana_complement}). Finally, it is important to note that no stability condition is required on $\Dt$ (as demonstrated in \ref{A:ana_complement}).}

% Figure: Weak Error Brownian Full IC(ra)=1,0,0 & IC(rb)=0.57,0.57,0.57
\begin{figure*}[ht!]
	\centering
	\subfloat[\label{subfig:weak_BF_p1}]{
		\includegraphics[width=0.32\textwidth]{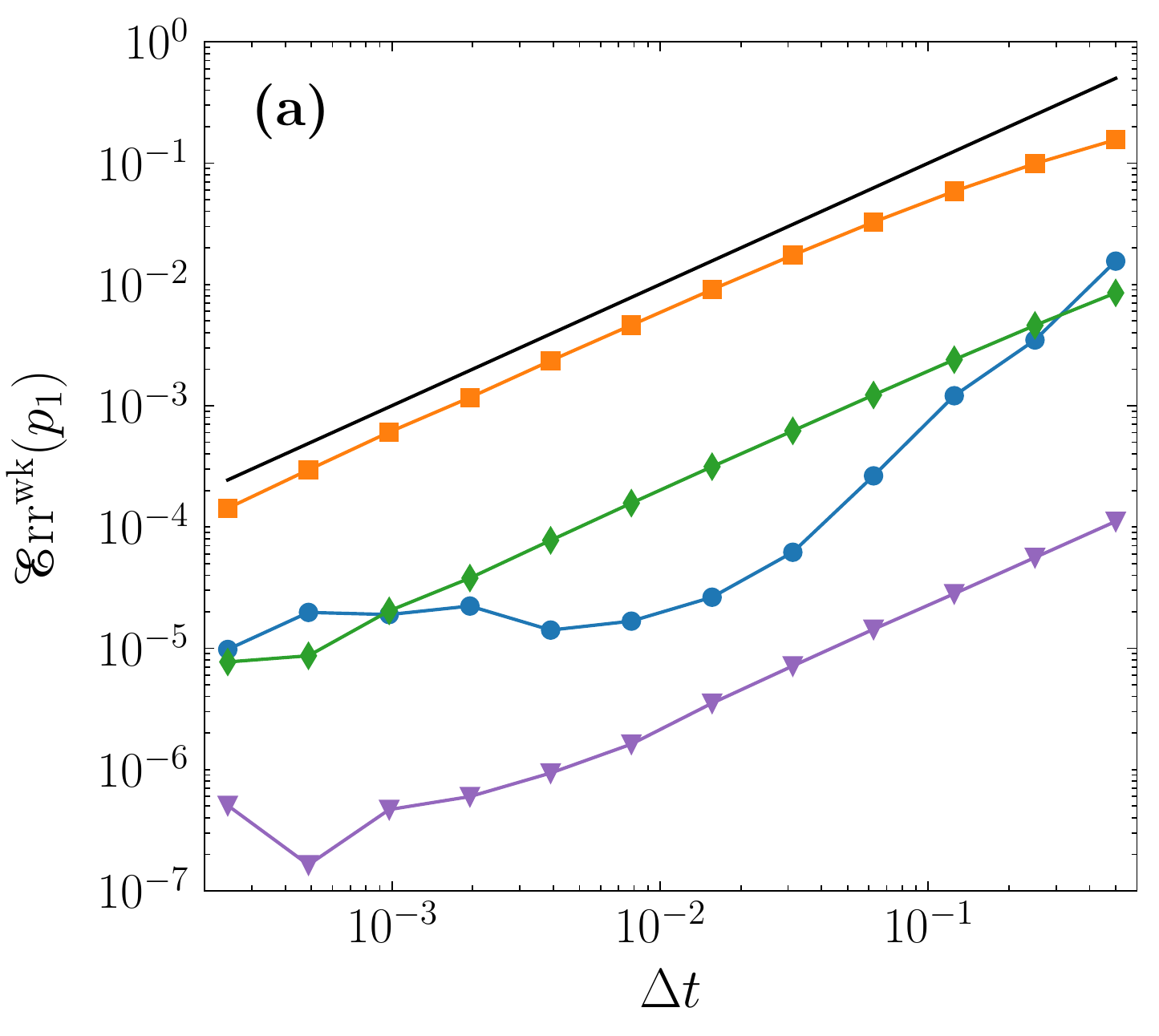}}\hfil
	\subfloat[\label{subfig:weak_BF_p1_2}]{
		\includegraphics[width=0.32\textwidth]{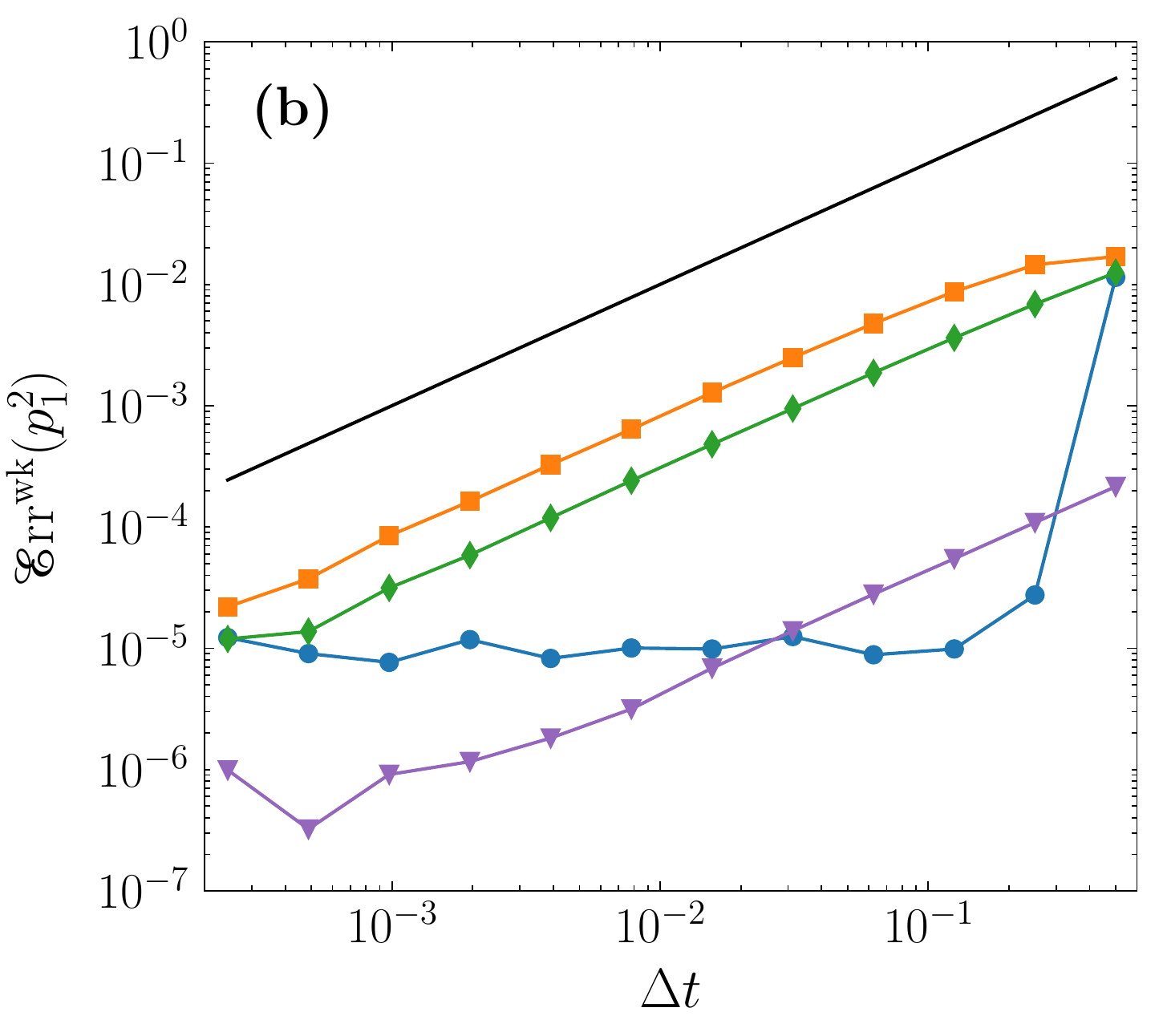}}\\
	\subfloat[\label{subfig:weak_BF_p1_3}]{
		\includegraphics[width=0.32\textwidth]{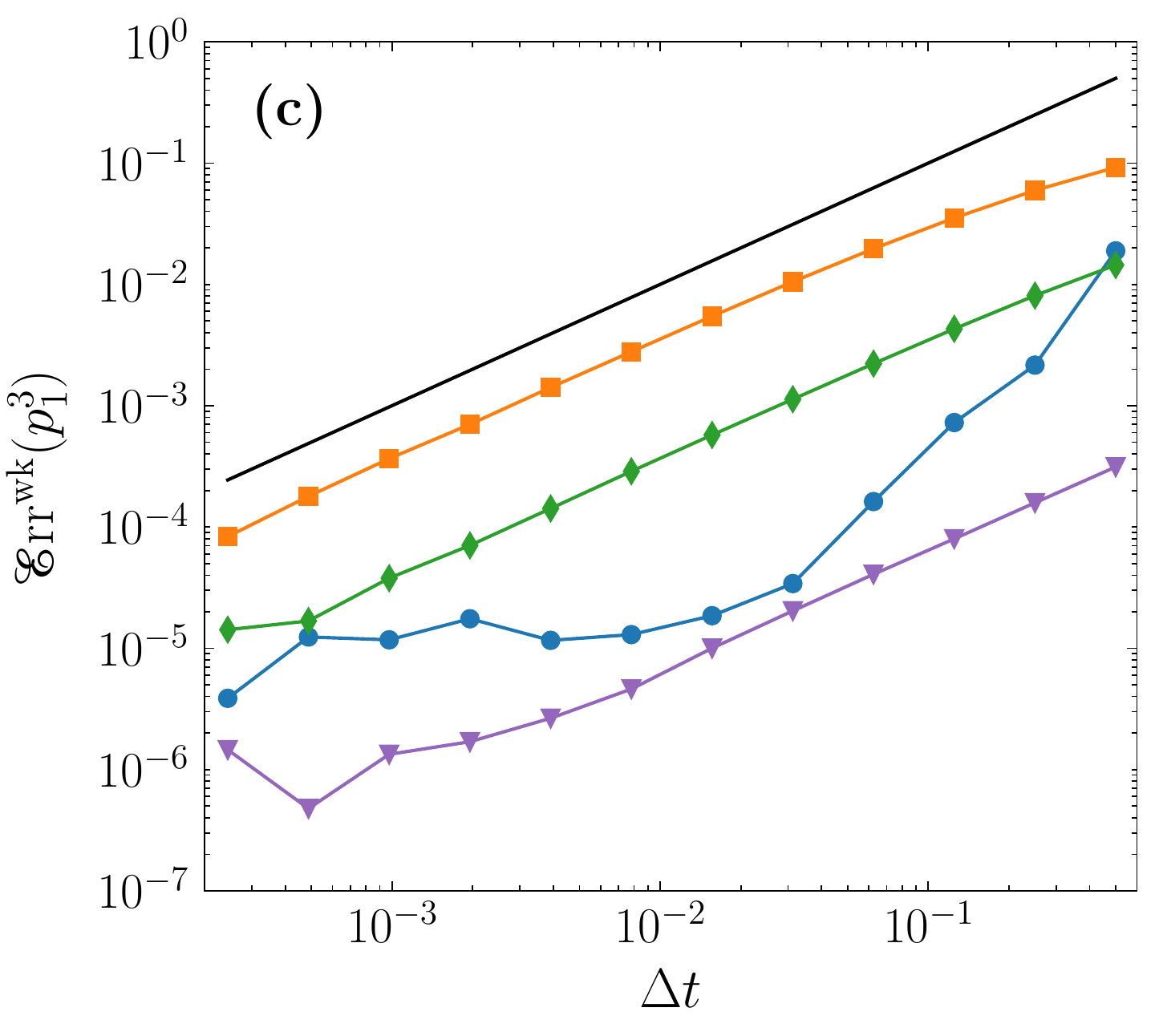}}\hfil
	\subfloat[\label{subfig:weak_BF_p1p2}]{
		\includegraphics[width=0.32\textwidth]{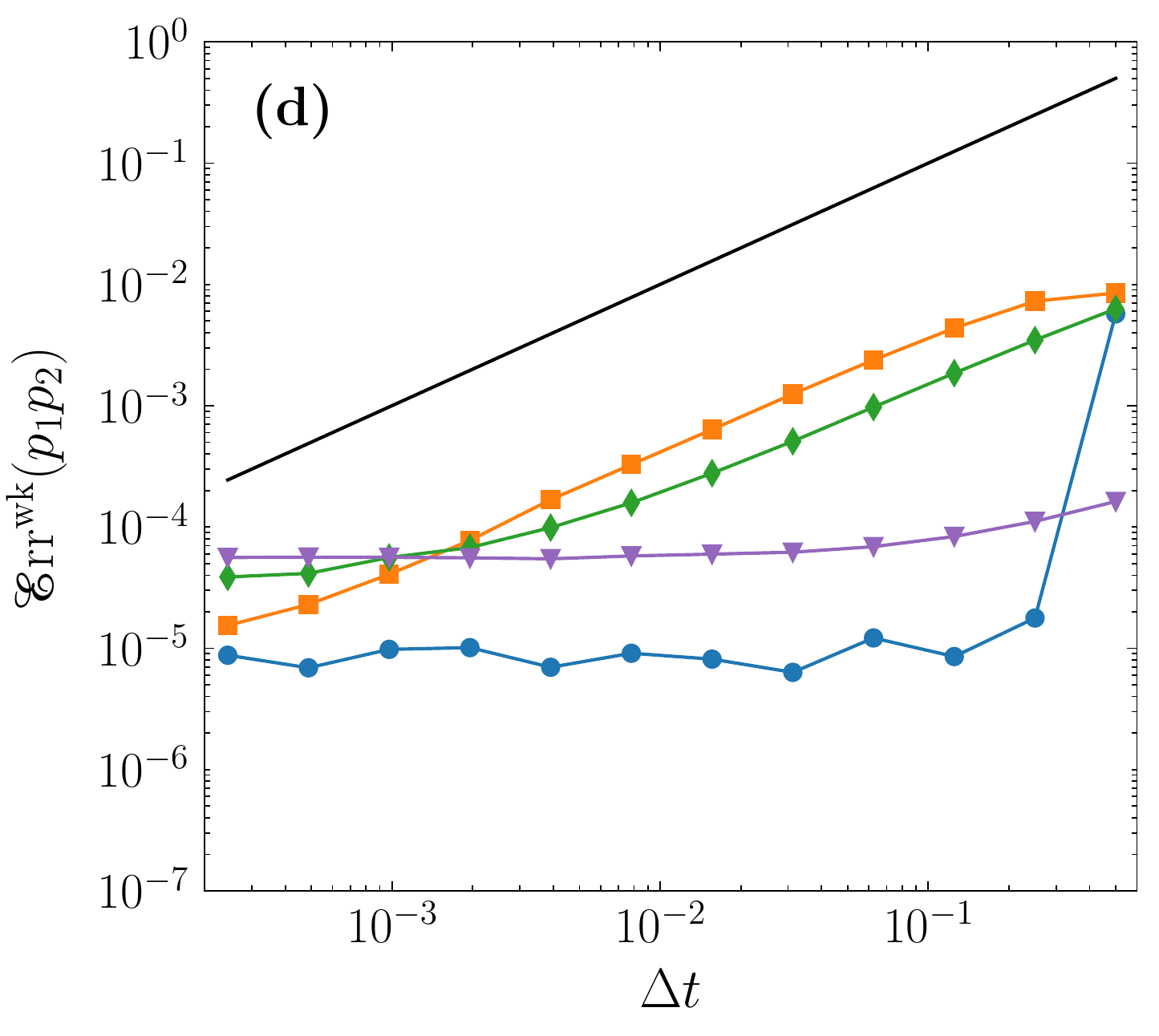}}
	\par\smallskip
	\centering
	\includegraphics[width=0.4\textwidth]{Legend_extern.pdf}\vspace*{-1mm}
	\caption{\label{fig:weak_BF} 
		Splitting scheme in HIT: Weak error ($\ErrW$) of the splitting algorithm against the time step  
		$\Delta t$ for different $\taukol$. 
		In \protect\subref{subfig:weak_BF_p1}, \protect\subref{subfig:weak_BF_p1_2}, and 
		\protect\subref{subfig:weak_BF_p1_3} the first three moments of the first component $\pb$ for 
		an initial condition of particle orientation $\bm{p}(0)=(1,0,0)$. In  
		\protect\subref{subfig:weak_BF_p1p2} the second cross moment $\EE[p_1 p_2]$ is starting 
		from  $\pb(0)=\tfrac{1}{\sqrt{3}}(1,1,1)$. Black line indicates the slope $1$. Simulations are performed with a shape parameter $\Lambda=1$, $N_p=5\times10^8$ particles, $\alpha=1$ 
and $T=0.5$. 
		The smallest $\Dt$ is $2^{-12}$ and reference trajectories are computed with $\Dt=2^{-13}$.}
\end{figure*}

The weak error as a function of the time step $\Delta t$, for different values of $\taukol$ is 
presented in Fig.~\ref{fig:weak_BF}. We observe a convergence of order one, confirmed by the 
convergence of order one for the two splitting sub-parts (see right panel in 
Fig.~\ref{fig:strong_weak_BR_BS} in~\ref{A:ana_complement}). In particular, the magnitude of the 
weak error for various moments are much closer to the one produced in the stretching sub-part 
compared to the rotation part. This indicates that the stretching contribution is dominant as for the 
strong convergence, which suggests that the numerical method used in the stretching sub-part 
should be refined here. The impact of $\taukol$ on the weak error is more balanced, between the 
previous effect and the ergodic convergence, the latter stabilizing exponentially fast around the 
constant solution \eqref{eq:pi_moments}, reducing the variance of the error and the impact of $\Dt$ 
(blue curve). The results on $\EE[p_1]$, $\EE[p_1^2]$ and $\EE[p_1^3]$ are obtained with an initial 
condition $(1,0,0)$, and are shown in Fig.~\ref{subfig:weak_BF_p1}, \ref{subfig:weak_BF_p1_2}, and 
\ref{subfig:weak_BF_p1_3} respectively. Alternatively, the initial condition $\tfrac{1}{\sqrt{3}}(1,1,1)$ 
has been used to show the convergence of $\EE[p_1 p_2]$ in Fig.~\ref{subfig:weak_BF_p1p2}.

\subsubsection*{Long time behavior of the PDF in HIT}
   \label{sec:num:valid:long-time}

\begin{figure}[htb!]
	\centering
	\subfloat[\label{subfig:pdf_phi_hit}]{
		\includegraphics[width=0.8\textwidth]{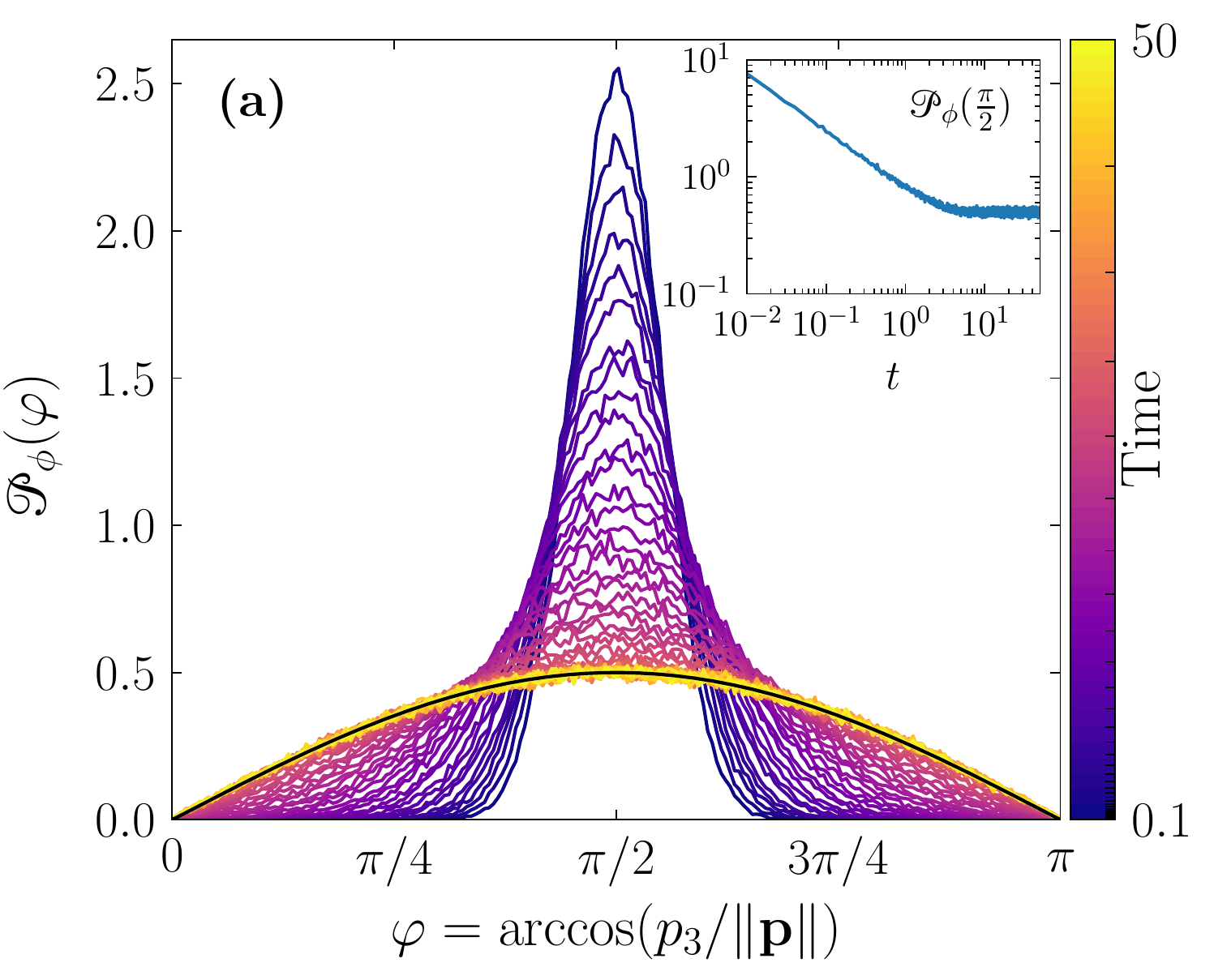}}\\
	\subfloat[\label{subfig:pdf_theta_hit}]{
		\includegraphics[width=0.8\textwidth]{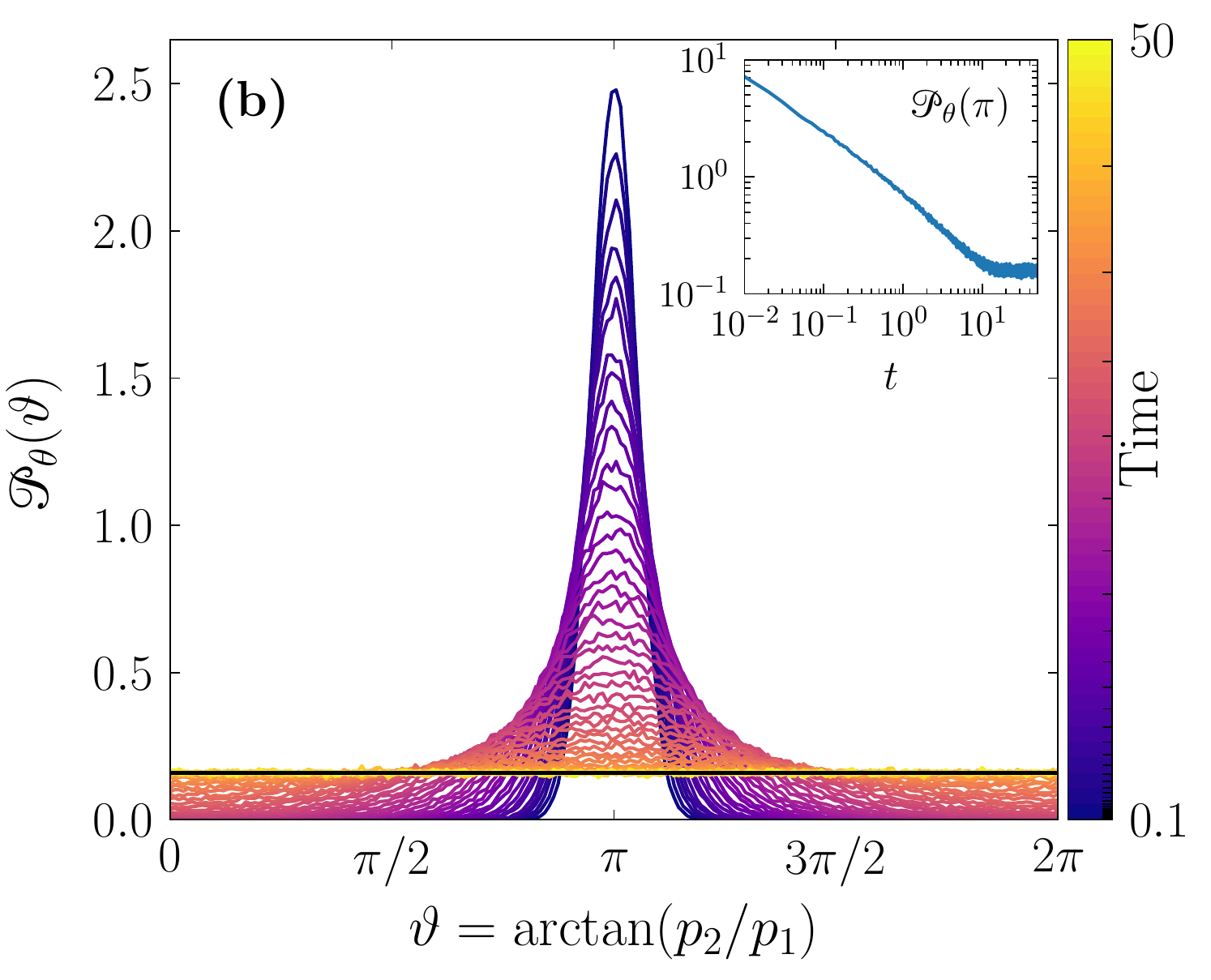}}
	\caption{\label{fig:pdf_time_hit} 
		Evolution in time of marginal empirical PDF for the angles $\phi$ 
		\protect\subref{subfig:pdf_phi_hit} and $\theta$ \protect\subref{subfig:pdf_theta_hit}, in HIT 
		case. The different curves correspond to different PDF at time instants spaced logarithmically 
		between  $t=0.1$ and $t=50$, starting from a deterministic initial position  $(1,0,0)$. In black 
		line, the theoretical equilibrium marginal PDF. Simulation performed with $N_p=10^5$ particles, 
		a time step $\Delta t=10^{-2}$, a flow time scale parameter $\taukol=1$ and a shape parameter 
		$\Lambda=1$. \textbf{Inset:} The evolution in time,  from the first time step to $t=50$, of the 
		PDF value $\Pdf_\phi(\tfrac{\pi}{2})$ displayed in log-log plot in 
		\protect\subref{subfig:pdf_phi_hit}. Same for $\Pdf_\theta(\pi)$  reported  in 
		\protect\subref{subfig:pdf_theta_hit}.}
\end{figure}

From Cartesian coordinates to spherical ones, we reduce the problem to two variables for the orientation $(\theta = \arctan(p_2/p_1), \phi = \arccos(p_3))$ to ease the plot of the associated  probability distribution function (PDF). Figure \ref{fig:pdf_time_hit} shows the time evolution of the empirical marginal distributions $\Pdf_\phi(t,\cdot)$ (see Fig.~\ref{subfig:pdf_phi_hit}) and $\Pdf_\theta(t,\cdot)$ (see Fig.~\ref{subfig:pdf_theta_hit}), starting from $\pb(0) = (1,0,0)$, or equivalently $(\theta, \phi)=(0,0)$.   
% Figure: PDF in time Phi and Theta sigma=0
The different curves indicate different instants of time and are logarithmically spaced starting from $t=0.1$ to $t=50$. As time increases (from violet to yellow lines), the PDF approaches the equilibrium distribution. These equilibrium marginal distributions (plotted in black continuous line) correspond to the uniform distribution on the 2d-sphere. Theoretically, we expect an exponentially fast convergence to equilibrium (see \cite{bensoussan1978asymptotic}). This is confirmed in the inset of Fig.~\ref{fig:pdf_time_hit} for $\phi$ (inset of Fig.~\ref{subfig:pdf_phi_hit}) and $\theta$ distributions (inset of Fig.~\ref{subfig:pdf_theta_hit}) evaluated at the median.

\subsubsection*{Tumbling and spinning in HIT}
 \label{sec:ass:hit:TS}
When the flow is spatially homogeneous, means and variances of the tumbling and spinning motion through the stochastic model are (quasi)-analytical quantities that can also be used to assess the model and it numerical approximation. Tumbling rate (TuR) and spinning rate (SpiR) definitions deviate from the DNS framework, because of the Brownian irregularity contained in $\pb$ which prevents deriving in time without first taking a statistic. Considering first and second moments statistics for orthogonal (tumbling) and parallel (spinning) angles \eqref{eq:phi_orthogonal_devel} and \eqref{eq:phi_parallel_devel}, we define rates for means and variances with 
\begin{align}\label{eq:TR_SR}
\begin{aligned}
\tumbE(t) =& \frac{d}{dt} \left\|\EE\left[\bmphiO(t)\right]\right\|, 
%\label{eq:mean_DPhi_orthogonal}
\\
\spinE(t) =& \frac{d}{dt} \EE\left[\phiP(t)\right], 
%\label{eq:mean_DPhi_parallel}\\
\\
\tumbV(t) =& \frac{d}{dt} \left(
\EE\left\|\bmphiO(t)\right\|^2 
-\left\|\EE\left[\bmphiO(t) \right]\right\|^2\right),
%\label{eq:var_DPhi_orthogonal} \\
\\
\spinV(t) =&\frac{d}{dt} \left(
\EE\left[\phiP^2(t)\right] - \EE\left[\phiP(t) \right]^2\right). 
%\label{eq:var_DPhi_parallel}
\end{aligned}
\end{align}

From Eq.~\eqref{eq:p_hit_devel}, these expressions are computed in the case of HIT in \ref{A:TumbSpin_HIT}, leading to the following (constant in time) analytical rates: 
\begin{align}\label{eq:DPhi_orthogonal_hit_nu}
\begin{aligned}
\spinV^{\textrm{hit}} & = \tfrac{1}{2}{\nua^2},  & \spinE^{\textrm{hit}}  &=0\\
\tumbV^{\textrm{hit}} &= \nua^2 +\nus^2 \Lambda^2, & \tumbE^{\textrm{hit}} &=0 \ . 
\end{aligned}
\end{align} 
Similar qualitative results have been obtained by~\citet{parsa2012rotation} using different arguments. 

\begin{figure}[htb!]
	\centering
	\includegraphics[width=0.85\textwidth]{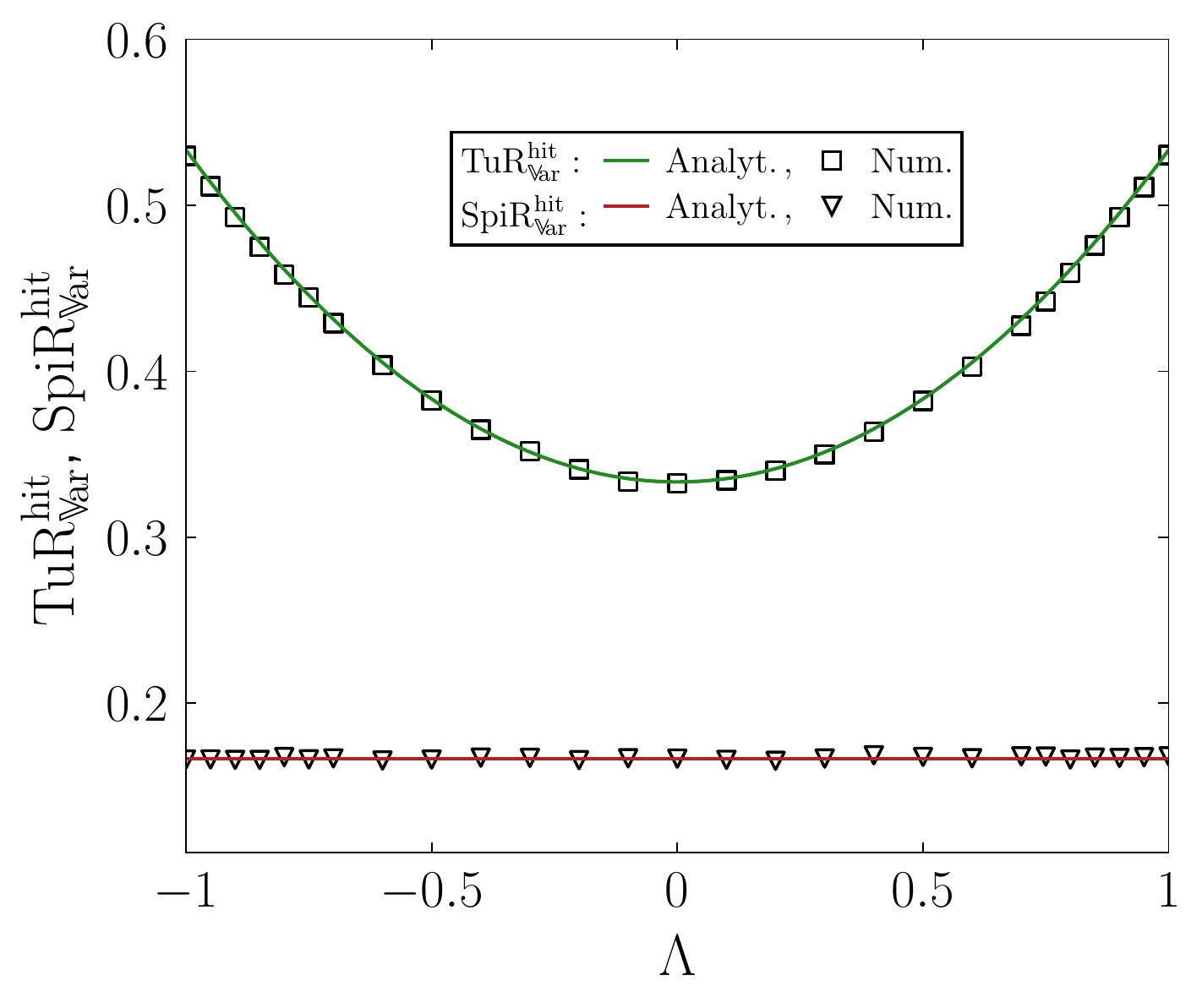}
	\caption{\label{fig:DPhi_hit}
		Analytical tumbling rate (green line) and spinning rate as a function of the particle shape 
		parameter $\Lambda$ for HIT case. Markers show results of numerical simulations. Simulations 
		are performed with a number of particles $N_p=10^5$, time step $\Delta t=10^{-3}$, final time 
		$T=1000$, time scale parameter $\taukol=1$ and using as initial condition a uniform distribution 
		on a sphere.  Estimators \eqref{eq:estimator_tumb} are computed with a step $h=1$ and a time 
		average from $t=100$ to $1000$. }
\end{figure}

In Fig.~\ref{fig:DPhi_hit}, we compare these theoretical values with the numerical ones, computed with estimators of the form 
\begin{align}\label{eq:estimator_tumb}
\begin{aligned}
{\htumbV} =\frac{1}{h} \left( \frac{1}{N_p} 
\sum_{j=1}^{\lfloor \frac{T-t_0}{h}\rfloor} \sum_{i=1}^{N_p} \|
\widehat{\bm{\phi}}^i_{\perp p}(T-h (j+1)h)\|^2 \right.\\
\left.\qquad\qquad\qquad\qquad-\|\widehat{\bm{\phi}}^i_{\perp p}(T-jh)\|^2 \right)
\end{aligned}
\end{align} 
with $h$ fixed to $10\times \Dt$, and using the numerical solutions \eqref{eq:scheme_phiO} and 
\eqref{eq:scheme_phiP} to compute independent draws for $\widehat{\bm{\phi}}^i_{\perp p}$ and 
$\widehat{\phi}_{\parallel p}$. Figure \ref{fig:DPhi_hit} shows the very good agreement between 
exact and numerical estimations. 

In particular, we can see that the tumbling and spinning rates are symmetric with respect to the 
change $\Lambda \to -\Lambda$. Note that this comes from the spatial homogeneity here, which 
makes the stochastic model for $\bm{p}(t)$ statistically time-reversible, as the Brownian motion 
involved. As a consequence, the change $\Lambda \to -\Lambda$ in Eq.~\ref{eq:p_hit_devel} gives 
that rod-like particles ($\Lambda>0$) have the same statistical behavior of disk-like particles 
($\Lambda<0$).
% Figure: DPhi for the model in HIT

%******************************************************************************
% Assessing the proposed model in specific cases (pure shear)
%******************************************************************************
\newpage
\section{Assessing the proposed model against ideal shear flow cases}
 \label{sec:assess}
 
In this section, the proposed splitting scheme for the SDE~\eqref{eq:model_jeffery_ito} is assessed in the case of a pure homogeneous shear flow case (HST). This case has been chosen since it allows to evaluate how the model captures the orientation of spheroids exposed to a simple shear flow without any boundary effects (which are out of the scope of this paper).

 \subsection{Strong convergence}
 \label{sec:assess:analyt}

The aim of this section is to study the behavior of the splitting scheme in the presence of a mean 
velocity gradient. Here, the focus is more on testing the coupling between the stochastic and 
deterministic parts than on analyzing the convergence as it has been done before. Moreover, this 
case introduces additional difficulties because of the presence of deterministic drift terms (\ie mean 
stretching and rotation). Indeed, an exact estimation for the moments is no longer accessible; from a 
numerical point of view, additional parameter such as the shear rate $\sigma=\braket{\partial_2 
{U_{\! f,1}}}$ affects the convergence error. The total mean contribution of the velocity gradient is 
\begin{equation}\label{eq:matrix_hsf}
	\braket{\Ac} =\small{
	\begin{pmatrix}
		0 & \sigma & 0 \\
		0 & 0 & 0 \\
		0 & 0 & 0 	
	\end{pmatrix}}
\end{equation} 
and $\avomega=(0,0,\sigma)$. Here, we restrict ourselves to assessing the strong error convergence by fixing $\taukol=1$ and considering two different values of the shear rate $\sigma=0.5, 8$ for $\braket{\Sc}$ and $\braket{\Oc}$. 
% Figure: Strong Error Shear IC(ra)=1,0,0
\begin{figure*}[ht!]
	\centering
	\subfloat[\label{subfig:strong_hsf_p1}]{
		\includegraphics[width=0.32\textwidth]{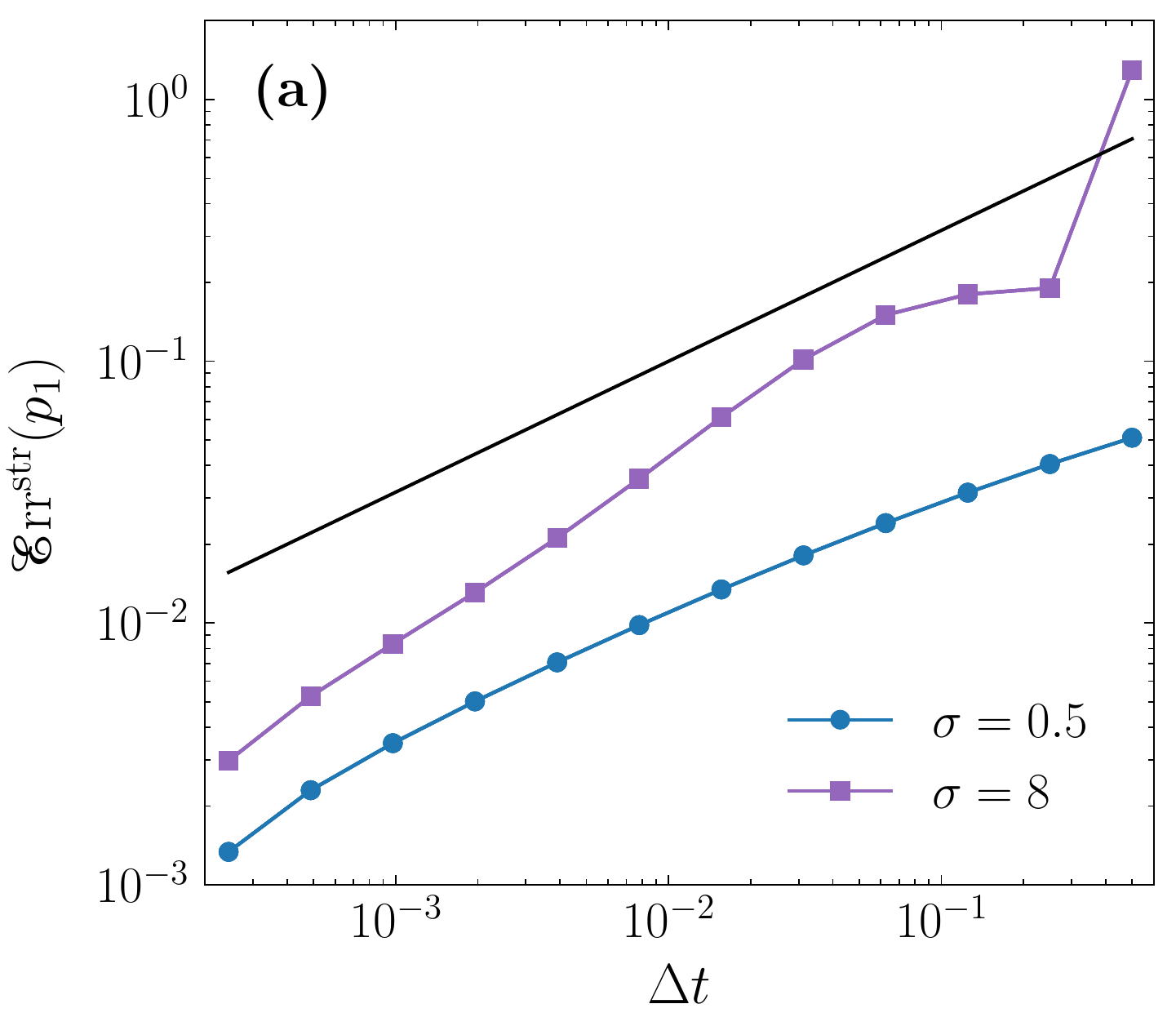}}
	\subfloat[\label{subfig:strong_hsf_phiO1}]{
		\includegraphics[width=0.32\textwidth]{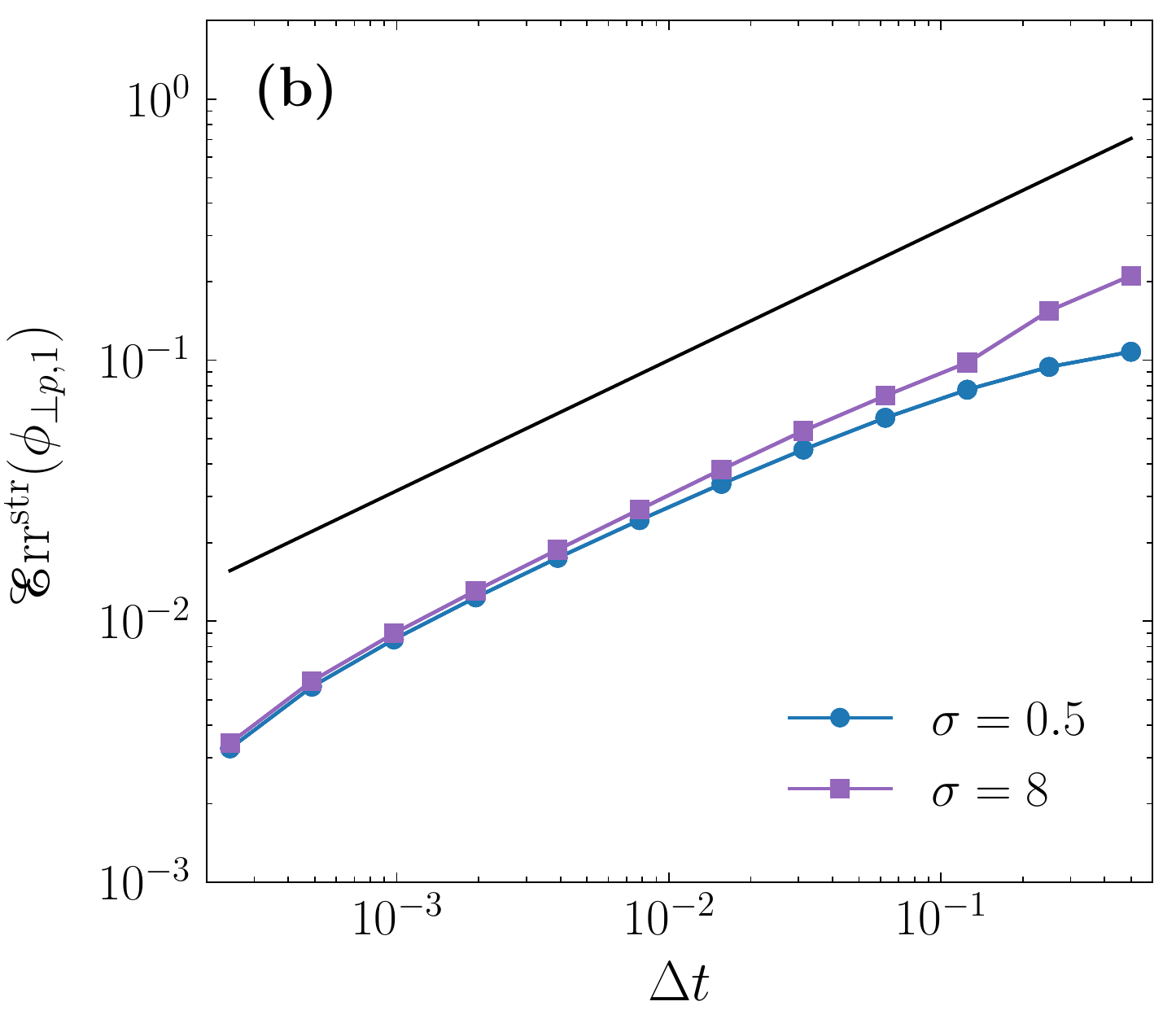}}
	\subfloat[\label{subfig:strong_hsf_phiP}]{
		\includegraphics[width=0.32\textwidth]{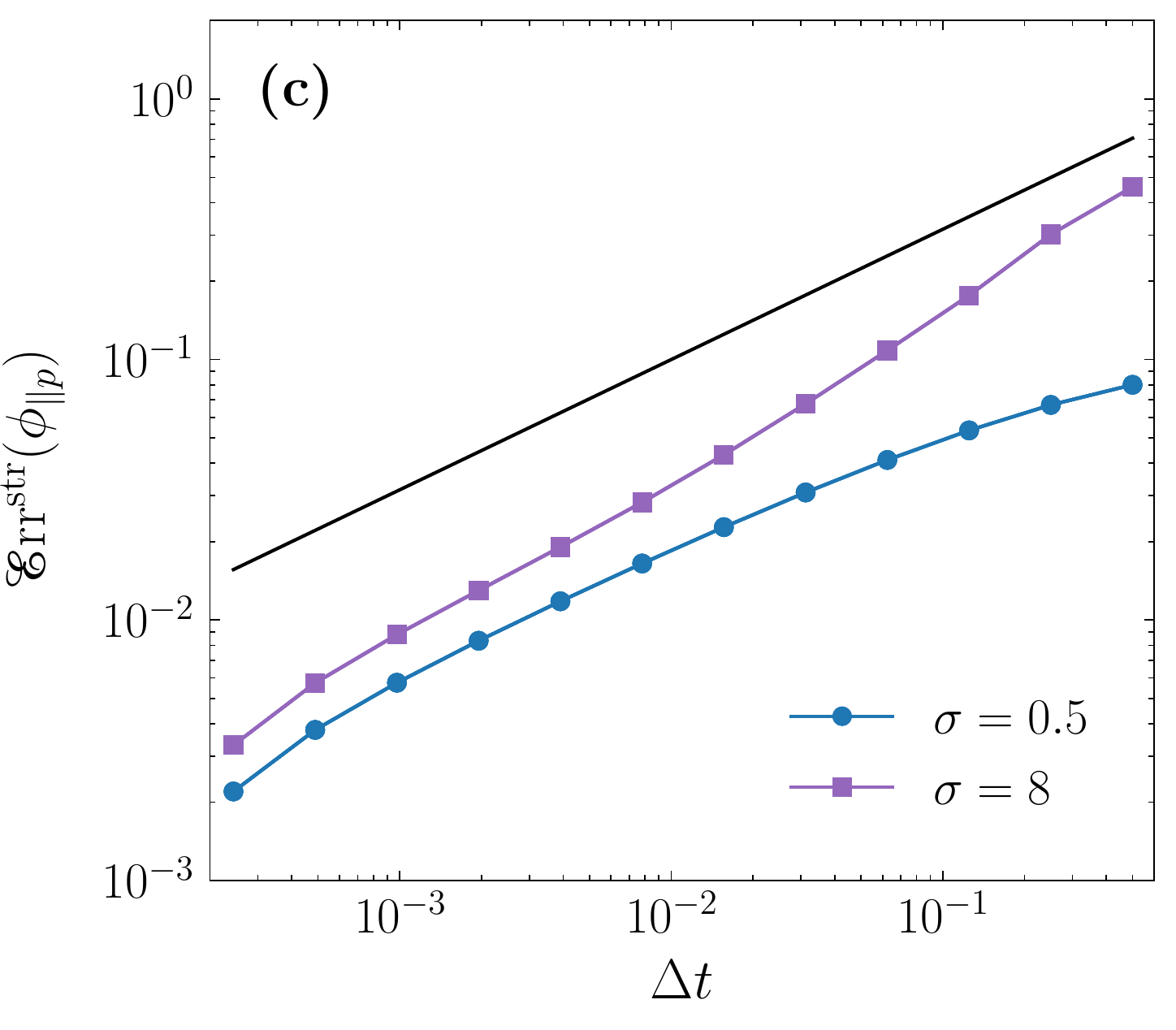}}
	\caption{\label{fig:strong_shear}
		Splitting scheme in HST: Strong error ($\ErrS$) of the splitting algorithm (for $\taukol=1$) 
		against the time step $\Delta t$ for two different values of shear parameter $\sigma$. In 
		\protect\subref{subfig:strong_hsf_p1} the first component of $\bm{p}$, 
		\protect\subref{subfig:strong_hsf_phiO1} the first component $\bmphiO$ of tumbling and 
		\protect\subref{subfig:strong_hsf_phiP} the spinning $\phiP$. Black line indicates the slope 
		$\tfrac{1}{2}$; the initial condition of particle orientation is $\bm{p}(0)=(1,0,0)$. Simulation 
		performed with a shape parameter $\Lambda=1$, $N_p=5\times10^8$ particles, $\alpha=1$ and $T=0.5$. 
		The smallest $\Dt$ is $2^{-12}$ and reference trajectories are computed with $\Dt=2^{-13}$.}
\end{figure*}

Figure~\ref{fig:strong_shear} shows the strong error as a function of the time step $\dt$ for the two 
values of $\sigma$. The first component $p_1$ is shown in Fig.~\ref{subfig:strong_hsf_p1} ($p_2$, 
$p_3$ have the same behavior, not shown here). It can be seen that the strong error converges with 
a slope slightly greater than $\tfrac{1}{2}$ when $\dt < 10^{-2}$ for both values of $\sigma$. This 
could be explained by the use of high order schemes for the mean stretching together with an exact 
solution for the mean rotation; it means that the splitting scheme marginally mixes the different 
orders of convergence.

The magnitude of the error strongly depends on the values of the shear rate, but does not deviate significantly from the ones observed for the strong convergence in a HIT case (see Section~\ref{sec:num:valid}), when $\dt$ becomes sufficiently small. Similar observations can be made for tumbling $\phi_{\perp p, 1}$ (see Fig.~\ref{subfig:strong_hsf_phiO1}) and spinning $\phi_{\parallel p}$ (see Fig.~\ref{subfig:strong_hsf_phiP}). For these two quantities, the magnitude of the error does not increase significantly moving from $\sigma=0.5$ to $\sigma=8$.

\subsection{Impact of shear on the long-time equilibrium for the particle orientation PDF}\label{sec:assess:orient}

Similarly to what has been described in Section~\ref{sec:num:valid:long-time}, the asymptotic 
behavior is analyzed here in a HST flow. 
 
For that purpose, we focus again on the marginal probability distribution function (PDF) expressed in spherical coordinates. In the case of a constant shear flow, an analytical solution for the stationary marginals PDFs of $\varphi$ and $\vartheta$ is not available and only numerical results are reported here. 

% Figure: PDF in time Phi and Theta sigma=8
\begin{figure*}[ht!]
	\centering
	\subfloat[\label{subfig:pdf_phi_hsf}]{
		\includegraphics[width=0.4\textwidth]{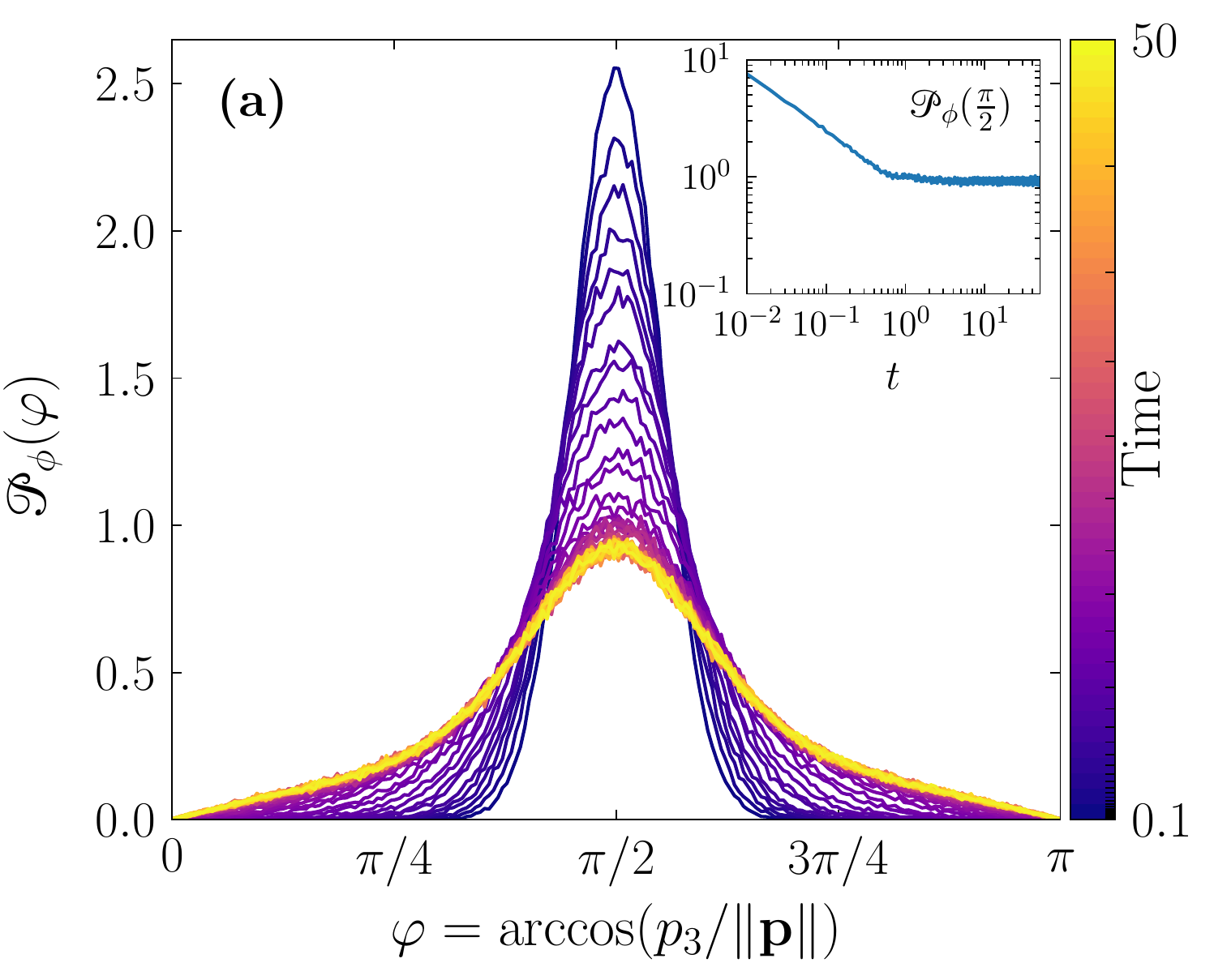}}
	\subfloat[\label{subfig:pdf_theta_hsf}]{
		\includegraphics[width=0.4\textwidth]{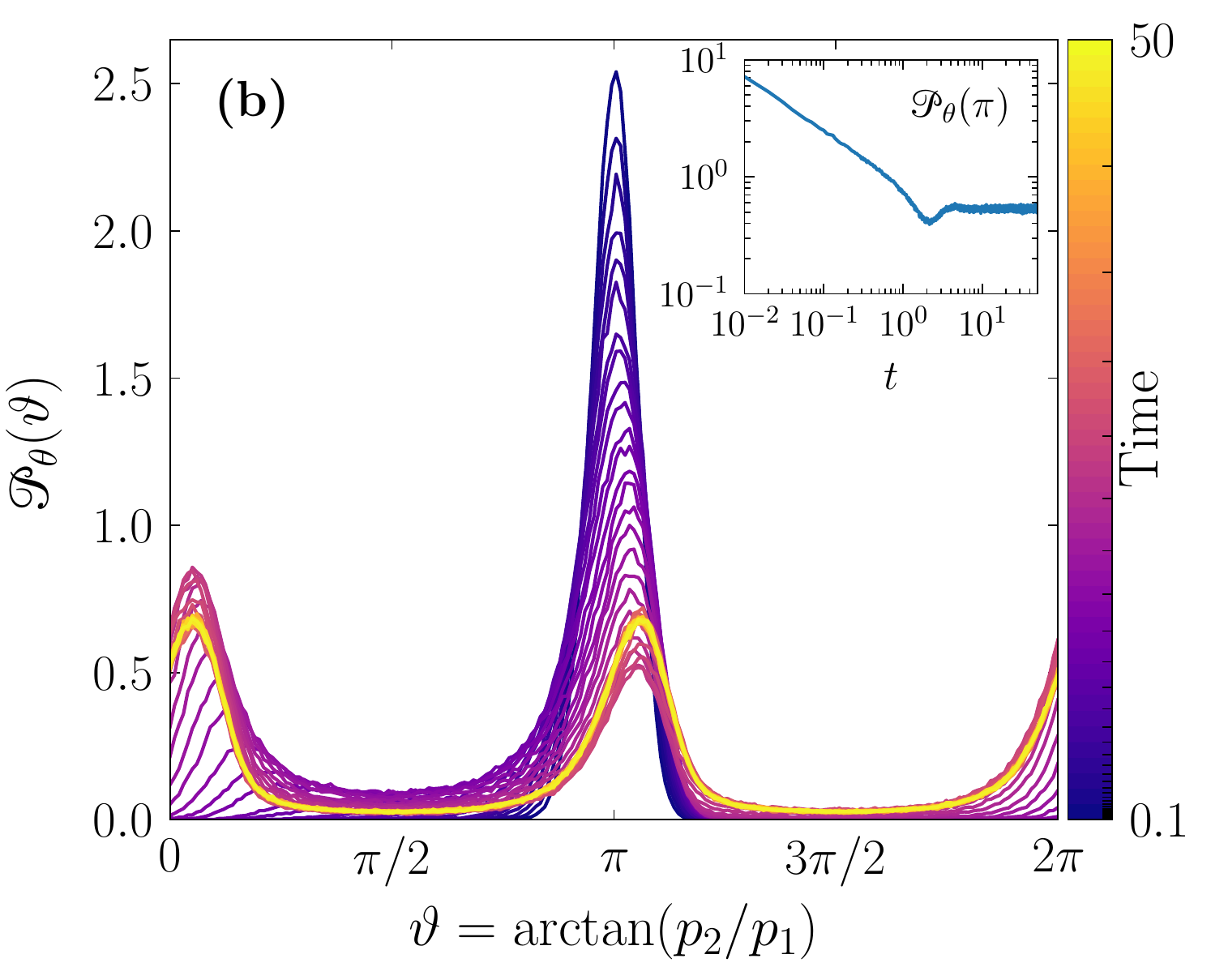}}
	\caption{\label{fig:pdf_time_hsf} 
		Evolution in time of marginal empirical PDF for the angles $\phi$ 
		\protect\subref{subfig:pdf_phi_hsf} and $\theta$ \protect\subref{subfig:pdf_theta_hsf} for 
		HST case with shear rate $\sigma =8$. The different curves correspond to PDF at time instants 
		spaced logarithmically between $t=0.1$ and $t=50$ and starting from a deterministic initial 
		position $(1,0,0)$. Simulation performed with $N_p=10^5$ particles, a time step $\Delta 
		t=10^{-2}$, a flow time scale parameter $\taukol=1$ and a shape parameter $\Lambda=1$. 
		\textbf{Inset:} The evolution in time, from the first time step to $t=50$, of the PDF value	
		$\Pdf_\phi(\tfrac{\pi}{2})$ displayed in log-log plot in \protect\subref{subfig:pdf_phi_hsf}. Same 
		for $\Pdf_\theta(\pi)$ reported in \protect\subref{subfig:pdf_theta_hsf}.}
\end{figure*}

Figure~\ref{fig:pdf_time_hsf} shows the evolution in time of the empirical marginal distributions $\Pdf_\phi(t,\cdot)$ (Fig.~\ref{subfig:pdf_phi_hsf}) and $\Pdf_\theta(t,\cdot)$ (Fig.~\ref{subfig:pdf_theta_hsf}) starting from an initial condition $\pb(0)=(1,0,0)$. The different curves indicate different instants of time and are logarithmically spaced between $t=0.1$ and $t=50$. As time increases, the numerical PDF approaches again (from violet to yellow lines) the invariant marginal PDF. Compared to the HIT case, where the orientation at equilibrium is uniformly distributed on a sphere, the presence of a mean shear rate tends to align the vector $\pb$ along its direction, which corresponds here to $\varphi=\pi/2$ and $\vartheta=0, \pi, 2\pi$. These two numerical experiments confirm the convergence of the scheme towards a unique invariant measure. Again, the convergence towards the invariant measure for the two marginal PDFs is exponentially fast as shown in the inset of Fig.~\ref{fig:pdf_time_hsf} for $\phi$ (inset in Fig.~\ref{subfig:pdf_phi_hsf}) and $\theta$ (inset in Fig.~\ref{subfig:pdf_theta_hsf}).

\subsection{Impact of shear on the tumbling and spinning rates}
\label{sec:assess:tumb-spin}
 
Inherent to the asymptotic behavior of the numerical scheme, it is interesting to compute 
numerically quantities related to tumbling and spinning, as defined in \eqref{eq:TR_SR}, with 
estimators of the form \eqref{eq:estimator_tumb}. 

In the HST case, we identify the drift part of the angles Eqs.~\eqref{eq:phi_orthogonal_devel} and~\eqref{eq:phi_parallel_devel}
$$(\bm{g}^{\perp}(\bm p), \ g^{\parallel}(\bm{p}) ) = \big(\small{\tfrac{1}{2}(\Id -\bm{p}\bm{p}^\intercal) 
\braket{\bm{\omega}} +\Lambda \bm{p} \times \braket{\bm S} \bm{p}, \ \tfrac{1}{2}\bm{p} \cdot \braket{\bm{\omega}}}\big),$$
to be 
\begin{equation}\label{eq:drifts_hsf}
\begin{aligned}
  \bm{g}^{\perp}(\pb) &=\tfrac{1}{2}{\sigma}\small{
  \begin{pmatrix}
   -(\Lambda-1) p_1p_3 \\
   (\Lambda+1) p_2p_3  \\
   (\Lambda-1) p_1^2 -(\Lambda+1) p_2^2	
  \end{pmatrix}},
  \\
  g^{\parallel}(\bm{p}) &=\tfrac{1}{2}\sigma p_3.
\end{aligned}
\end{equation}
When $\Lambda=0$ (for spheres), the tumbling and spinning statistics \eqref{eq:TR_SR} can be exactly computed within the model (see \ref{A:TumbSpin_HIT}). By symmetry argument $\EE[p_3(t)]=0$, and thus $\spinE^{\textrm{hst}}=0$ for all $\sigma$ values and all $t$. These are the only analytical results that we were able to compute exactly in HST cases.

Nevertheless, using the ergodicity of the orientation process $\pb$, and denoting $\pb_{\textrm{Eq}}$ the orientation distributed at equilibrium, we can extract a reliable long-time approximation, with the following computation: first 
\begin{align*}
\begin{aligned}
  & \frac{d}{dt} \left(\|\EE[\int_{0}^{t}\bm{g}^{\perp}(\bm{p}(s)) ds]\|^2\right)^{\frac{1}{2}} \\
 & \quad = \frac{\sum_{i}\EE[\int_{0}^{t} g_i^{\perp}(\bm{p}(s)) ds]
\EE[g_i^{\perp}(\bm{p}(t))]}{(\|\EE[\int_{0}^{t}\bm{g}^{\perp}(\bm{p}(s)) 
ds]\|^2)^{\frac{1}{2}}.}
\end{aligned}
\end{align*}
Taking the limit $t \to \infty$, $\EE[g_i^{\perp}(\bm{p}(t))] \to \EE[g_i^{\perp}(\pb_{\textrm{Eq}})]$ and 
\[
\int_{0}^{t}\bm{g}^{\perp}(\bm{p}(s)) ds \simeq t \EE[\bm{g}^{\perp}(\bm{p}_{\textrm{Eq}})], 
\]
leading to 
\[
\lim_{t \to \infty} 
\frac{d}{dt} \left(\|\EE[\int_{0}^{t}\bm{g}^{\perp}(\bm{p}_s) ds]\|^2\right)^{\frac{1}{2}} 
= (\sum_{i=1}^{3}\EE\left[g_i^{\perp}(\bm{p}_{\textrm{Eq}})\right]^2 )^{\frac{1}{2}}.\]
Second 
\begin{align*}
\frac{d}{dt} \EE\left[\left\|\int_{0}^{t}\bm{g}^{\perp}(\bm{p}_s)ds\right\|^2\right]
=2\sum_{i=1}^{3}\EE\left[\int_{0}^{t} g_i^{\perp}(\bm{p}(s))g_i^{\perp}(\bm{p}(t)) ds\right]. 
\end{align*}
For a time $\Tlong$ large enough such that the distribution of $\pb(\Tlong)$ approximates $\pb_{\textrm{Eq}}$ (which is happening exponentially fast), the autocovariance  
\begin{align*}
\mathcal{R}_{g_i^{\perp}} (\tau) = 
\EE\left[g_i^{\perp}(\bm{p}(\Tlong)) \ g_i^{\perp}(\bm{p}(\Tlong+\tau))\right] 
-\EE\left[g_i^{\perp}(\bm{p}_{\textrm{Eq}} )\right]^2,
\end{align*}
and we deduce from the ergodic property of $\pb$ that 
\begin{align*}
\begin{aligned}
&\lim_{t \to \infty} 
2\sum_{i=1}^{3}\EE\left[\int_{0}^{t} g_i^{\perp}(\bm{p}(s))g_i^{\perp}(\bm{p}(t)) ds\right]\\
&\qquad \qquad \simeq  2\sum_{i=1}^{3} \int_{t}^{\infty} \mathcal{R}_{g_i^{\perp}} (\tau) d\tau +
t \EE\left[g_i^{\perp}(\bm{p}_{\textrm{Eq}})\right]^2.
\end{aligned}
\end{align*}
Following the same technique to get a semi-analytical expression for the norm squared of the 
parallel projection $\phi_{\parallel p}$, we finally get some long-time statistical expressions 
summarized as 
{\small{
\begin{align}
\tumbE^{\textrm{hst}}(\infty)  &= \|  \EE\left[\bm{g}^{\perp}(\bm{p}_{\textrm{Eq}})\right]^2 \|	\label{eq:mean_DPhi_orthogonal_hsf_infty} \\
\tumbV^{\textrm{hst}}(\infty) &= \,
	2\sum_{i=1}^{3} \int_{\Tlong}^{\infty} \mathcal{R}_{g_i^{\perp}} (\tau) d\tau 
	+ \tumbV^{\textrm{hit}}
	\label{eq:DPhi_orthogonal_hsf_infty} \\
\spinV^{\textrm{hst}}(\infty) & =  \small{
	\int_{\Tlong}^{\infty} \EE\left[g^{\parallel}(\bm{p}(\Tlong)) \  g^{\parallel}(\bm{p}(\tiny{\Tlong+\tau}))\right] d\tau  + \spinV^{\textrm{hit}}. }
	\label{eq:DPhi_parallel_hfs_infty}
\end{align}
}}
These semi-analytical expressions are used as diagnostic tools for the numerical scheme. The numerical parameters are the same as those used in the HIT case (see Section~\ref{sec:num:valid:HIT}), \ie final time $T=1000$. In addition, we compute dedicated estimators for \eqref{eq:mean_DPhi_orthogonal_hsf_infty},\eqref{eq:DPhi_orthogonal_hsf_infty}, \eqref{eq:DPhi_parallel_hfs_infty} to compare with the direct estimators for statistics in \eqref{eq:TR_SR} and similar to \eqref{eq:estimator_tumb}, using $t_0 = \Tlong=100$.

% Figure: Mean Dphi HST
\begin{figure}[ht!]
	\centering
	\subfloat[\label{subfig:mean_DPhiP_hsf}]{
		\includegraphics[width=0.85\textwidth]{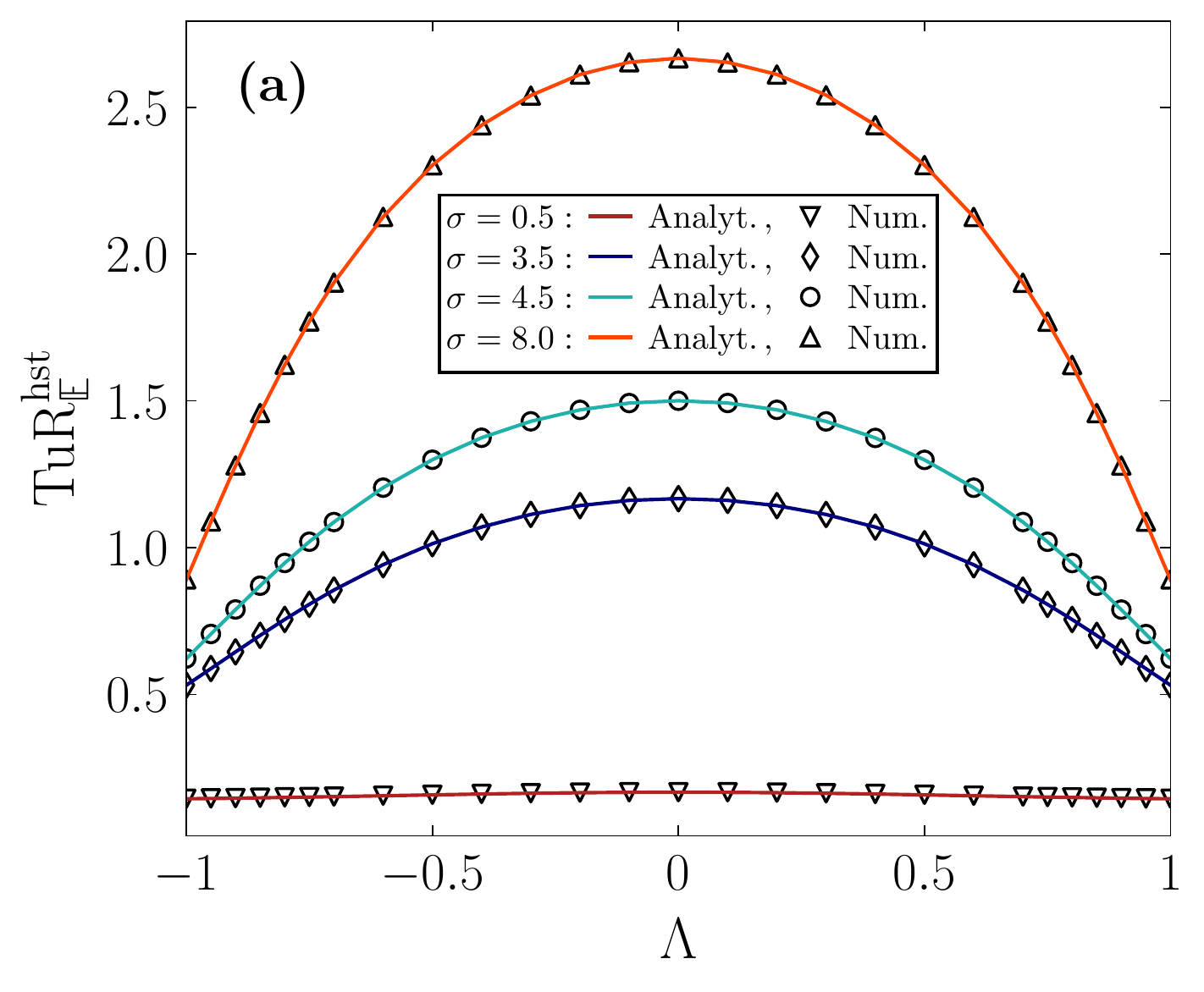}}\\
	\subfloat[\label{subfig:mean_DPhiO_hsf}]{
		\includegraphics[width=0.85\textwidth]{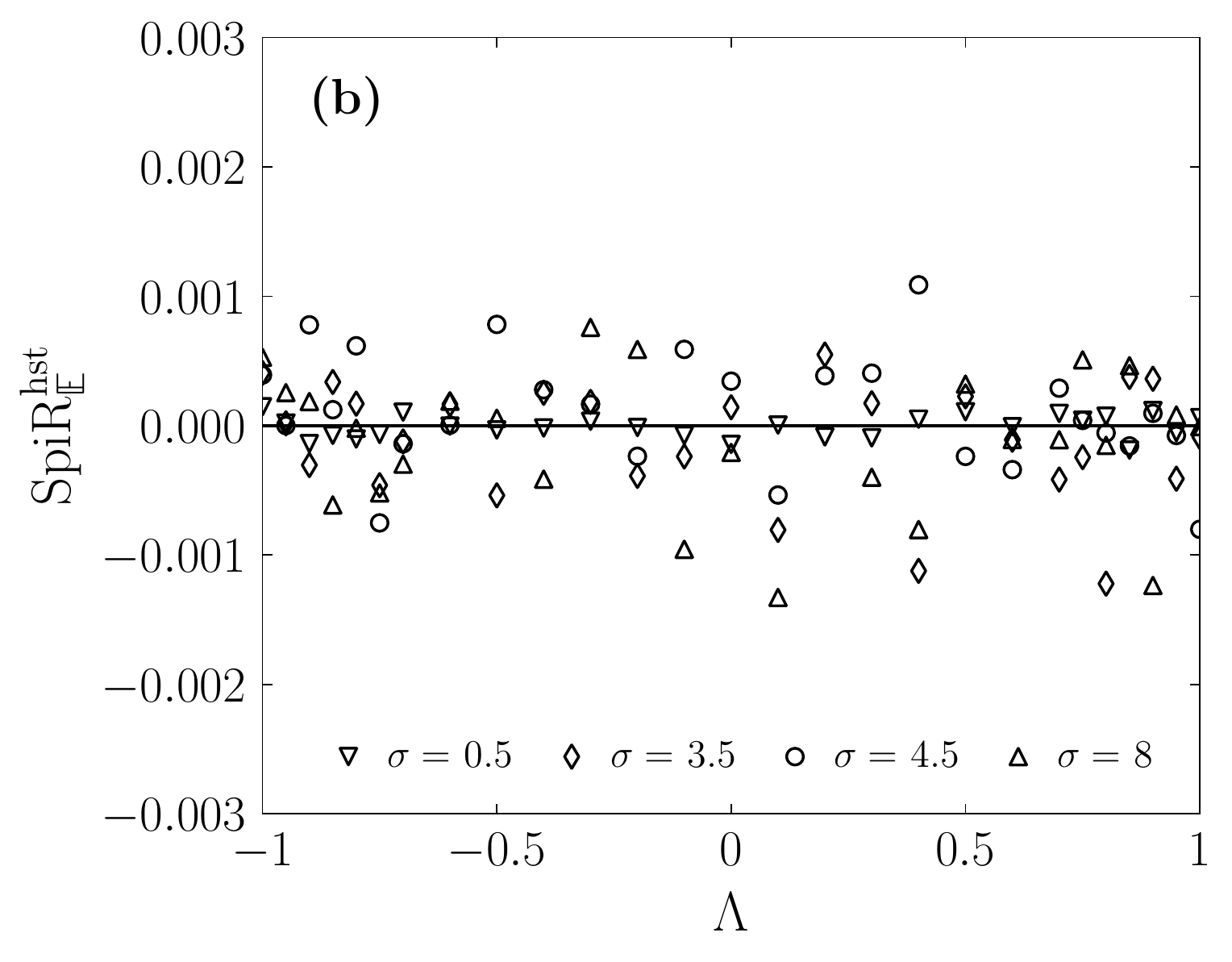}}
	\caption{\label{fig:mean_DPhi_hsf} 
		The mean tumbling rate in \protect\subref{subfig:mean_DPhiP_hsf} and spinning rate in  
		\protect\subref{subfig:mean_DPhiO_hsf} as a function of the particle shape parameter 
		$\Lambda$ in HST for different values of the shear rate parameter $\sigma$ (see figure legend). 
		Markers show results of numerical estimators such as \eqref{eq:estimator_tumb}. 
		Semi-analytical Eq.~\eqref{eq:mean_DPhi_orthogonal_hsf_infty} (lines in 
		\protect\subref{subfig:mean_DPhiP_hsf}) and analytical result (black line in 
		\protect\subref{subfig:mean_DPhiO_hsf}). Simulations are performed with a number of particles 
		$N_p=10^5$, time step $\Delta t=10^{-3}$, final time $T=1000$, time scale parameter 
		$\taukol=1$ and using as initial condition a uniform distribution on a sphere.}
\end{figure}

Figure~\ref{subfig:mean_DPhiP_hsf} shows the mean tumbling rate and Fig.~\ref{subfig:mean_DPhiO_hsf} the mean spinning rate as a function of the particle shape parameter $\Lambda$. Numerical results (markers) are compared to semi-analytical results (lines) and the analytical results for spheres (black cross). Four different values of the shear rate parameter $\sigma$ are used (see figure legend). It can be seen that both $\tumbE$ and $\spinE$ evaluated numerically match with their semi-analytical/analytical values. The maximum of the error between these two approaches can be observed in Fig.~\ref{subfig:mean_DPhiO_hsf}. It shows that the maximum of the error passes from order $2\times10^{-3}$ for $\sigma=8$ to $10^{-4}$ for $\sigma=0.5$. Thus, the error values increase with $\sigma$, which is reasonable compared to strong convergence results. For Fig.~\ref{subfig:mean_DPhiP_hsf}, a similar behaviour is observed.

\begin{figure*}[ht!]
	\centering
	\subfloat[\label{subfig:var_DPhiP_hsf}]{
		\includegraphics[width=0.4\textwidth]{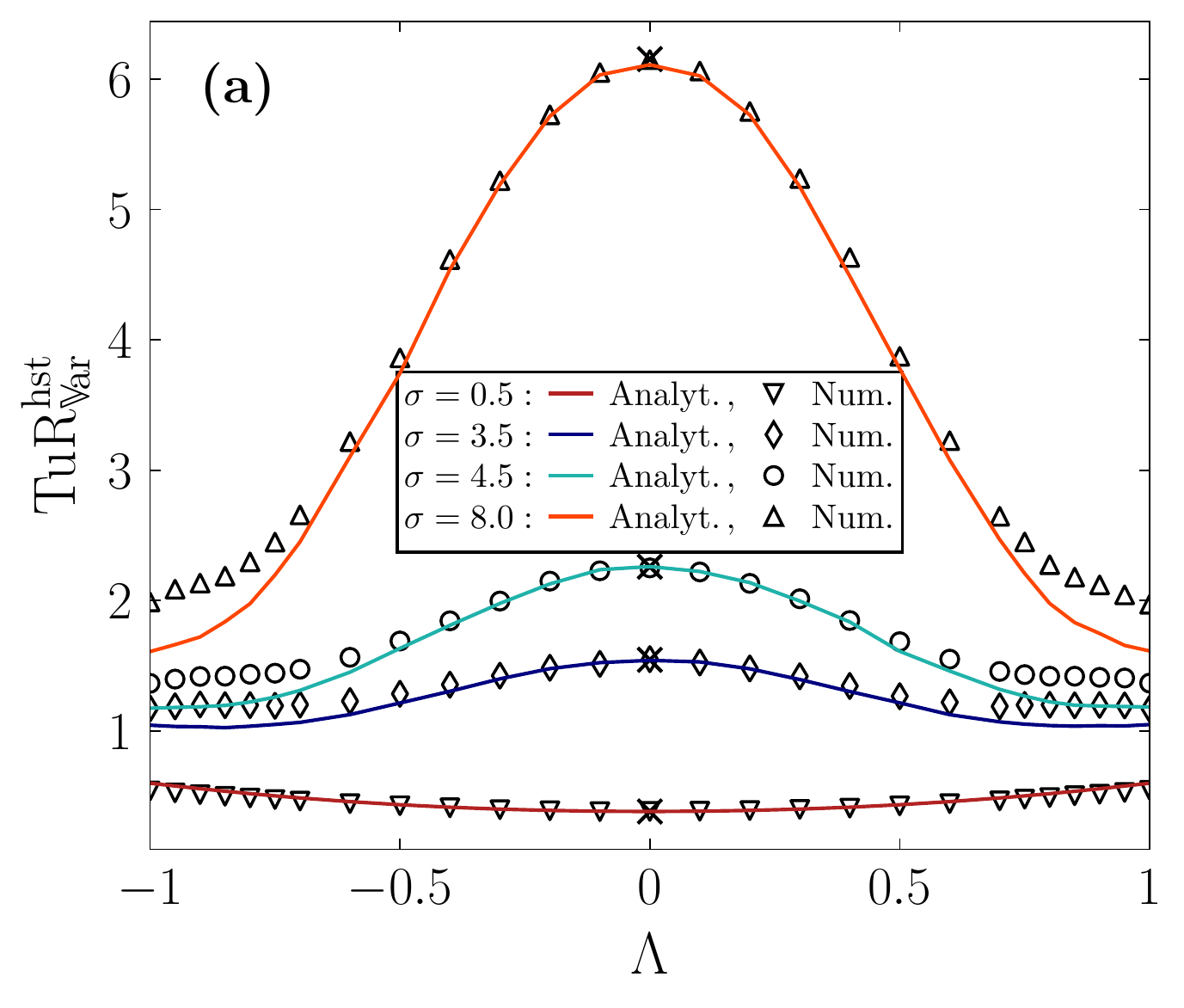}}
	\subfloat[\label{subfig:var_DPhiO_hsf}]{
		\includegraphics[width=0.4\textwidth]{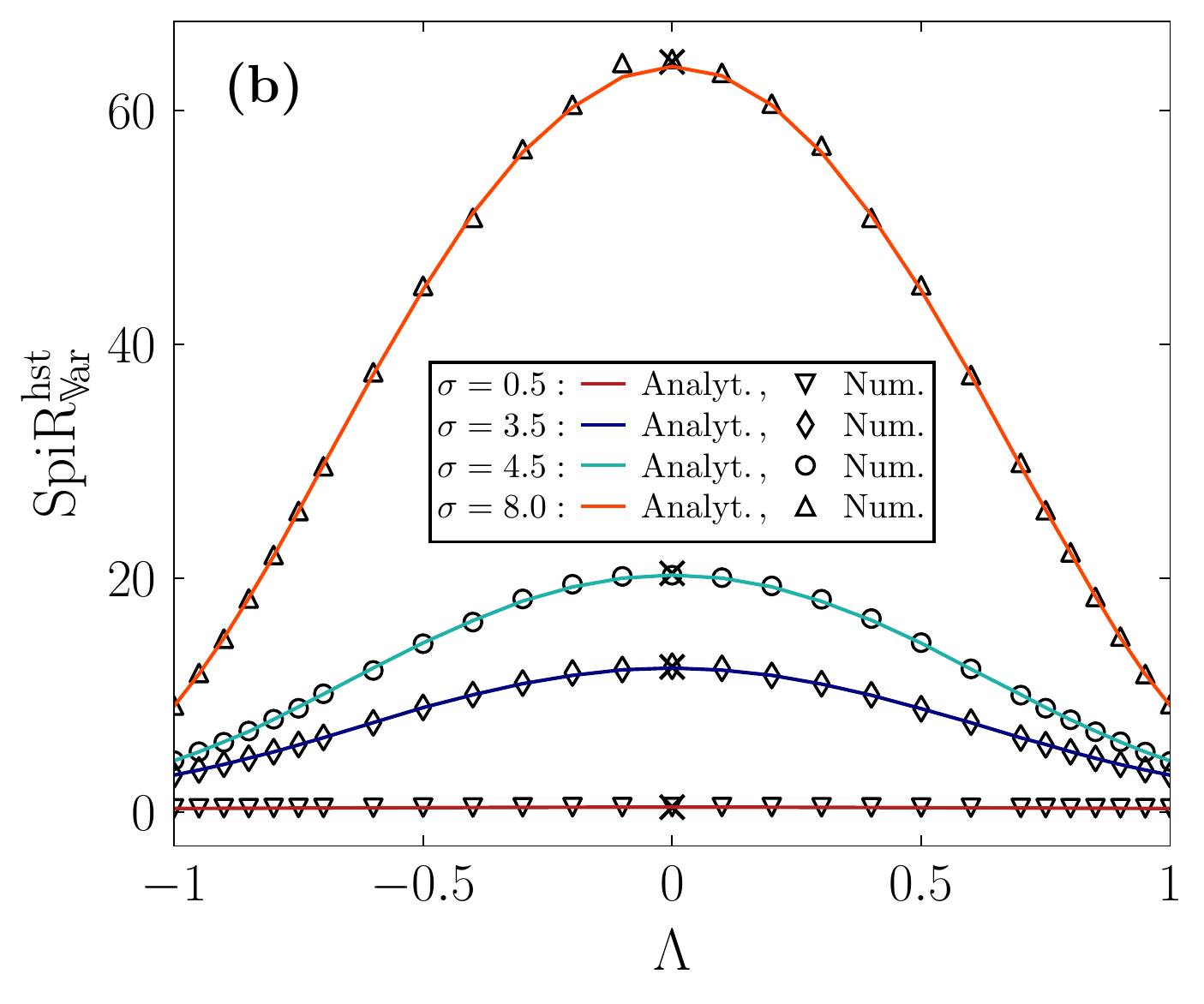}}
	\caption{\label{fig:var_DPhi_hsf} 
		The variance of tumbling rate in \protect\subref{subfig:var_DPhiP_hsf} and spinning rate in 
		\protect\subref{subfig:var_DPhiO_hsf} as a function of the particle shape parameter $\Lambda$ 
		in HST for different values of the shear rate parameter $\sigma$ (as in the caption). Markers 
		show results of numerical simulations in HST. Black cross markers are the analytical results 
		obtained for the case of sphere ($\Lambda=0$) (see~\ref{A:TumbSpin_HIT}).
		Semi-analytical tumbling rate in Eq.~\eqref{eq:DPhi_orthogonal_hsf_infty} (lines) 
		and semi-analytical spinning rate in Eq.~\eqref{eq:DPhi_parallel_hfs_infty} (lines)
		are reported. Simulations are performed with a number of particles $N_p=10^5$, time step
		$\Delta t=10^{-3}$, final time $T=1000$ , time scale parameter $\taukol=1$ and using as initial 
		condition a uniform distribution on a sphere.}
\end{figure*}

\Rev{Figure~\ref{fig:var_DPhi_hsf} displays the evolution of the variance of the tumbling rate (a) and 
the variance of the spinning rate (b) as a function of the particle shape parameter $\Lambda$, for 
different values of the shear rate $\sigma$ (see legend in the figure). The comparison between the 
numerical results obtained (markers) to semi-analytical solutions (in lines) reveals two important 
features. First, numerical results for the spinning rate reproduce the semi-analytical solutions for all 
values of the particle aspect ratio $\Lambda$. In fact, the maximum relative error is about $0.5\%$ 
(it is obtained for the case of spherical particles, \ie $\Lambda=0$). This demonstrates that the 
unfolded statistics of the spinning rate are well reproduced by the numerical splitting scheme. 
Second, numerical results for the tumbling rate match the semi-analytical solutions for small values 
of the shape parameter $\Lambda$. Hence, these results confirm that the present scheme already 
properly captures most of the long-time behavior for spheroidal particles exposed to a constant 
shear (including the orientation, the spinning and tumbling rates).

However, one can notice that the numerical results obtained for the tumbling rate differ from the semi-analytical solutions close to the extreme cases of flat disks ($\Lambda \to - 1$) and long rods ($\Lambda \to 1$). The discrepancy becomes even more pronounced when the value of the shear rate $\sigma$ is increased. One possible origin for this bias could be the method used for the computation of the semi-analytical results for the tumbling. In fact, it leads to evaluate integral correlations of several components, such as $\mathcal{R}_{p_1p_3}(\tau)$, $\mathcal{R}_{p_2p_3}(\tau)$, $\mathcal{R}_{p_1^2}(\tau)$, $\mathcal{R}_{p_2^2}(\tau)$ and $\mathcal{R}_{p_1p_2}(\tau)$. It is a numerically difficult task, more difficult than directly computing $\tumbV$. Hence, we can suggest that high-order moments need to be computed. In comparison with spinning rates (where only second order moments are needed), tumbling rates require the computation of fourth-order moments. In addition, spinning only involves the component along the spheroids' main axis $p_3$, while tumbling gathers contributions from all components of $\pb$. Hence, complex covariance terms should be added in the semi-analytic computations (possibly by resorting to PDE methods).
}

\section{Conclusion and perspectives}
 \label{sec:concl}

In this paper, a new Lagrangian stochastic model for the orientation of spheroids in 3D turbulent flows has been proposed. In particular, this model draws on a recent study for the orientation of elongated particle in 2D turbulence \cite{campana2022stochastic}. The extension to 3D cases has been developed in the context of hybrid formulations that are typically used in CFD software, whereby a finite-volume solver for the fluid phase (here based on a RANS turbulence model) is coupled to a tracking module (here a Lagrangian stochastic method). \Rev{The present model proposes to track the orientation $\pb$ of a spheroid using a stochastic Jeffery equation that reproduces the key features of spheroid orientation}. In addition, a new formulation has been suggested to derive information about the rotation statistics (namely the tumbling and spinning rates) even when relying on the present SDE for the time-evolution of the spheroid orientation.

A specific numerical method has then been designed and implemented in an existing CFD software. The algorithm is based on a semi-implicit splitting method, which includes four different sub-parts: the first two contributions correspond to the deterministic stretching and rotation related to the mean velocity gradient, while the others two contributions are represented by the stochastic Brownian stretching and Brownian rotation due to fluctuations in the velocity gradient. In particular, a semi-implicit scheme for the Brownian rotation sub-part has been developed by adapting the deterministic quaternion dynamics to the stochastic case. This allowed to find a convergent scheme avoiding to introduce more intricate numerical methods. 

The numerical method has been analyzed in simple flow cases. First, the semi-implicit splitting 
method has been proven to be mean-square convergent of order $1/2$ and weakly convergent of 
order $1$. Second, we showed that, in contrast with classical \EM type methods, the proposed 
splitting method is able to preserve the geometric features of the SDE under study (namely, the 
non-extensibility constraint, \ie that the orientation vector should lie on a manifold $SO(3)$). This 
makes our scheme applicable at reasonable computational costs. Further, we pointed out that our 
method is stable on long times, which is a highly valuable feature for applications in realistic cases. 
Third, the statistics of tumbling and spinning rates were evaluated: numerical results showed that 
the model is able to reproduce the analytical solution obtained in the case of homogeneous isotropic 
turbulence. Fourth, the numerical scheme was assessed in the case of a homogeneous shear flow. 
This case confirmed the effectiveness of the numerical scheme for the study of spheroid orientation 
in turbulent flows.

These promising results open the way for a number of improvements that will be addressed in future 
studies. First, the numerical scheme for each sub-part of the splitting scheme can be revisited 
relying on a formulation based on quaternions for all of them. Although more complex to handle, the 
interest of quaternion formulations is that they naturally ensure that the norm of the orientation is 
conserved. This would lead to a more advanced numerical method. Second, new developments are 
foreseen regarding the physical model itself. This includes a refined stochastic model for the 
translational dynamics, which could provide a more accurate coupling between the translational and 
rotational dynamics of spheroids. In addition, the model can be extended to treat the case of inertial 
particles, especially to include inertia effects in the SDE for the spheroid orientation. Last but not 
least, another avenue for advancements is to apply this model to more complex cases, especially to 
wall-bounded turbulence like channel flows. This would require to extend the model to include 
boundary effect. Furthermore, the statistics for tumbling and spinning rates would become much 
more complex to handle than in the idealized flows considered here. In fact, these statistics would 
depend on the distance to the boundary (due to the anisotropy of the flow). As a result, the statistics 
on tumbling or spinning rates should be conditioned on the starting/arriving position of spheroids, 
which would lead to more intricate analysis.

\section*{Acknowledgement}
We acknowledge J{\'e}r{\'e}mie Bec, Martin Ferrand and Jean Pierre Minier for useful and constructive discussions on the model validation.

The authors are grateful to the OPAL infrastructure from Universit\'{e} C\^{o}te d'Azur and Inria Sophia Antipolis - M\'{e}diterran\'{e}e "NEF" computation platform for providing resources and support. This work has been supported by EDF R\&D (projects PTHL of MFEE and VERONA of LNHE) and by the French government, through the Investments for the Future project UCAJEDI ANR-15-IDEX-01 (grant no.~ANR-21-CE30-0040-01) managed by the Agence Nationale de la Recherche.

%----------------------------------------------------------------------------------------
%	APPENDIX
%----------------------------------------------------------------------------------------
\appendix

%----------------------------------------------------------------------------------------
%	FLUCTUATIONS ORIENTATION
%----------------------------------------------------------------------------------------
\section{Dynamics of particles with inertia}
   \label{app:A}

\subsection{Dynamics of inertial spheres}
   \label{app:A:spheres}

\Rev{The Lagrangian module in \CS is based on a stochastic Lagrangian model for the translational dynamics of spheres. It describes the evolution in time of a number of variables associated to each particle, which constitute the state vector $\Zp=(\Xp, \Up, \Us)$, with $\Xp$ the particle position, $\Up$ its velocity and $\Us$ the fluid velocity seen. The corresponding system of stochastic differential equations (SDEs) for the time-evolution of these variables is (more details in \cite{minier2001pdf, minier2016statistical}):
\begin{subequations}
 \label{eq:SLM}
 \begin{align}
  d\Xpi =& \, \Upi \, dt \label{eq:SLM_inertia_xp} \\
  d\Upi =& \, \frac{\Usi-\Upi}{\tau_p} dt \label{eq:SLM_inertia_up} \\
  d\Usi =& \ \textit{Stochastic model.}
 \end{align}
\end{subequations}
This set of SDEs has been obtained considering only the hydrodynamic drag force acting on a sphere (\ie neglecting contributions from Basset history, added-mass force, buoyancy). The drag force is expressed in terms of the particle relaxation time $\tau_p$, which is a measure of particle inertia and is defined as
\begin{equation}
 \label{eq:taup}
 \tau_{p}=\frac{\rho_p}{\rho_{\! f}}\frac{4\,d_p}{3\,C_{\!D} \vert \,\Us-\Up\, \vert}~,
\end{equation}
with $\rho_{\! f}$ (resp. $\rho_p$) the fluid density (resp. particle density), $d_p$ the particle diameter and $C_{\!D}$ the drag coefficient. 

The model used for the drag force requires knowledge on the instantaneous fluid velocity sampled 
at the particle position $\Uf(\Xp(t),t)$. The main difficulty here is that the stochastic Lagrangian 
model is not coupled to fully resolved simulations of a turbulent flow, where $\Us(t)=\Uf(\Xp(t),t)$ 
(as in DNS). Instead, it is coupled to RANS simulations which only provide mean-field information on 
the fluid velocity. RANS models are based on a Reynolds decomposition of the fluid velocity in terms 
of its average and fluctuating part, \ie $\braket{\Uf}+\bm{u}'_{\! f}$. This implies that the 
instantaneous fluid velocity at any point in space is not directly available but has to be modeled. As 
detailed elsewhere \cite{minier2001pdf}, the fluid velocity seen $\Us$ is generally modeled by a 
Langevin model of the form:
\begin{align}
 d\Usi(t) =& \, - \frac{1}{\rho_{\! f}} \frac{\partial \braket{P_{\!f}}}{\partial x_i} dt
  + \left(\braket{\Upj} - \braket{\Ufj}\right) \frac{\partial \braket{\Ufi}}{\partial x_j} dt \nonumber \\
  & + G^*_{ij} \left(\Usj - \braket{\Ufj}\right) dt + B_{s,ij} \, dW_j(t),
 \label{eq:SLM_inertia_us}
\end{align}
where $W_j(t)$ is a vector of independent Wiener processes, $P_{\!f}$ is the fluid pressure and quantities under bracket $\braket{\cdot}$ represent conditional means (\ie mean field quantities evaluated at time $t$ in the region of the particle position $\Xp(t)$). The precise expressions used to close the matrices $G^{*}_{ij}$ and $B_{s,ij}$ have been detailed elsewhere \cite{minier2001pdf,minier2016statistical}. Details on the numerical implementation of such a stochastic Lagrangian approach are available in \cite{peirano2006mean} as well as in the open-source software \CS. 

\paragraph{Limit case of inertialess spheres} In the present paper, we focus on the case of inertialess particles, \ie when $\tau_p=0$. At this stage, it is important to note that the stochastic model for inertial particles described in Eq.~\eqref{eq:SLM} naturally gives Eq.~\eqref{eq:SLM_tracer} for inertialess spheres when $\tau_p$ is set to zero. As detailed in \cite{peirano2006mean}, the algorithm implemented in \CS properly handles the case of vanishing particle relaxation time $\tau_p$ (thanks to the use of exponential formulations).
}

\subsection{Dynamics of inertial spheroids} 
   \label{app:A:spheroid}

\Rev{
When dealing with spheroidal particles, the state vector is usually extended to track not only the translational motion but also the rotational motion of spheroids. For instance, assuming point-particles moving due to Stokes drag only, the equations of rigid-body motion are given by (see \eg \cite{mortensen2008dynamics, voth2017anisotropic}):
\begin{align}\label{eq:translat_inert_ellips}
\begin{aligned}
 \frac{d\Xp}{dt} &=\Up  \\
 \frac{d\Up}{dt} &= \frac{\nuf \ \rho_{\! f}}{m_p} \Mc^{-1}  \Ktilde \ \Mc \left(\Us - \Up\right) \\
 \frac{d\left(\Ic \cdot \Rtilde\right)}{dt} & + \Rtilde \times \left(\Ic \cdot \Rtilde\right)= \Ntilde.
\end{aligned}
\end{align}
The equations for the translational dynamics (first two lines) resemble those for sphere (see Eq.~\eqref{eq:SLM}), except that the drag force has now different components along each direction due to the spheroid orientation with respect to the fluid. In fact, the tensor $\Mc$ is a rotation matrix allowing to change the frame of reference from the global one ($x,y,z$) to the local one attached to the spheroid ($\widehat{x},\widehat{y},\widehat{z}$), as displayed in Fig.~\ref{fig:spheroids}. In the local coordinate system, the particle resistance tensor $\Ktilde$ is a purely diagonal matrix. The exact expressions entering this system are detailed elsewhere (e.g., see previous use in DNS simulations with point-particle spheroids in \cite{mortensen2008dynamics, siewert2014orientation}). Note that, when Jeffery equation are coupled to DNS simulations, the instantaneous fluid velocity $\Us$ is directly equal to the fluid velocity at the particle position (as usually written in papers relying on DNS simulations, \eg \cite{marchioli2010orientation, mortensen2008dynamics}).

The last equation for the rotational dynamics is also expressed in the local frame of reference attached to a spheroid. This allows to have a rotational inertia tensor $\Ic$ that remains constant in time (for rigid particles). The evolution of the local rotational velocity $\Rtilde$ depends on the torques acting on spheroids $\Ntilde$: expressions for these torques were provided by Jeffery \cite{jeffery1922motion} and they are directly related to the fluid velocity gradient (see also \cite{marchioli2016relative, mortensen2008dynamics} for more details).

% In the framework of stochastic Lagrangian method, we have chosen here to extend the previous SDEs for spheres to the case of spheroids by replacing the particle relaxation time with the particle resistance tensor $\Ktilde$ (expressed in the local coordinate system associated to each spheroid). This choice amounts to fixing the particle orientation during the time step $\dt$ in order to compute its trajectory. This choice is consistent with other choices made, in particular concerning the coupling to the fluid velocity seen by particles (based on a $\rm{P0}$ interpolation of the average velocity given by RANS computations).  
% Improved models for the translational dynamics which account for the change of orientation during the time step are left out of the scope of this paper (this leaves roads for future model improvements that will be recalled in the conclusion).

}

%----------------------------------------------------------------------------------------
%	FLUCTUATIONS ORIENTATION
%----------------------------------------------------------------------------------------
\section{Isotropic tensor for the velocity gradient's fluctuation}\label{A:D_tensor}
 The expression of the single-time second order 
tensor function $\braket{\partial_j u'_{\! f,i}(0) \partial_l u'_{\! f,k}(0)}$ can be obtained imposing  
incompressibility (trace-free), homogeneous, and general isotropic form~\cite{pumir2017structure}:
\begin{equation}\label{Aeq:correl_grad}
	\braket{\partial_j u'_{\! f,i}(0) \partial_l u'_{\! f,k}(0)} = \tfrac{1}{30}
	\frac{\varepsilon}{\nuf}\left( 4 \delta_{ik}\delta_{jl} -\delta_{ij}\delta_{kl} 
	-\delta_{il}\delta_{jk}\right). 
\end{equation}
This expression depends  only on the correlation time scale of 
the velocity gradient tensor, $\varepsilon/\nuf = \taukol^{-2}$. 
Finally $\Cijkl$ is recovered from~\eqref{Aeq:C_tens} and~\eqref{Aeq:correl_grad} with 
\begin{equation*}
	2\Cijkl = \tfrac{1}{15}\frac{\alpha}{\taukol} \left( 4 \delta_{ik}\delta_{jl} -\delta_{ij}\delta_{kl} 
	-\delta_{il}\delta_{jk}\right). 
\end{equation*}
The parameter $\alpha=\tau_{\textrm{I}}/\taukol$ is often interpreted as a Kubo number and links the 
instantaneous properties of the flow (entailed in $\taukol$) to the long-term effect of gradients that 
the noise with correlations $\Cijkl$ is expected to reproduce. It remains to solve $\Dc_{ijmn} \Dc_{klmn} = 2\Cijkl$. Using the general 
isotropic form $\Dc_{ijkl} = d_1 \delta_{ik}\delta_{jl} +d_2 \delta_{ij}\delta_{kl} +d_3 
\delta_{il}\delta_{jk}$, the tensor contractions yield to the following system 
\begin{equation*}
	\begin{aligned}
		\tfrac{4}{15}\frac{\varepsilon}{\nuf} &= d_1^2 +d_3^2,  \\
		-\tfrac{1}{15}\frac{\varepsilon}{\nuf} &= 2d_2d_1 +3d_2^2 +2d_2d_3, \\
		-\tfrac{1}{15}\frac{\varepsilon}{\nuf} &= 2d_1d_3, \\
	\end{aligned}
\end{equation*}
which is solved in
\begin{equation}\label{Aeq:D_tensor}
	\begin{aligned}
	\sqrt{\tfrac{\taukol}{\alpha}}	\  \Dc_{ijkl} = & \tfrac{1}{2}  \left(\sqrt{\tfrac{1}{5}} 
		+\sqrt{\tfrac{1}{3}}\right)\delta_{ik}\delta_{jl}
		-\tfrac{1}{3}\sqrt{\tfrac{1}{5}}\delta_{ij}\delta_{kl} \\
		&+\tfrac{1}{2}\left(\sqrt{\tfrac{1}{5}}-\sqrt{\tfrac{1}{3}}\right)\delta_{il}\delta_{jk}.		
	\end{aligned}
\end{equation}
 
%----------------------------------------------------------------------------------------
%	ITO-STRATONOVICH
%----------------------------------------------------------------------------------------
\section{Stochastic diffusion in the model}\label{A:ito_stra}
We detail the computation of the fluctuations contribution in  
Eq.~\eqref{eq:model_elongation_strato}, knowing the $\Dc_{ijkl}$ from \ref{A:D_tensor}. We further 
compute the  \Stra to \Ito  drift term  for the elongation SDE.  

Considering the noise part of Eq.~\eqref{eq:sde_jeffery_linear_strat} together  
with~\eqref{Aeq:D_tensor}, we have
\begin{align}
\begin{aligned}
&\Dc_{ijkl} \Wm_{kl} = d_1 \delta_{ij}  \ \Wm_{\ell \ell} \\
& + \begin{pmatrix*}[r]
(d_2 +d_3) \Wm_{11}   & 
d_2 \Wm_{12} + d_3 \Wm_{21} & 
d_2 \Wm_{13} + d_3 \Wm_{31} \\
d_2 \Wm_{21} + d_3 \Wm_{12}  & 
(d_2 +d_3)  \Wm_{22}  & 
d_2 \Wm_{23} + d_3 \Wm_{32} \\
d_2 \Wm_{31} + d_3 \Wm_{13} & 
d_2 \Wm_{32} + d_3 \Wm_{23} & 
(d_2 +d_3) \Wm_{33} 
\end{pmatrix*}%$}
\end{aligned}
\end{align}
and we easily identify 
\begin{align*}
\begin{aligned}
(\Dc_{ijkl} \Wm_{kl})^a  = 
(d_2 - d_3) \Wma_{ij}  = \nua \  \Wma_{ij}\\ 
(\Dc_{ijkl} \Wm_{kl})^s  = 
(d_2 + d_3) \Wms_{ij} + d_1 \delta_{ij} \Wm_{\ell \ell} \\
		 = \nus \big(\Wms_{ij}  - \tfrac{1}{3} \Tr(\Wm) \delta_{ij}\big) \\
\text{with} \quad \nus = \sqrt{\tfrac{\alpha}{5 \taukol}}=\sqrt{\tfrac{\alpha}{5}}\left(\frac{\varepsilon}{\nuf}\right)^{\frac{1}{4}} \\
\text{and} \quad \nua = \sqrt{\tfrac{\alpha}{3 \taukol}}=\sqrt{\tfrac{\alpha}{3}}\left(\frac{\varepsilon}{\nuf}\right)^{\frac{1}{4}}.
	\end{aligned}
\end{align*}

To compute the orientation model in its \Ito form, we  rewrite first the \Stra diffusion term as a $3 \times 9$ matrix against the 9d standard Brownian motion $\Wb$  which  maps  $\Wc_t^{mn}={W}_t^{3(m-1)+n}$: 
\begin{align*}
\Wb := (\Wc_{11}, &\Wc_{12}, \Wc_{13},
\Wc_{21},\Wc_{22},\Wc_{23}, 
\Wc_{31},\Wc_{32},\Wc_{33})^\intercal, 
\end{align*}  
and such that 
\begin{align}\label{Aeq:def_diffusion}
	\begin{aligned}
		&\left(\mathcal{D}_{ijkl} \partial \Wm_{kl}\right)^{a}  r_j
		+\Lambda \left(\mathcal{D}_{ijkl} \partial \Wm_{kl}\right)^{s}  r_j \\
		&= \nua \partial \Wma_{ij}  r_j
		+\Lambda  \nus \big(\partial \Wms_{ij}  - \frac{1}{3} \Tr(\partial \Wc) \delta_{ij} r_j \big) \\
		&=\left(\Gc^{a}_{ik}(\qb) +\Lambda \; \Gc^{s}_{ik}(\qb)\right) \partial \bm{W}_{k}
	\end{aligned}
\end{align}
where we identify
\small{
\begin{subequations}
	\begin{align*}%\label{eq:bm_rotation_matrix}
		\Gc\supa(\qb) &= \frac{\nua}{2}
		\resizebox{.7\hsize}{!}{$
		\begin{pmatrix*}[c]
			0 & r_2 & r_3 & -r_2  & 0 & 0	 &-r_3  & 0 	& 0 \\ 	
			0 &-r_1 & 0	   & r_1   & 0 & r_3 &  0	 & -r_3 & 0 \\
			0 & 0	 &-r_1 &  0 	& 0 &-r_2 & r_1  & r_2  & 0 
		\end{pmatrix*}$}, 
	\end{align*}
	and $\Gc\sups = \bar{\Gc}\sups  - \tilde{\Gc}\sups$, with 
	\begin{align*}%\label{eq:bm_stretc_matrix}
		\overline{\Gc}\sups(\qb)  = \frac{\nus}{2}
		\resizebox{.7\hsize}{!}{$
		\begin{pmatrix*}[c]
			2r_1 & r_2 & r_3 & r_2 & 0 		& 0		& r_3 & 0 	  & 0 	\\ 	
			0  & r_1  & 0	 & r_1 & 2r_2 & r_3 &  0	& r_3 & 0 	\\
			0  & 0	  & r_1  &  0 	& 0   & r_2 & r_1 & r_2  & 2r_3 
		\end{pmatrix*}$}, \\
		\widetilde{\Gc}\sups(\qb)  =\frac{\nus}{3}
		\resizebox{.7\hsize}{!}{$
		\begin{pmatrix*}[c]
			r_1 & 0 &  0 & 0 & r_1 & 0 & 0 & 0 & r_1  \\ 	
			r_2 & 0 & 0	& 0  & r_2 & 0 & 0 & 0 & r_2 \\
			r_3 & 0	& 0 & 0  & r_3 & 0 & 0 & 0 & r_3 
		\end{pmatrix*}$}.
	\end{align*}
\end{subequations}
}
We apply the 
conversion rule from \Stra to \Ito integrals (see e.g.~\cite{oksendal2003stochastic}) on the   
stochastic term
\begin{align*} %\label{eq:ito_stratonovich}
\begin{aligned}
\Gc_{ik} (\qb (t)) \ \partial W_k =& \Gc_{ik} (\qb (t)) \ dW_k \\
&+ \tfrac{1}{2}  \Gc_{jk}(\qb (t)) \frac{\partial \Gc_{ik}}{\partial r_j}(\qb (t)) dt,
\end{aligned}
\end{align*}
where $\Gc(\qb) = \Gc\supa(\qb) + \Lambda \Gc\sups(\qb)$.  We further identify the additional \Ito 
term for the $i$th component as 
$$\tfrac{1}{2} \Gc_{jk}(\qb) \frac{\partial \Gc_{ik}}{\partial r_j}(\qb) 
= \left(\Lambda^2 \tfrac{5}{6} \nus^2 - \tfrac{1}{2} \nua^2\right)r_i.
$$
In this way,  Eq.~\eqref{eq:model_elongation_strato} rewritten in the \Ito convention is,
\begin{equation}\label{Aeq:model_elongation_ito}
	\begin{aligned}		
	dr_i(t) = & \left(\braket{\Oc_{ij}} + \Lambda \braket{\Sc_{ij}} 
	+ (\Lambda^2 \tfrac{5}{6} \nus^2- \tfrac{1}{2}  \nua^2)\delta_{ij}\right) \, r_j(t) \, dt \\
	&+ \Gc_{ik}(\qb (t))  dW_{k}.
	\end{aligned}
\end{equation}
This linear SDE is well posed and admits a strong solution at any time, with pathwise uniqueness 
(see \eg \cite{pages2018numerical}). 

%----------------------------------------------------------------------------------------
%	ITO LEMMA ORIENTATION 
%----------------------------------------------------------------------------------------

\section{It\^o's Lemma for the orientation}\label{A:ito_lemma}
The SDE~\eqref{Aeq:model_elongation_ito} for the separation $\qb $ shortly rewrites  
\begin{equation}\label{Aeq:r_ito_lemma}
	d\qb (t) = \bm{\mu}(t) \qb (t)\, dt +\Gc(\qb(t)) \, d\Wb_t
\end{equation}
with $\bm{\mu}(t)=\Braket{\Oc} +\Lambda \Braket{\Sc} +(\Lambda^2 \frac{5}{6} \nus^2- \frac{1}{2}  
\nua^2) \Id$.

In order to derive the stochastic version of Jeffery equation~\eqref{eq:model_jeffery_ito0}, we apply 
the \Ito's lemma on the renormalization  function (or projection function on the sphere) $q\mapsto 
\pb =F(\qb ) = \frac{1}{\|\qb \|}(r_1, r_2, r_3)^\intercal$ applied to the solution of SDE 
\eqref{Aeq:model_elongation_ito} getting  
\begin{align}\label{Aeq:multi_ito_lemma}
	\begin{aligned}
		d\bm{p}(t) =& \Big(J_F(\qb (t)) \, \bm{\mu}(t) \qb (t)
		+ \bm{h}_{\Tr}(\qb (t)) \Big)dt \\
		&+ \bm{J}_{F}(\qb (t)) \, \Gc(\qb (t)) \; d\Wb_t 	
	\end{aligned}
\end{align}
where $\bm{J}_F(\qb)$ is the Jacobian matrix of the renormalization
\begin{equation*}
\bm{J}_F(\qb )={\|\qb \|^{-1}}\Id - {\|\qb \|^{-3}} \Sigma(\qb ),
\end{equation*}
with $\Sigma(\qb) = \qb \qb^\intercal$, and $\bm{h}_{\Tr}(\qb)$ denotes the additional \Ito term 
obtained from its Hessians
\begin{equation*}
\small{
	\bm{h}_{\Tr}(\qb)=\tfrac{1}{2}
	\begin{pmatrix}
%		\Tr\Big[\Gc^\intercal \; \Hess[F_1](\qb) \Gc(\qb) \Big] \\
%		\Tr\Big[\Gc^\intercal \; \Hess[F_2](\qb) \Gc(\qb) \Big] \\
		\Tr\big[\Gc^\intercal \; \Hess[F_i](\qb) \Gc(\qb) \big], i 
	\end{pmatrix}^\intercal.}
\end{equation*}

\paragraph{The diffusion matrix in \eqref{Aeq:multi_ito_lemma}} 
We observe that $\Sigma(\qb ) \Gc\supa(\qb)=0$ and then 
$$\bm{J}_F(\qb ) \Gc\supa (\qb )=\Gc\supa (F(\qb )) = \Gc\supa (\pb), $$ 
which translates the fact that rotation preserves the projection onto the unit sphere.   The  diffusion in \eqref{Aeq:multi_ito_lemma} writes 
$$\bm{J}_F (\qb ){\Gc}(\qb ) = \Gc\supa (\pb) + \Lambda \bm{J}_F(\qb ) {\Gc}^s(\qb).$$ 
We observe next that $\bm{J}_F(\qb) \tilde{\Gc}\sups(\qb ) = 0$, and 
%$J_F(\qb)\Gc\sups(\qb ) = \bm{J}_F(\qb )\overline{\Gc}\sups(\qb)$, with  
\begin{align*}
	\begin{aligned}
		&\bm{J}_F(\qb) \Gc\sups (\qb) = \overline{\Gc}\sups(\pb) - \nus B(\pb), \qquad\text{with} \\
		& B(\pb)=
		\resizebox{.8\hsize}{!}{$
		\begin{bmatrix*}[r]
			p_1^3 & p_1^2p_2 & p_1^2 p_3  
			&  p_1^2 p_2 & p_1 p_2^2 &  p_1 p_2 p_3 
			&  p_1^2 p_3 &   p_1 p_2 p_3  &  p_1 p_3^2 \\
			p_1^2 p_2  &  p_2^2 p_1 &  p_1 p_2 p_3 
			& p_1 p_2^2 & p_2^3 & p_2^2 p_3 & p_1 p_2 p_3 
			& p_2^2 p_3 & p_2 p_3^2  \\
			p_1^2 p_3 & p_1 p_2 p_3 & p_1p_3^2 & p_1 p_2 p_3 & p_2^2 p_3 & p_2 p_3^2 & p_1 
			p_3^2 & p_2 p_3^2 & p_3^3
		\end{bmatrix*}$}.
	\end{aligned}
\end{align*}
So, introducing the matrix $B$ above, we recognize the diffusion in \eqref{Aeq:multi_ito_lemma} with Brownian motions $\Wb$ (or $\Wc$ considering~\eqref{Aeq:def_diffusion}), to be 
\begin{align*}
\bm{J}_F(\qb) \Gc\sups (\qb) d\bm{W}_t &= \big( \Gc\supa(\pb) + \Lambda \overline{\Gc}\sups(\pb)   
	- \nus \Lambda B(\pb) \big) d\bm{W}_t \\
	&= \nua d\Wc^a_t \pb + \Lambda \nus d\Wc^s_t \pb 
	- \Lambda \nus (\pb^\intercal d\Wc^s_t \pb) \pb.
\end{align*}

\paragraph{The It\^o term in \eqref{Aeq:multi_ito_lemma}}   
It remains to compute $\bm{h}_{\Tr}(\qb)$ by identifying 
\begin{align*}
\Hess[F_k]_{ij}  &= 
- {\|\qb\|}^{-3} ( r_j\delta_{ik} + r_k\delta_{ij} 
+ r_i\delta_{kj})  + 	{3}{\|\qb\|^{-5}} r_ir_jr_k, 
\end{align*}
and \eqref{Aeq:multi_ito_lemma} rewrites as 
\begin{equation}\label{Aeq:model_jeffery_ito}
	\begin{aligned}
		dp_i(t) =& \braket{\Oc_{ij}} p_j 
		+\Lambda \left(\braket{\Sc_{ij}} ) p_j -p_i p_k \braket{\Sc_{kl}} p_l\right) \\
		&-\frac{\nus^2}{2}  \Lambda^2 p_i dt 
		+\nus \Lambda \left(d\Wcs_{ij}p_j -p_i p_k d\Wcs_{kl} p_l \right) \\
		&-\frac{\nua^2}{2} p_i dt +\nua d\Wca_{ij} \ p_j,
	\end{aligned}
\end{equation}
which is the component-by-component  version of \eqref{eq:model_jeffery_ito0}.

%******************************************************************************
% Appendix: Ito Isometry
%******************************************************************************
\section{Stochastic Tumbling and Spinning}\label{A:TumbSpin_HIT}
We first detail the computation of the tumbling and spinning rates in Eq.~\eqref{eq:DPhi_orthogonal_hit_nu} in the HIT flow case.  Eqs.~\eqref{eq:phi_orthogonal_devel}-\eqref{eq:phi_parallel_devel} become
\begin{align}\label{Aeq:phi_devel} 
\begin{aligned}
\bmphiO(t)	=&\tfrac{1}{2}{\nua}\int_0^{t} 
		(\Id -\pb \pb^\intercal) d\wa_s
		+\nus \Lambda \int_0^{t}\pb \times d\Wcs_s \pb,	\\		\phiP(t) =&\tfrac{1}{2}{\nua}\int_0^{t} \pb \cdot d\wa_s.
\end{aligned}
\end{align}
From the martingale property of the \Ito integral, we immediately deduce that $\EE[\bmphiO(t)]=0$ and $\EE[\phiP(t)]=0$. 
To compute the expectation of the norm squared of $\bmphiO$ and $\phiP$, we need to apply the \Ito isometry for the stochastic integrals appearing in the right-hand side of \eqref{Aeq:phi_devel}. 

The \Ito isometry \cite{oksendal2003stochastic} has the following property: 
if $(\bm{B}_t, t\geq 0)$ is a standard  $n$-dimensional Brownian motion and $(\bm{M}_t,t\geq 0)$ is a matrix-$\mathbb{R}^{d \times n}$-valued  stochastic process (adapted to the natural filtration of the Brownian motion), then
	\[
	\EE[\|\int_{0}^{t} \bm{M}_s \, d\bm{B}_s\|^2]
	=\EE[\int_{0}^{t} \|\bm{M}_s\|_{\mathcal{F}}^2 \, ds],
	\]  
where the norm $\|\cdot\|_{\mathcal{F}}$ denotes the Frobenius norm for matrix. With the standard 9d Brownian vector $\Wb$ defined in  \ref{A:ito_stra}, with the matrix process $\bm{M}$ defined below, the expectation of the norm squared for $\bmphiO$ can be written as,
\begin{align*}
	\begin{aligned}
		&\EE [\| \bmphiO(t) \|^2] = \EE [\| \int_0^{t} \left(
		\tfrac{1}{2}{\nua}(\Id -\pb \pb^\intercal) d\wa
		+\nus \Lambda \pb \times d\Wcs \pb \right)\|^2] \\
		&= \EE [\| \int_0^t \bm{M}_s d\bm{W}_s \|^2]
		= \EE [\int_0^{t}\|\bm{M}_s\|_{\mathcal{F}}^2 ds ]=(\nua^2 +\nus^2 \Lambda^2 ) t.
	\end{aligned}
\end{align*}
From which we deduce that $\tumbV^{\textrm{hit}} =\nua^2 +\nus^2 \Lambda^2$ for any time. 
To ease the identification of the ${3 \times 9}$ matrix $\bm{M}$, we define it by blocks of 3 by 3 matrices:
$$
\bm{M}_{1:3,1:9} = \tfrac{1}{2} (\bm{M}_{1:3,1:3} \, \bm{M}_{1:3,4:6} \, \bm{M}_{1:3,7:9})
$$
with, using the cross notation \eqref{eq:cross_notation}, using $\ell^\pm = \nua \pm \Lambda \nus$, 
$$
\resizebox{0.8\hsize}{!}{$
\bm{M}_{1:3,1:3}=	\begin{pmatrix*}
	0 & \ell^- p_1p_3 & -\ell^- p_1p_2   \\
	2\Lambda \nus p_1p_3 & \ell^+ p_2p_3 &  \ell^+ p_3^2   \\
	-2\Lambda \nus p_1p_2 & - \ell^+ p_2^2 & - \ell^+ p_2p_3  \\
\end{pmatrix*} - \ell^-p_1^2 
\begin{bmatrix*}
1\\
0\\
0
\end{bmatrix*}_\times
$}
$$
$$
\resizebox{0.8\hsize}{!}{$
\bm{M}_{1:3,4:6}=	\begin{pmatrix*}
	-\ell^+ p_1p_3  & -2\Lambda \nus p_2p_3 
	& -\ell^+ p_3^2	  \\
	- \ell^- p_2p_3 & 0 & \ell^- p_1p_2 3  \\
	 \ell^+ p_1^2 & 2\Lambda \nus p_1p_2 & \ell^+ p_1p_3 \\
\end{pmatrix*} - \ell^-p_2^2 
\begin{bmatrix*}
0\\
1\\
0
\end{bmatrix*}_\times$}
$$
$$
\resizebox{0.8\hsize}{!}{$
\bm{M}_{1:3,7:9}=	\begin{pmatrix*}
	\ell^+ p_1p_2 &  \ell^+ p_2^2 
	& 2\Lambda \nus p_2p_3  \\
	- \ell^- p_1^2 
	& - \ell^+ p_1p_2 & - 2\Lambda \nus p_1p_3  \\
	\ell^- p_2p_3 & - \ell^- p_1p_3 & 0  \\
\end{pmatrix*} - \ell^-p_3^2 
\begin{bmatrix*}
0\\
0\\
1
\end{bmatrix*}_\times$}.
$$
With the same technique,  we compute the expectation of the norm squared for $\phi_{\parallel p}$:
\begin{equation*}
\begin{aligned}
\EE [ \phi_{\parallel p}(t) ]^2 =	\EE [ \int_0^{t} \tfrac{1}{2}{\nua} \pb\cdot d\wa ]^2 =\tfrac{1}{2}{\nua^2} \EE [
		\int_0^{t} \|\pb\|_{\mathcal{F}}^2 ds]=\tfrac{1}{2}{\nua^2} t,
	\end{aligned}
\end{equation*}
since $\pb$ has a unitary norm and considering the explicit form of $\wa$ 
in~\eqref{eq:wa_vorticity_fluctuation}. From this, we can deduce that $\spinV^{\textrm{hit}} =\tfrac{1}{2}{\nua^2}$ for any time.

In the HTS  case, the possibility to derive analytic formula in the model is much more limited. In the case $\Lambda=0$ (spherical particles), with the help of \eqref{eq:drifts_hsf}, the drifts in  Eqs.~\eqref{eq:phi_orthogonal_devel}-\eqref{eq:phi_parallel_devel} become
\begin{equation}
\begin{aligned}
  \bm{g}^{\perp}(\pb)= \tfrac{1}{2}{\sigma}(
   p_1p_3, 
   p_2p_3,  
   p_3^2-1 )^\intercal, \qquad g^{\parallel}(\bm{p})=\tfrac{1}{2}\sigma p_3.
\end{aligned}
\end{equation}
and 
\begin{align*}
\begin{aligned}	
d\bmphiO(t) &=  \bm{g}^{\perp}(\pb) +\tfrac{1}{2}{\nua}(\Id -\bm{p}\bm{p}^\intercal) d\wa\\
d\phiP(t) &= \tfrac{1}{2}\sigma p_3 dt+\tfrac{1}{2}{\nua} \bm{p} \cdot d\wa. 
\end{aligned}
\end{align*}
The computation of statistics such as \eqref{eq:mean_DPhi_orthogonal_hsf_infty} relies on the computation of cross time second-order terms such as $\EE[p_i(t)p_j(t)p_i(s)p_j(s)]$. 
But, except in the \Ito integrals involved in \eqref{eq:model_jeffery_ito0}, with $\Lambda=0$, the structure of the orientation equation \eqref{eq:model_jeffery_ito0} is linear, leading to a linear system that can be solved explicitly, after taking the expectation in 
\small{
\begin{align*}
&d(p_3^2(s)p_3^2(t)) = \frac{\nua^2}{2} (p_3^2(s)-3(p_3^2(s)p_3^2(t))) dt 
	+\text{\Ito integral}\\		
&dp_1p_3(t) =  \frac{\sigma}{2} p_2p_3(t) dt -\frac{3}{2} \nua^2 p_1p_3(t) dt +\text{\Ito integral}, \\
&dp_2p_3(t) = -\frac{\sigma}{2} p_1p_3(t) dt -\frac{3}{2} \nua^2 p_2p_3(t) dt +\text{\Ito integral}\\
&d(p_1p_3(s)p_1p_3(t)) \\
&=  \frac{\sigma}{2} p_1p_3(s)p_2p_3(t) dt
		-\frac{3}{2} \nua^2 p_1p_3(s)p_1p_3(t) dt +\text{\Ito integral}, \\
&d(p_2p_3(s)p_2p_3(t)) \\
&= -\frac{\sigma}{2} p_2p_3(s)p_1p_3(t) dt
		-\frac{3}{2} \nua^2 p_2p_3(s)p_2p_3(t) dt +\text{\Ito integral}.
\end{align*}
}
We refer to \cite{campana2022phd} for the detailed computation and we jump to its conclusion 
\begin{align*}
\spinV^{\textrm{hst}}\big|_{\{\Lambda=0\}}(\infty)&= \tfrac{1}{3}\frac{\sigma^2}{\nua^2}  +\tfrac{1}{2}{\nua^2}, \\
\tumbV^{\textrm{hst}}\big|_{\{\Lambda=0\}}(\infty)&= \tfrac{2}{5}\sigma^2
	\big((\frac{\nua^2}{\sigma^2+9\nua^2}) 
	+\tfrac{2}{27}\nua^{-2}\big) + \nua^2.
\end{align*}
%--------------------------------------------------------------------------
% Appendix: Analytical moments 
%--------------------------------------------------------------------------
\section{Statistical moments' equation}\label{A:moments_analytical}
Considering the SDE~\eqref{Aeq:model_jeffery_ito} in the absence of a mean gradient contribution
\begin{equation}
\begin{aligned}
dp_i(t) = &-\tfrac{1}{2} ({\nus^2}\Lambda^2 + \nua^2)  p_i dt  \\
& +\nus \Lambda \left(d\Wcs_{ij}p_j -p_i p_k d\Wcs_{kl} p_l \right) +\nua d\Wca_{ij} \ p_j,
\end{aligned}
\end{equation}
denoting $\kappa = {\nus^2}\Lambda^2 + \nua^2$, and taking expectation on both sides, we get immediately 
\[
\frac{d}{dt}\EE[p_i](t) 	= - \tfrac{1}{2} {\kappa}\EE[p_i]\quad\textrm{ or }\quad
\EE[p_i](t) = \EE[p_i](0) e^{-\frac{\kappa}{2} t}.\]
Second- or three-order moments computation requires to apply the \Ito's lemma  in 3D   to the 
function $\pb\mapsto p_ip_j, p_i^3$.  The application of the formula is a simple calculation, but it 
takes up space. We will not detail it here to avoid being too long and we refer to 
\cite{campana2022phd}. Then, computing the expectation, we get 
\begin{align}
\frac{d}{dt}\EE[p_ip_j](t)  =& \tfrac{1}{2}{\kappa} \left(\delta_{ij} -3\EE[p_ip_j](t)\right) \\
	\frac{d}{dt}\EE[p_i^3](t)  =& \tfrac{3}{2} \kappa \left(\EE[p_i](t) -2\EE[p_i^3](t)\right),
\end{align} 
that are easily integrated in \eqref{eq:pi_moments}.
 
\section{Complement on the splitting component schemes}\label{A:ana_complement}

\subsection{Solving the stochastic stretching scheme}\label{A:ana_bs}

We give some additional comments and results on the scheme \eqref{eq:scheme_BS_nonlinear.vec}, at least from the viewpoint of the strong convergence.  

It is known that, in such a situation of linear drift with inner sign in the equation \eqref{eq:BS_nonlinear}, the implicitation of this particular term in the scheme avoids introducing further constraint on the choice of the time-step $\Dt$, whereas an explicit version on that term converges with the same rate, but under the additional condition that $\Dt < ( \Lambda^2 \nus^2)^{-1}$, a threshold which in our case is proportional to $\tau_\eta$.

We evaluate the mean deviation of  $\|\nbpbs(t_{k+1})\|^2$  from unity: using Eq.~\eqref{eq:scheme_BS_nonlinear.vec}, we can rewrite the prediction step as 
\begin{align}\label{eq.appendix.scheme_BS_nonlinear.vec}
\begin{aligned}
&(1+\tfrac{1}{2}{\nus^2}\Lambda^2 \Dt)\nbpbs(t_{k+1}) \\
& = \hnbpbs(t_k) +  \Lambda\nus \,  \hnbpbs(t_k) \times \left(\Delta\Wcs_{t_{k+1}} \hnbpbs(t_k) \times    \hnbpbs(t_k) \right).   
\end{aligned}
\end{align}
Taking the expectation in \eqref{eq.appendix.scheme_BS_nonlinear.vec}, and since $\|\hnbpbs(t_k)\| =1$,  we immediately observe that 
\begin{align*}%\label{eq:moment_1_ap}
\|\EE\nbpbs(t_k)\| = (1+\tfrac{1}{2}{\nus^2} \Lambda^2 \Dt)^{-1}.
\end{align*}
Now, using the vectorial product property
\begin{align*}
&\|\hnbpbs(t_k) +  \Lambda\nus \,  \hnbpbs(t_k) \times \left({\Delta\Wc}^s_{t_{k+1}} \hnbpbs(t_k) \times    \hnbpbs(t_k) \right)    \|^2\\
& = 1 + \Lambda^2\nus^2 \| \Delta\Wc^s_{t_{k+1}} \hnbpbs(t_k) \times    \hnbpbs(t_k)|^2, 
\end{align*}
showing that the norm of right-hand side of  \eqref{eq.appendix.scheme_BS_nonlinear.vec} is never  smaller than 1, and so  that the scheme is always well defined.  Next, a straightforward computation shows that $\EE\| \Delta\Wc^s_{t_{k+1}} \hnbpbs(t_k)\times    \hnbpbs(t_k)\|^2 = \Dt$, and we get 
\begin{align*}
\EE [\|\nbpbs(t_{k+1})\|^2] =  (1 + {\nus^2} \Lambda^2  \Dt)(1+\tfrac{1}{2}{\nus^2} \Lambda^2 \Dt)^{-2}.
\end{align*}
Since the map $x\mapsto\frac{(1+2x)}{(1+x)^2}$ expands around zero as $1 -x^2 + 
\mathcal{O}(x^3)$, we get for the norm of the semi-implicit \EM scheme for Brownian stretching 
before the renormalization step 
\begin{align}\label{eq:norm_control_appendix}
\EE [\|\nbpbs(t_{k+1})\|^2] = 1 - \tfrac{1}{4}{\nus^4} \Lambda^4 \Dt^2 + \mathcal{O}(\Dt^3). 
\end{align}

For the analysis of the $L^2$-strong error of the scheme, we need also to evaluate the order of the quantity $\EE[(\|\nbpbs(t_{k+1})\| -1)^2]$. From the above, and using the bound ${1 + x^2} \leq (1 + \frac{x^2}{2} + \frac{x^4}{8})^2$, 
\begin{align*}
\begin{aligned}
& (\|\nbpbs(t_{k+1})\| -1)^2 \\
& \leq 
\tfrac{1}{4}\Lambda^4\nus^4(1 + \tfrac{1}{2}\Lambda^2   \nus^2  \Dt)^{-2}\\
&\quad\times \left( 
 \| \Delta\Wc^s_{t_{k+1}} \hnbpbs \times \hnbpbs(t_k)  \|^2  \right. \\
 & \qquad\qquad + \left. \Lambda^2\tfrac{1}{4}{\nus^2}\| \Delta\Wc^s_{t_{k+1}} \hnbpbs\times\hnbpbs(t_k)\|^4 + \Dt
\right)^2.
\end{aligned}
\end{align*}
We observe that, for $n\geq 1$ (by Jensen inequality) 
\begin{align*}
\begin{aligned}
& \EE[ \|\Delta\Wc^s_{t_{k+1}} \hnbpbs(t_k) \times    \hnbpbs(t_k)  \|^{2n}] \leq  \EE[ \|\Delta\Wc^s_{t_{k+1}} \hnbpbs(t_k)\|^{2n}]  \\
& \leq \EE \big( \EE[ \sum_{i,j} (\Delta\Wc^s_{ij,t_{k+1}})^2  \hnpbsj(t_k)^2 \big/ \mathcal{F}_{t_k}]
\big)^n = ( 2\Dt )^n. 
\end{aligned}
\end{align*}
Above, we have used the sub-conditioning with respect to the set of events  $\mathcal{F}_{t_k}$ of all Brownian motions past until time $t_k$, for whom $\hnbpbs(t_k)$ is adapted to. Coming back to 
$\EE[(\|\nbpbs(t_{k+1})\| -1)^2]$, we have obtained that 
\begin{align}\label{eq:appen:final_dt_norm_bs}
\begin{aligned}
\EE[(\|\nbpbs(t_{k+1})\| -1)^2] \leq {\Lambda^4\nus^4} \frac{9}{4} \Dt^2 + \mathcal{O}(\Dt^3).
\end{aligned}
\end{align}
Considering the mean square error at time $t_{k+1}$, we decompose it in two parts
\begin{align*}%\label{eq:two_part_str_er_bs}
&\EE[\|\hnbpbs(t_{k+1}) -\pb\subbs(t_{k+1})\|^2] \\
& \leq  \EE[\|\nbpbs(t_{k+1}) -\pb\subbs(t_{k+1})\|^2] + \EE[\|\hnbpbs(t_{k+1})-\nbpbs(t_{k+1}) \|^2].
\end{align*}
We immediately bound the second term, as
\begin{align*}
&\EE[\|\hnbpbs(t_{k+1})-\nbpbs(t_{k+1}) \|^2] \\
& = \EE[(\|\nbpbs(t_{k+1})\|-1)^2] \leq 2 {\nus^4} \Lambda^4 \Dt^2.
\end{align*} 
In particular, the terms  above are summable over $k$,  and the resulting sum still decreases with $\Dt$ with a rate of order 1.  For the first part, using the SDEs \eqref{eq:BS_nonlinear}  we get
\begin{align*}
\begin{aligned}
&(1 +  \tfrac{1}{2} \nus^2\Lambda^2 \Dt) (\nbpbs(t_{k+1}) -  \pb\subbs(t_{k+1})) \\
& = \hnbpbs(t_k) -  \pb\subbs(t_{k}) - \int_{t_k}^{t_{k+1}}  \tfrac{1}{2} \Lambda^2   \nus^2  (\pb\subbs(t_{k+1})  - \pb\subbs(r))  dr \\
& \quad  +    \nus\Lambda  \int_{t_k}^{t_{k+1}}  \Big(\hnbpbs(t_k) {d\Wc}^s_r \,\hnbpbs(t_k)\Big)\hnbpbs(t_k) \\
 &\qquad\qquad\qquad\qquad\qquad -  \Big(\pb\subbs(r){d\Wc}^s_r \,\pb\subbs(r)\Big)\pb\subbs(r)
\end{aligned}
\end{align*}
from which we deduce that 
\begin{align*}
\begin{aligned}
&(1 +  \tfrac{1}{2} \Lambda^2   \nus^2 )^2 \EE\|\nbpbs(t_{k+1}) -  \pb\subbs(t_{k+1})\|^2 \\
& \leq 2\EE\|\hnbpbs(t_k) -  \pb\subbs(t_{k})\|^2  + 2 \mathcal{E}({t_k},{t_{k+1}}). 
\end{aligned}
\end{align*}
The last term, usually called the local error, takes the particular form 
\begin{align*}
\begin{aligned}
\mathcal{E}({t_k},{t_{k+1}})
& =  \EE\| 
 \int_{t_k}^{t_{k+1}}  \tfrac{1}{2} \Lambda^2   \nus^2  (\pb\subbs(t_{k+1})  - \pb\subbs(r))  dr  \\
&\qquad +  \Lambda\nus  \int_{t_k}^{t_{k+1}}  \Big(\hnbpbs(t_k) {d\Wc}^s_r \,\hnbpbs(t_k)\Big)\hnbpbs(t_k)  \\
&\qquad \qquad \qquad-  \Big(\pb\subbs(r){d\Wc}^s_r \,\pb\subbs(r)\Big)\pb\subbs(r)\|^2
\end{aligned}
\end{align*}
The first term bounds easily with $C \nus^4\Lambda^4    \Dt^2$. With the \Ito isometry (see \ref{A:TumbSpin_HIT}), the second  bounds with $ C \Dt \EE\|\hnbpbs(t_k) -  \pb\subbs(t_{k})\|^2 + C\Dt^2$. 

Coming back to the error estimation, we have obtained
\begin{align*}%\label{eq:two_part_str_er_bs2}
&\EE[\|\hnbpbs(t_{k+1}) -\pb\subbs(t_{k+1})\|^2] \\
& \leq  ( \tfrac{1}{(1 +  \nus^2\Lambda^2\Dt) } + C\Dt) \EE\|\hnbpbs(t_k) -  \pb\subbs(t_{k})\|^2  +  C \nus^4\Lambda^4    \Dt^2.
\end{align*}
Thus we obtain the standard $L^2$-norm (strong) convergence 
\begin{align}\label{eq:bs_strong_error}
\EE[\|\hnbpbs(t_k) -\pb\subbs(t_{k})\|^2] \leq   C(t_k) \nus^4\Lambda^4 \Dt,
\end{align}
for an updated constant $C$ independent of $\Dt$. 

\subsection{On the Brownian rotation scheme} \label{A:ana_q}

Stochastic diffusion models on the sphere \eqref{eq:BR_part} or \eqref{eq:BR_cross_strat} arise in several contexts (transport of biological substances at 
cellular level \cite{krishna2000brownian}, swimming of bacteria motion \cite{li2008amplified}, 
polymer systems \cite{snook2006langevin} and global migration patterns of marine mammals 
\cite{brillingerparticle}). Simulation algorithms for Brownian on the sphere  exist already. 
Some  are based on the approximation of its transition density 
(see \eg \cite{nissfolk2003brownian, carlsson2010algorithm}), some are based on exact transformations \cite{mijatovic2018note}, while others are using the distribution of solid angles \cite{krishna2000brownian, brillingerparticle}.  These methods are mainly based on \textit{intrinsic} representation of $\pb\subba(t)$ on the manifold. However, our splitting approach forces us to maintain the natural  dependence between the symmetric and antisymmetric increments of the $(3\times 3)$-- $\Wc$ matrix, prohibiting any replacement of \eqref{eq:BR_cross_strat}, by an independent random draw of the Brownian increment distribution on the sphere.

\medskip

From the scheme definition in \eqref{eq:scheme_quaternions}, 
\begin{align*}
\EE [\|\Delta\tilde {\qu}_{k+1}\|^2]
 = \frac{1 + \tfrac{3}{8} \nua^2 \Dt  }{(1+\tfrac{3}{16} \nua^2 \Delta t)^2}.
\end{align*}
Following what was done for \eqref{eq:norm_control_appendix}, we obtain 
\begin{align}
\EE [\|\Delta\tilde {\bm\qu}_{k+1}\|^2] =  1 - \frac{9 \nua^4}{16^2} \Dt^2 + \mathcal{O}(\Dt^3). 
\end{align}

\medskip
~

\begin{minipage}{\linewidth}
	\begin{center}
		\centering
		\includegraphics[width=0.5\textwidth]{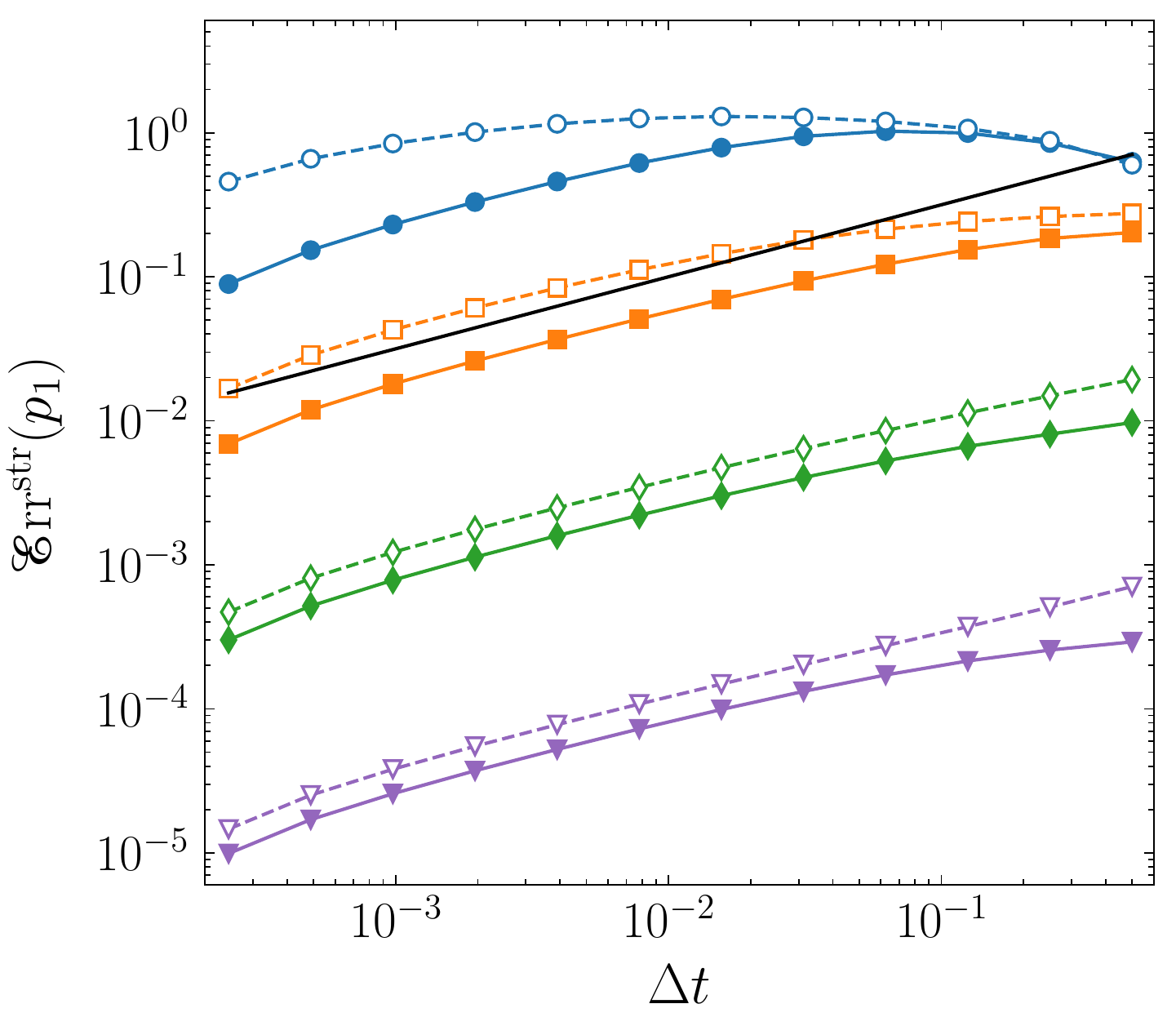}\hfil
		\includegraphics[width=0.5\textwidth]{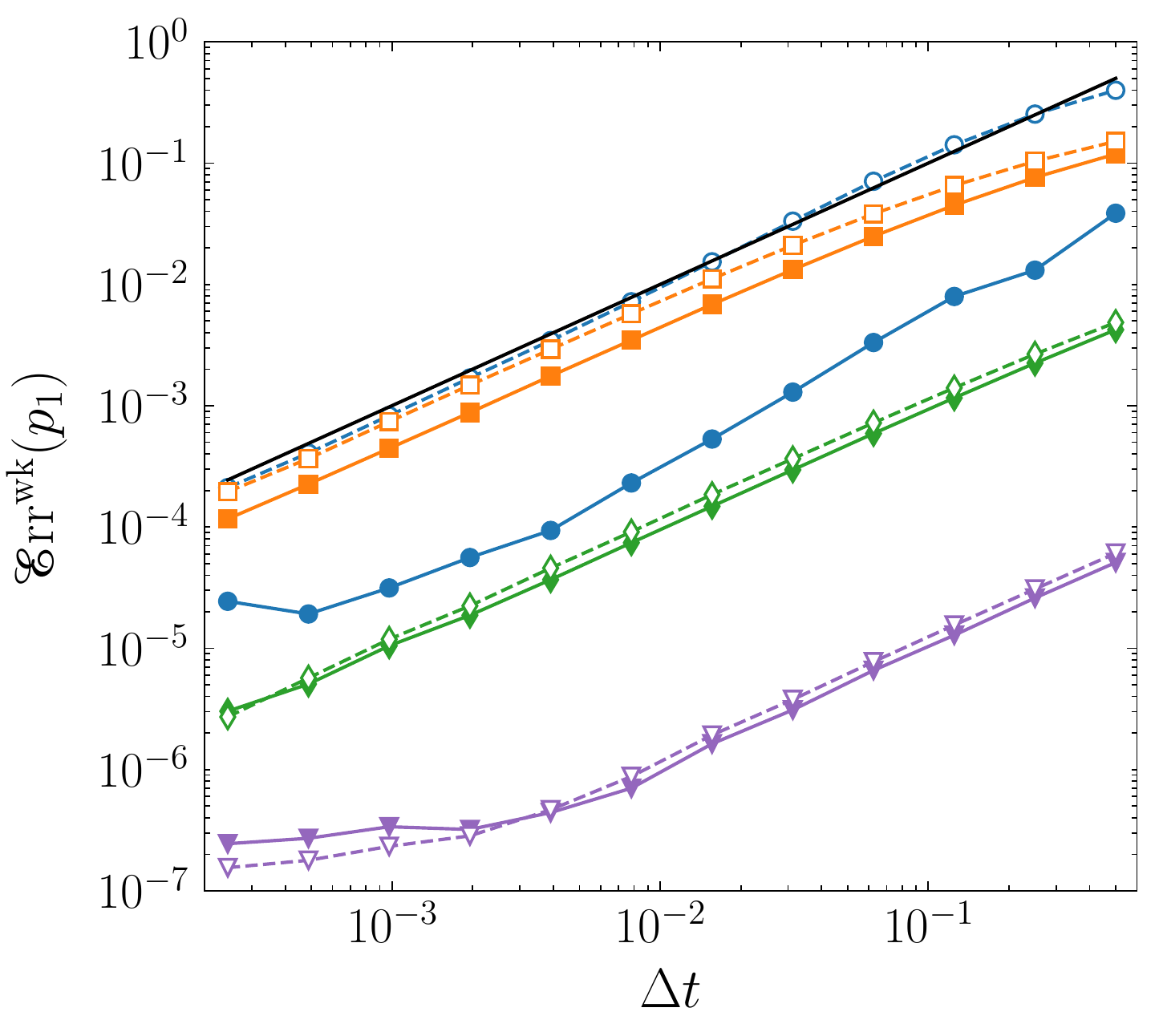}\\
		\captionof{figure}{\label{fig:strong_weak_BR_BS} 
			Comparison between the stochastic stretching (empty markers dotted lines) in 
			Eq.~\eqref{eq:scheme_BS_nonlinear.vec} and stochastic rotation sub-step (marked lines) in 
			Eq.~\eqref{eq:euler_rodriguez} for the strong error ($\ErrS$) (\textbf{left panel}) and 
			weak error ($\ErrW$) (\textbf{right panel}) of $p_1$ against the time step $\Delta t$. 
			Colors correspond to different values of $\taukol$: $0.01$ (\textcolor{pyblue}{blue}), $0.1$ 
			(\textcolor{pyorange}{orange}), $1$ (\textcolor{pygreen}{green}), $10$ 		
			(\textcolor{pyviolet}{violet}). The initial condition of particle orientation is 
			$\bm{p}(0)=(1,0,0)$. Simulation performed with a shape parameter $\Lambda=1$, 
			$N_p=5\times10^8$ particles, $\alpha=1$ and $T=0.5$. The smallest $\Dt$ is $2^{-12}$ and reference 
			trajectories  are computed with $\Dt=2^{-13}$.}
	\end{center}
\end{minipage}

%******************************************************************************
% Bibliography
%******************************************************************************

%%       \citet{<label>} ==> Jones et al. [21]
%%       \citep{<label>} ==> [21]
%%

\bibliographystyle{elsarticle-num-names}
%\bibliography{biblio}

%%%%%%%%%%%%%%%%%%%%%%%%%%%%%%%%%%%%%%%%%%%%%%%%%%%%%%%%%%%%%%%%%
%%%%%%%%%%%%%%%%%%%%%%%%%%%%%%%%%%%%%%%%%%%%%%%%%%%%%%%%%%%%%%%%%
\end{document}